# Asymptotic Completeness for Rayleigh Scattering

J. Fröhlich[1*], M. Griesemer[2†]and B. Schlein[1‡]

1. Theoretical Physics, ETH–Hönggerberg,
CH–8093 Zürich, Switzerland

2. Department of Mathematics, University of Alabama at Birmingham,
Birmingham, AL 35294

30 March, 2001

### Abstract

It is expected that the state of an atom or molecule, initially put into an excited state with an energy *below* the ionization threshold, relaxes to a groundstate by spontaneous emission of photons which propagate to spatial infinity. In this paper, this picture is established for a large class of models of non-relativistic atoms and molecules coupled to the quantized radiation field, but with the simplifying feature that an (arbitrarily tiny, but positive) infrared cutoff is imposed on the interaction Hamiltonian.

This result relies on a proof of *asymptotic completeness* for Rayleigh scattering of light on an atom. We establish asymptotic completeness of Rayleigh scattering for a class of model Hamiltonians with the features that the atomic Hamiltonian has point spectrum coexisting with absolutely continuous spectrum, and that either an infrared cutoff is imposed on the interaction Hamiltonian or photons are treated as massive particles.

We show that, for models of *massless* photons, the spectrum of the Hamiltonian strictly below the ionization threshold is *purely continuous*, except for the groundstate energy.

## 1 Introduction

Ever since the inception of the quantum theory of atoms interacting with the quantized radiation field, theoreticians have expected that when an atom (with an infinitely heavy nucleus) in a state where all electrons are bound to the nucleus is targeted by a finite number of photons in such a way that the total energy of the composed system remains below the ionization threshold of the atom the following physical processes unfold: First, some of the electrons in the shells of the atom are lifted into an excited state by absorbing incoming photons; but, since the total energy is below the ionization threshold, they remain bound to the nucleus. As time goes on, the excited state relaxes to a groundstate of the atom by spontaneous emission of photons, which propagate essentially freely to spatial infinity. Thus, asymptotically, the state of the total system, atom plus quantized radiation field, describes an atom in its ground state and a cloud of photons escaping to infinity with the velocity of light.

---

*juerg@itp.phys.ethz.ch

†marcel@math.uab.edu

‡schlein@itp.phys.ethz.ch

Relaxation of an excited initial state to a groundstate by emission of outgoing radiation is the simplest example of an *"irreversible process"*, accompanied by information loss at infinity, occuring in an open quantum system with infinitely many degrees of freedom. It would seem worthwhile to attempt to understand this process mathematically precisely; (see Sect. 10).

The picture described above suggests that the scattering operator describing Rayleigh scattering of light off an atom with a static nucleus, i.e., the scattering operator restricted to the subspace of states with energies below the ionization threshold of the atom, is *unitary*; (see Sect. 9 and 10). If true - one says that *"asymptotic completeness"* (AC) is valid for Rayleigh scattering.

In attempting to establish this picture mathematically, one faces the problem that, in the scattering of light at an atom, an arbitrarily large number of soft photons of arbitrarily small total energy can, in priciple, be produced (in processes of high order in the feinstructure constant). Perturbative calculations of scattering amplitudes suggest, however, that in the analysis of Rayleigh scattering of light at an atom with a static nucleus, one does *not* encouter a genuine infrared catastrophe of the kind first described by Bloch and Nordsieck. Yet, the mathematical problems connected with controlling very large numbers of very soft photons in a mathematically rigorous, non-perturbative way are quite substantial and have not been fully mastered, yet.

In order to simplify matters to a manageable size, we propose to study Rayleigh scattering and the phenomenon of relaxation to a ground state for models of massive photons and for models of *massless* photons with an infrared cutoff. In this paper, results in this direction are proven.

In order to avoid inessential technical complications, we consider models of *"scalar photons"*, *bosons*. But our analysis can be extended to the quantum electrodynamics of non-relativistic electrons (bound to a static nucleus) interacting with the quantized electromagnetic field, provided we work within the dipole approximation and impose an (arbitrarily tiny) infrared cutoff on the electron-field interaction.

We plan to study more difficult scattering problems in similar models and their consequences for "irreversible phenomena" in future papers.

Next, we describe our main results in more detail. To avoid starting with a list of assumptions, we formulate our results for a concrete, simple model, which is physically relevant and captures the main features of the problems we propose to solve. For precise assumptions and other models see Section 3.

Consider $N$ non-relativistic electrons subject to a potential $V$, which may also include two–body interactions. The electrons are linearly coupled to a quantized field of relativistic bosons. The Hamilton operator of this system is

$$H = K \otimes 1 + 1 \otimes d\Gamma(\omega) + \phi(G) \tag{1}$$

and acts on the Hilbert space $\mathcal{H} = \mathcal{H}_{el} \otimes \mathcal{F}$, where $\mathcal{H}_{el} = \wedge^N L^2(\mathbb{R}^3; \mathbb{C}^2)$ is the anti-symmetric tensor product of $N$ copies of $L^2(\mathbb{R}^3; \mathbb{C}^2)$, and $\mathcal{F}$ is the bosonic Fock space over $L^2(\mathbb{R}^3)$. The operator $K$ describes the time evolution of the electrons without radiation and is given by $K = -\Delta + V$, where $\Delta$ denotes the Laplacian on $\mathbb{R}^{3N}$. (In our units, $\hbar$ and the electron mass are equal to one.) Electron spin will be neglected henceforth.

We assume that $V_-$, the negative part of $V$, is infinitesimally form–bounded with respect to $-\Delta$, that $K$ is essentially self-adjoint on $C_0^\infty(\mathbb{R}^{3N})$, and that

$$\inf \sigma_{ess}(K) > \inf \sigma(K),$$

i.e., that $K$ has bound states. In particular, $\inf \sigma(K)$ is an isolated eigenvalue of $K$.

The operator $d\Gamma(\omega)$ describes the energy of free bosons. Formally

$$d\Gamma(\omega) = \int_{\mathbb{R}^3} dk\, \omega(k) a^*(k) a(k)$$

where $a(k)$ and $a^*(k)$ are the usual annihilation- and creation operators (operator–valued distributions) depending on the wave vector $k$, and $\omega(k) = \sqrt{k^2 + m^2}$ is the energy of a relativistic particle with momentum $k$ and mass $m \geq 0$. The operator

$$\phi(G) = \int dk \left( \overline{G_x(k)} a(k) + G_x(k) a^*(k) \right)$$

describes the interaction between electrons and bosons. In this introduction, we choose $G_x(k) = \sum_{i=1}^N e^{-ik \cdot x_i} \kappa(k)$, and $\kappa(k) \in C_0^\infty(\mathbb{R}^3)$. If $m = 0$ we cut off the infrared modes from the interaction by assuming that $\kappa(k) = 0$, for $|k|$ small. Thus, in any case

$$\inf_{k \in \mathrm{supp}\,\kappa} \omega(k) > 0. \tag{2}$$

Without this assumption, it is presently not known how to control the number of soft bosons produced in the course of the time evolution.

Let $\Sigma$ denote the ionization threshold. This is the smallest energy the system can reach when one or several electrons have been moved to infinity. States with energy below $\Sigma$ are exponentially localized w.r. to the electron coordinates. More precisely

$$\| e^{\alpha|x|} E_\Delta(H) \| < \infty$$

if $\Delta \subset (-\infty, \Sigma)$ and $\sup \Delta + \alpha^2 < \Sigma$. Clearly $\Sigma \geq \inf \sigma(H)$, and, for one-electron atoms and if the coupling is weak enough, it is known that $\Sigma > \inf \sigma(H)$ [GLL00, BFS98]. Of course, one expects $\Sigma > \inf \sigma(H)$ for all neutral atoms and molecules. Note that $\Sigma = \infty$ if $V(x) \to \infty$, for $|x| \to \infty$, i.e., when $\sigma(K)$ is discrete. This situation is included in our analysis, but one of our main points is to prove results on Rayleigh scattering when $\Sigma < \infty$.

For states $\varphi \in \mathcal{H}$ with energy below $\Sigma$, that is $\varphi = E_{(-\infty, \Sigma)}(H)\varphi$, one expects that $\varphi_t = e^{-iHt}\varphi$ is well approximated, in the distant future, by a linear combination of states of the form

$$a^*(h_{1,t}) \dots a^*(h_{n,t}) e^{-iEt} \varphi_b \tag{3}$$

where $\varphi_b$ is a bound state of $H$, $H\varphi_b = E\varphi_b$, $h_{j,t} = e^{-i\omega t}h_j$, $j = 1, \dots, n$, are one–particle wave functions of freely propagating bosons, and $a^*(h) = \int \overline{h(k)}a(k)\,dk$, $a(h) := (a^*(h))^*$. This property is called *asymptotic completeness (AC) for Rayleigh scattering*. It asserts, in particular, that the asymptotic dynamics of escaping photons is well approximated by their free dynamics. This requires that the strength of the interaction between ballistically moving bosons and electrons decays at an integrable rate. In our model of non-relativistic electrons bound to static nuclei this is true, thanks to the spatial localization of the electrons. For massless bosons, it follows more generally from the fact that the propagation velocity of electrons is strictly smaller than the velocity of the bosons, i.e., the velocity of light, [FGS00].

To give a mathematically more precise formulation of AC, let us introduce asymptotic creation operators $a^*_+(h)$. Let $\varphi = E_\lambda(H)\varphi$, $h_j \in L^2(\mathbb{R}^3)$, $j = 1, \dots, n$ and $M_j = \sup\{\omega(k)|h_j(k) \neq 0\}$. Then

$$a^*_+(h_1) \dots a^*_+(h_n)\varphi = \lim_{t \to \infty} e^{iHt} a^*(h_{1,t}) \dots a^*(h_{n,t}) e^{-iHt}\varphi \tag{4}$$

exists if

$$\lambda + \sum_{i=1}^n M(h_i) < \Sigma. \tag{5}$$

Asymptotic completeness of Rayleigh scattering is the statement that linear combinations of states of the form (4), (5), with $\varphi \in \mathcal{H}_{pp}(H)$, are dense in $E_{(-\infty, \Sigma)}(H)\mathcal{H}$. Assuming (2), we prove AC for all $m \geq 0$, with an infrared cutoff imposed when $m = 0$.

If $m = 0$ one can show more: If the Pauli principle for the electrons is neglected then $H$ has a unique ground state $\varphi_0$ [BFS98, GLL00] but no other stationary states with energy below $\inf \sigma_{ess}(K) - \varepsilon$, $\varepsilon > 0$, provided $g$ is sufficiently small and the life times of all excited states of $H_{g=0}$, as computed by Fermi's Golden Rule, are finite. This follows from results of Bach et al. [BFSS99] together with an argument given in the present paper, which excludes eigenvalues close to $\inf \sigma(H)$. As a consequence, states of the form

$$a^*_+(h_1) \dots a^*_+(h_n)\varphi_0, \qquad \inf \sigma(H) + \sum M(h_i) \leq \inf \sigma_{ess}(K) - \varepsilon \tag{6}$$

are dense in $E_{\inf \sigma_{ess}(K) - \varepsilon}(H)\mathcal{H}$, for some $\varepsilon > 0$ depending on the coupling constant. Moreover, we show that every state $\psi_t \in E_{\inf \sigma_{ess}(K) - \varepsilon}(H)\mathcal{H}$ eventually relaxes to the ground state $\varphi_0$, in the following sense: Let $\mathcal{A}$ denote the $C^*$ algebra generated by the Weyl operators $e^{i\phi(h)}$ where $\phi(h) = a(h) + a^*(h)$, and $h \in \mathcal{S}(\mathbb{R}^3)$, the Schwartz space of test functions. By taking sums of tensor products of operators in $\mathcal{A}$ with arbitrary bounded operators acting on the $N$–electron Hilbert space one obtains a $C^*$ algebra $\tilde{\mathcal{A}}$.

"Relaxation of $\psi_t$ to the ground state $\varphi_0$" is the statement that

$$\lim_{t\to\infty} \langle \psi_t, A\psi_t \rangle = \langle \varphi_0, A\varphi_0 \rangle \langle \psi, \psi \rangle, \tag{7}$$

for all operator $A \in \tilde{\mathcal{A}}$, and for all $\psi \in E_{\inf \sigma_{ess}(K)-\varepsilon}(H)\mathcal{H}$. This is our second main result. It essentially follows from asymptotic completeness and, of course, from the absence of eigenvalues in $(\inf \sigma(H), \inf \sigma_{ess}(K) - \varepsilon]$.

Asymptotic Completeness for *massive bosons* was previously established by Derezinski and Gérard, for confined electrons (i.e., $\Sigma = \infty$) and under a somewhat unphysical short-range assumption [DG99]. From this important paper, and from [DG00], we have learned how to translate techniques from $N$-body quantum theory to quantum field theory. Before [DG99], Arai had established AC in the standard model of non-relativistic QED in the dipole–approximation and with $V(x) = x^2$, a model which is explicitly soluble [Ara83]. Later, Spohn extended this result, using the Dyson series, to include potentials which are small perturbations of $x^2$ [Spo97].

Our proof of asymptotic completeness adapts methods and techniques from the scattering theory of $N$-particle Schrödinger operators to the present situation. In particular, we use a Mourre estimate and propagation estimates, and we rely on localization techniques in bosonic configuration space. As in the more recent papers on $N$-body quantum scattering, we derive AC from the fact that the mean square diameter of $\langle d\Gamma(y^2) \rangle_{\psi_t}$ of a given state $\psi_t$, with $y$ the position operator in bosonic configuration space, diverges like $t^2$ if $\psi$ is away, in energy, from thresholds and eigenvalues. Correspondingly, a central object in our proof is an asymptotic observable $W$ that measures the square of the asymptotic velocities of the escaping photons. That is,

$$\begin{aligned} \langle W \rangle &= \lim_{t\to\infty} \langle d\Gamma(y^2/2t^2) \rangle_t \\ &= \lim_{t\to\infty} \frac{d}{dt} \langle d\Gamma(y^2/2t) \rangle_t \end{aligned} \tag{8}$$

Thanks to the ballistic escape property mentioned above, $W$ is positive and thus invertible, on suitable spectral subspaces. We construct a Deift-Simon wave operator $W_+$ with the property that $W_+ W^{-1}$ is a right-inverse of an extended wave operator on a dense subspace of $\mathcal{H}_{cont}(H) \cap E_{(-\infty, \Sigma)}(H)\mathcal{H}$, the orthogonal complement of all eigenvectors. The proof is completed with an inductive argument explained further below.

Let us temporarily assume that the interaction $\phi(G)$ vanishes, in order to explain the ideas underlying the construction of $W$ and $W_+$ in their purest form. The main observation is that

$$D^2 \frac{y^2}{2t} = \frac{1}{t}(\nabla \omega - y)^2 \geq 0, \tag{9}$$

where $D$ denotes the Heisenberg derivative $[i\omega, .] + \partial/\partial t$. As a consequence the time derivative of the expression (8), whose limit is $\langle W \rangle$, is non-negative. Since $d/dt\langle d\Gamma(y^2/2t) \rangle_t$ is bounded uniformly in time, it follows that $d^2/dt^2\langle d\Gamma(y^2/2t) \rangle_t = \langle d\Gamma(D^2[y^2/2t]) \rangle_t$ is integrable. This *propagation estimate*, with small modifications to accommodate the interaction, proves existence of $\langle W \rangle$ and suffices to establish existence of $W$ as a strong limit. Existence of $W_+$ requires, in addition, some geometry in bosonic configuration space.

Once these asymptotic operators are constructed, AC follows by induction in the number of energy intervals of length $m$, the smallest energy of a boson. Assuming that AC holds on the spectral subspace of $H$ corresponding to energies below $\min(\Sigma, m(n-1))$, we prove AC for energies below $\min(\Sigma, mn)$. Roughly speaking, the positivity of $W$ on suitable spectral subspaces $E_\Delta(H)\mathcal{H}$, where $\Delta \subset \min(\Sigma, mn)$, allows us to show that at least one boson of a given state $\varphi \in E_\Delta(H)\mathcal{H}$ escapes to infinity. It thus carries away an energy of at least $m$. The energy distribution of the remaining system is contained in $(-\infty, m(n-1))$, where asymptotic completeness holds by assumption, and hence $\varphi_t$, for large $t$, is of the form (3). Obviously the positivity of the boson mass, or condition (2), in the case of more general dispersion relations, is absolutely crucial in this argument.

Our strategy and the constructions of $W$ and $W_+$ are strongly inspired by ideas and constructions developed by Graf and Schenker for $N$-body quantum scattering theory [GS97]. The Mourre estimate we use is essentially the one of Derezinsky and Gérard [DG99]. We follow closely the notation of [DG99]; but, otherwise, there are only few similarities between our approach to AC and the one in [DG99].

Our paper is organized as follows.

In Sect. 2, we consider the quantum theory of the bosons. We briefly review the standard formalism of second quantization and introduce some basic notions that are useful in scattering theory.

In Sect. 3, we describe the physical systems and define the models studied in this paper. We formulate some basic assumptions on the Hamiltonians generating the dynamics in these models which will be important to gain mathematical control over Rayleigh scattering. We describe some concrete examples of models.

In Sect. 4, we review and prove results on spectral properties of the Hamiltonians of our models. In particular, we recapitulate a theorem on the existence of a ground state and on the location of the essential spectrum. We state a Mourre estimate and use it to establish properties of the continuous and point spectrum and to prove a virial theorem.

In Sect. 5, we construct the Møller wave operators of our models on spectral subspaces corresponding to bound electrons, using a variant of Cook's argument; see also [FGS00] for more detailed results.

Sect. 6 contains our basic propagation estimates needed for the construction of an asymptotic observable and of a Deift-Simon wave operator.

An asymptotic observable, $W$, is constructed in Sect. 7 and shown to be selfadjoint, positive on appropriate spectral subspaces, and to commute with the Hamiltonian $H$.

In Sect. 8, a Deift-Simon wave operator is constructed and shown to invert, with respect to $W$, an extended variant of the Møller wave operator on spectral subspaces where $W$ is positive.

By combining the results of previous sections, asymptotic completeness for Rayleigh scattering is established in Sect. 9 with the help of an inductive argument in the number of asymptotic bosons.

In Sect. 10, models of *massless* bosons with an infrared cutoff are analyzed, and the phenomenon of relaxation to a groundstate is exhibited. A novel positive-commutator estimate is proven which, together with results in [BFSS99], excludes the existence of point spectrum above the groundstate energy (and below the ionization threshold).

Our most difficult and innovative results appear in Sects. 6 through 10. Various technical arguments are deferred to appendices, the most important ones being Appendices E through G.

*Acknowledgements.* We thank V. Bach, Chr. Gérard, G.-M. Graf and I.M. Sigal for many useful discussions of problems related to those studied in this paper.

# 2    Fock Space and Second Quantization

The natural Hilbert space of states of the radiation field is the Fock space. Let $\mathfrak{h}$ be a complex Hilbert space and let $\otimes_s^n \mathfrak{h}$ denote the $n$-fold symmetric tensor product of $\mathfrak{h}$. The bosonic Fock space over $\mathfrak{h}$

$$\mathcal{F} = \mathcal{F}(\mathfrak{h}) = \oplus_{n \geq 0} \otimes_s^n \mathfrak{h}$$

is the space of sequences $\varphi = (\varphi_n)_{n \geq 0}$, with $\varphi_0 \in \mathbb{C}$, $\varphi_n \in \otimes_s^n \mathfrak{h}$, and with the scalar product given by

$$\langle \varphi, \psi \rangle := \sum_{n \geq 0} (\varphi_n, \psi_n),$$

where $(\varphi_n, \psi_n)$ denotes the inner product in $\otimes^n \mathfrak{h}$. The vector $\Omega = (1, 0, \dots) \in \mathcal{F}$ is called the vacuum. By $\mathcal{F}_0 \subset \mathcal{F}$ we denote the dense subspace of vectors $\varphi$ for which $\varphi_n = 0$, for all but finitely many $n$. The number operator $N$ on $\mathcal{F}$ is defined by $(N\varphi)_n = n\varphi_n$.

## 2.1    Creation- and Annihilation Operators

The creation operator $a^*(h)$, $h \in \mathfrak{h}$, on $\mathcal{F}$ is defined by

$$a^*(h)\varphi = \sqrt{n}\, S(h \otimes \varphi), \qquad \text{for } \varphi \in \otimes_s^{n-1} \mathfrak{h},$$

and extended by linearity to $\mathcal{F}_0$. Here $S \in \mathbf{B}(\otimes^n \mathfrak{h})$ denotes the orthogonal projection onto the subspace $\otimes_s^n \mathfrak{h} \subset \otimes^n \mathfrak{h}$. The annihilation operator $a(h)$ is the adjoint of $a^*(h)$ restricted to $\mathcal{F}_0$. Creation- and annihilation operators satisfy the canonical commutation relations (CCR)

$$[a(g), a^*(h)] = (g, h), \qquad [a^\#(g), a^\#(h)] = 0.$$

In particular $[a(h), a^*(h)] = \|h\|^2$, which implies that the graph norms associated with the closable operators $a(h)$ and $a^*(h)$ are equivalent. It follows that the closures of $a(h)$ and $a^*(h)$ have the same domain. On this common domain we define

$$\phi(h) = \frac{1}{\sqrt{2}}(a(h) + a^*(h)). \tag{10}$$

The creation- and annihilation operators, and thus $\phi(h)$, are bounded relative to the square root of the number operator:

$$\|a^{\#}(h)(N+1)^{-1/2}\| \leq \|h\|$$
$$\|(N+1)^{-1/2}a^{\#}(h)\| \leq \|h\|. \tag{11}$$

More generally, for any $p \in \mathbb{R}$ and any integer $n$

$$\|(N+1)^p a^{\#}(h_1)\ldots a^{\#}(h_n)(N+1)^{-p-n/2}\| \leq C_{n,p}\|h_1\| \cdot \ldots \cdot \|h_n\|. \tag{12}$$

This follows from $a^*(h)N = (N-1)a^*(h)$, $a(h)N = (N+1)a(h)$, and from (11).

## 2.2   The Functor $\Gamma$

Let $\mathfrak{h}_1$ and $\mathfrak{h}_2$ be two Hilbert spaces and let $b \in \mathbf{B}(\mathfrak{h}_1, \mathfrak{h}_2)$. We define

$$\Gamma(b) \;:\; \mathcal{F}(\mathfrak{h}_1) \to \mathcal{F}(\mathfrak{h}_2)$$
$$\Gamma(b){\restriction} \otimes_s^n \mathfrak{h}_1 = b \otimes \ldots \otimes b.$$

In general $\Gamma(b)$ is unbounded but if $\|b\| \leq 1$ then $\|\Gamma(b)\| \leq 1$. From the definition of $a^*(h)$ it easily follows that

$$\Gamma(b)a^*(h) = a^*(bh)\Gamma(b), \qquad h \in \mathfrak{h}_1. \tag{13}$$
$$\Gamma(b)a(b^*h) = a(h)\Gamma(b), \qquad h \in \mathfrak{h}_2. \tag{14}$$

where (14) is derived by taking the adjoint of (13). If $b^*b = 1$ on $\mathfrak{h}_1$ then these equations imply that

$$\Gamma(b)a(h) = a(bh)\Gamma(b) \qquad h \in \mathfrak{h}_1 \tag{15}$$
$$\Gamma(b)\phi(h) = \phi(bh)\Gamma(b) \qquad h \in \mathfrak{h}_1. \tag{16}$$

## 2.3   The Operator $d\Gamma(b)$

Let $b$ be an operator on $\mathfrak{h}$. Then

$$d\Gamma(b) \;:\; \mathcal{F}(\mathfrak{h}) \to \mathcal{F}(\mathfrak{h})$$
$$d\Gamma(b){\restriction} \otimes_s^n \mathfrak{h} = \sum_{i=1}^{n}(1 \otimes \ldots b \otimes \ldots 1).$$

For example $N = d\Gamma(1)$. From the definition of $a^*(h)$ we get

$$[d\Gamma(b), a^*(h)] = \;\;a^*(bh)$$
$$[d\Gamma(b), a(h)] = -a(b^*h),$$

where the second equation follows from the first one by taking the adjoint. If $b = b^*$ then

$$i[d\Gamma(b), \phi(h)] = \phi(ibh). \tag{17}$$

Note that $\|d\Gamma(b)(N+1)^{-1}\| \leq \|b\|$.

## 2.4 The Operator $\mathrm{d}\Gamma(a, b)$

Suppose $a, b \in \mathbf{B}(\mathfrak{h}_1, \mathfrak{h}_2)$. Then

$$\mathrm{d}\Gamma(a, b) \; : \; \mathcal{F}(\mathfrak{h}_1) \to \mathcal{F}(\mathfrak{h}_2)$$

$$\mathrm{d}\Gamma(a, b)\!\upharpoonright \otimes_s^n \mathfrak{h} = \sum_{j=1}^{n} (\underbrace{a \otimes \ldots a}_{j-1} \otimes b \otimes \underbrace{a \otimes \ldots a}_{n-j-1}).$$

For $a, b \in \mathbf{B}(\mathfrak{h})$ this definition is motivated by

$$\Gamma(a)\mathrm{d}\Gamma(b) = \mathrm{d}\Gamma(a, ab), \qquad \text{and} \qquad [\Gamma(a), \mathrm{d}\Gamma(b)] = \mathrm{d}\Gamma(a, [a, b]). \tag{18}$$

If $\|a\| \le 1$ then $\|\mathrm{d}\Gamma(a, b)(N+1)^{-1}\| \le \|b\|$.

## 2.5 The Tensor Product of two Fock Spaces

Let $\mathfrak{h}_1$ and $\mathfrak{h}_2$ be two Hilbert spaces. We define a linear operator $U : \mathcal{F}(\mathfrak{h}_1 \oplus \mathfrak{h}_2) \to \mathcal{F}(\mathfrak{h}_1) \otimes \mathcal{F}(\mathfrak{h}_2)$ by

$$\begin{aligned} U\Omega &= \Omega \otimes \Omega \\ Ua^*(h) &= [a^*(h_{(0)}) \otimes 1 + 1 \otimes a^*(h_{(\infty)})]U \qquad \text{for } h = (h_{(0)}, h_{(\infty)}) \in \mathfrak{h}_1 \oplus \mathfrak{h}_2. \end{aligned} \tag{19}$$

This defines $U$ on finite linear combinations of vectors of the form $a^*(h_1) \ldots a^*(h_n)\Omega$. From the CCRs it follows that $U$ is isometric. Its closure is isometric and onto, hence unitary. It follows that

$$Ua(h) = [a(h_{(0)}) \otimes 1 + 1 \otimes a(h_{(\infty)})]U. \tag{20}$$

Furthermore we note that

$$U\mathrm{d}\Gamma(b) = [\mathrm{d}\Gamma(b_0) \otimes 1 + 1 \otimes \mathrm{d}\Gamma(b_\infty)]U \quad \text{if} \quad b = \begin{pmatrix} b_0 & 0 \\ 0 & b_\infty \end{pmatrix} \tag{21}$$

For example $UN = (N_0 + N_\infty)U$ where $N_0 = N \otimes 1$ and $N_\infty = 1 \otimes N$.

Let $\mathcal{F}_n = \otimes_s^n \mathfrak{h}$ and let $P_n$ be the projection from $\mathcal{F} = \oplus_{n \ge 0} \mathcal{F}_n$ onto $\mathcal{F}_n$. Then the tensor product $\mathcal{F} \otimes \mathcal{F}$ is norm-isomorphic to $\oplus_{n \ge 0} \oplus_{k=0}^{n} \mathcal{F}_{n-k} \otimes \mathcal{F}_k$, the corresponding isomorphism being given by $\varphi \mapsto (\varphi_{n,k})_{n \ge 0, \, k=0..n}$ where $\varphi_{n,k} = (P_{n-k} \otimes P_k)\varphi$. In this representation of $\mathcal{F} \otimes \mathcal{F}$ and with $p_i(h_{(0)}, h_{(\infty)}) = h_{(i)}$, $U$ becomes

$$U\!\upharpoonright \otimes_s^n (\mathfrak{h} \oplus \mathfrak{h}) = \sum_{k=0}^{n} \binom{n}{k}^{1/2} \underbrace{p_0 \otimes \ldots \otimes p_0}_{n-k \text{ factors}} \otimes \underbrace{p_\infty \otimes \ldots \otimes p_\infty}_{k \text{ factors}}. \tag{22}$$

## 2.6 Factorizing Fock Space in a Tensor Product

Suppose $j_0$ and $j_\infty$ are linear operators on $\mathfrak{h}$ and $j : \mathfrak{h} \to \mathfrak{h} \oplus \mathfrak{h}$ is defined by $jh = (j_0 h, j_\infty h)$, $h \in \mathfrak{h}$. Then $j^*(h_1, h_2) = j_0^* h_1 + j_\infty^* h_2$ and consequently $j^* j = j_0^* j_0 + j_\infty^* j_\infty$. On the level of Fock spaces, $\Gamma(j) : \mathcal{F}(\mathfrak{h}) \to \mathcal{F}(\mathfrak{h} \oplus \mathfrak{h})$, and we define

$$\breve{\Gamma}(j) = U\Gamma(j) : \mathcal{F} \to \mathcal{F} \otimes \mathcal{F}.$$

It follows that $\breve{\Gamma}(j)^*\breve{\Gamma}(j) = \Gamma(j^*j)$ which is the identity if $j^*j = 1$. Henceforth $j^*j = 1$ is tacitly assumed in this subsection. From (13) through (16), (19) and (20) it follows that

$$\breve{\Gamma}(j)a^\#(h) = [a^\#(j_0h) \otimes 1 + 1 \otimes a^\#(j_\infty h)]\breve{\Gamma}(j) \tag{23}$$

$$\breve{\Gamma}(j)\phi(h) = [\phi(j_0h) \otimes 1 + 1 \otimes \phi(j_\infty h)]\breve{\Gamma}(j). \tag{24}$$

Furthermore, if $\underline{\omega} = \omega \oplus \omega$ on $\mathfrak{h} \oplus \mathfrak{h}$, then by (21)

$$\begin{aligned}
\breve{\Gamma}(j)\mathrm{d}\Gamma(\omega) &= U\Gamma(j)\mathrm{d}\Gamma(\omega) = U\mathrm{d}\Gamma(\underline{\omega})\Gamma(j) - U\mathrm{d}\Gamma(j, \underline{\omega}\, j - j\omega) \\
&= [\mathrm{d}\Gamma(\omega) \otimes 1 + 1 \otimes \mathrm{d}\Gamma(\omega)]\breve{\Gamma}(j) - \mathrm{d}\breve{\Gamma}(j, \underline{\omega}\, j - j\omega)
\end{aligned} \tag{25}$$

where the notation $\mathrm{d}\breve{\Gamma}(a, b) = U\mathrm{d}\Gamma(a, b)$ was introduced. In particular $\breve{\Gamma}(j)N = (N_0 + N_\infty)\breve{\Gamma}(j)$. Finally we remark that, by (22),

$$\breve{\Gamma}(j){\upharpoonright}\otimes_s^n \mathfrak{h} = \sum_{k=0}^n \binom{n}{k}^{1/2} \underbrace{j_0 \otimes \ldots \otimes j_0}_{n-k \text{ factors}} \otimes \underbrace{j_\infty \otimes \ldots \otimes j_\infty}_{k \text{ factors}}. \tag{26}$$

## 2.7 The "Scattering Identification"

An important role will be played by the scattering identification $I : \mathcal{F} \otimes \mathcal{F} \to \mathcal{F}$ defined by

$$\begin{aligned}
I(\Omega \otimes \Omega) &= \Omega \\
I\varphi \otimes a^*(h_1)\cdots a^*(h_n)\Omega &= a^*(h_1)\cdots a^*(h_n)\varphi, \qquad \varphi \in \mathcal{F}_0,
\end{aligned}$$

and extended by linearity to $\mathcal{F}_0 \otimes \mathcal{F}_0$. (Note that this definition is symmetric with respect to the two factors in the tensor product.) There is a second characterization of $I$ which will often be used. Let $\iota : \mathfrak{h} \otimes \mathfrak{h} \to \mathfrak{h}$ be defined by $\iota(h_{(0)}, h_{(\infty)}) = h_{(0)} + h_{(\infty)}$. Then

$$I = \Gamma(\iota)U^* \tag{27}$$

with $U$ as above. To see this consider states of the form

$$\begin{aligned}
\psi &= a^*(h_1)\cdots a^*(h_m)\Omega \otimes a^*(g_1)\cdots a^*(g_n)\Omega \\
&= \prod_{i=1}^m [a^*(h_i) \otimes 1] \prod_{j=1}^n [1 \otimes a^*(g_j)]\, \Omega \otimes \Omega
\end{aligned} \tag{28}$$

in $\mathcal{F}_0 \otimes \mathcal{F}_0$. By equations (19) and (13)

$$\Gamma(\iota)U^* [a^*(h_i) \otimes 1] = \Gamma(\iota)a^*(h_i, 0)U^* = a^*(h_i)\Gamma(\iota)U^* \tag{29}$$

and similarly $\Gamma(\iota)U^* [1 \otimes a^*(g_j)] = a^*(g_j)\Gamma(\iota)U^*$. Furthermore $\Gamma(\iota)U^*\Omega \otimes \Omega = \Omega = I(\Omega \otimes \Omega)$. This shows that $\Gamma(\iota)U^*\psi = I\psi$ for $\psi$ given by (28) which proves equation (27).

Since $\|\iota\| = \sqrt{2}$, the operator $I$ is unbounded.

**Lemma 1.** *For each positive integer $k$, the operator $I(N+1)^{-k} \otimes \chi(N \le k)$ is bounded.*

Let $j : \mathfrak{h} \to \mathfrak{h} \oplus \mathfrak{h}$ be defined by $jh = (j_0 h, j_\infty h)$ where $j_0, j_\infty \in \mathbf{B}(\mathfrak{h})$. If $j_0 + j_\infty = 1$, then $\check{\Gamma}(j)$ is a right inverse of $I$, that is,

$$I\check{\Gamma}(j) = 1. \tag{30}$$

Indeed $I\check{\Gamma}(j) = \Gamma(\iota)U^*U\Gamma(j) = \Gamma(\iota j) = \Gamma(1) = 1$.

# 3 Definition of the System and Basic Assumptions

## 3.1 The Electron System

The dynamics of the electron system (atom) is given by a self-adjoint operator $K$ on $L^2(X)$, where $X$ is a measure space. Typically $X = (\mathbb{R}^n, d^n x)$ or $X = \{1, \dots, n\}$ equipped with the counting measure, in which case $L^2(X) = \mathbb{C}^n$. We assume that $K$ is bounded from below and that

(H0) $$\inf \sigma_{\mathrm{ess}}(K) > \inf \sigma(K).$$

In other words, $\inf \sigma(K)$ is an isolated eigenvalue of $K$. We use $|x|$ to denote the norm of $x \in X$ if $X$ is a euclidean space. Otherwise $|x| := 0$.

## 3.2 The Radiation Field

Pure states of the radiation field are described by vectors in the bosonic Fock space $\mathcal{F}(\mathfrak{h})$ over $\mathfrak{h} = L^2(\mathbb{R}^d, dk)$. Their time evolution is generated by the Hamiltonian $d\Gamma(\omega)$, where $\omega$ denotes multiplication with a real-valued function $\omega(k)$ on $\mathbb{R}^d$. For easy reference, we summarize all further properties of $\omega$ in the following assumption.

(H1) $$\begin{cases} m := \inf \omega(k) > 0, \\ \omega \in C^\infty(\mathbb{R}^d), \text{ and } \partial^\alpha \omega \text{ is bounded for all } \alpha \ne 0, \\ \nabla \omega(k) \ne 0 \text{ if } k \ne 0. \end{cases}$$

*Remarks.* (i) The last condition ensures positivity of the Mourre constant away from thresholds. (ii) Typical examples we have in mind are $\omega(k) = \sqrt{k^2 + m^2}$ and smooth dispersions $\omega$ that differ from $|k|$ only for small $|k|$, where $\omega$ is chosen such that $\inf_k \omega(k) > 0$ (see Section 10). (iii) Bosons with non-zero spin or helicity, with $\mathfrak{h} = L^2(\mathbb{R}^d, dk) \otimes \mathbb{C}^s$, can also be handled. We then interpret $k$ as a pair $(k, \lambda) \in \mathbb{R}^d \times \{1, \dots, s\}$ and define integration over $k$ as $\sum_{\lambda=1}^s \int d^d k$.

Throughout this paper, $y$ denotes the position operator in $\mathfrak{h}$, i.e., $y = i\nabla_k$.

## 3.3 The Composed System

The dynamics of the composed system of matter (electron system) interacting with the radiation field is given by the Hamilton operator

$$H = K \otimes 1 + 1 \otimes d\Gamma(\omega) + \phi(G) = H_0 + \phi(G),$$

where

$$\phi(G) = \int_X^\oplus \phi(G_x) \, dx$$

and $\phi$ has been defined in (10). It acts on the Hilbert space $\mathcal{H} = L^2(X) \otimes \mathcal{F} \cong \int_X^\oplus \mathcal{F} \, dx$. For each $x \in X$, $G_x \in \mathfrak{h}$, and we assume that

$$\sup_x \|G_x\| < \infty. \tag{31}$$

It follows that $\phi(G)(N+1)^{-1/2}$ and hence that $\phi(G)(H_0+i)^{-1/2}$ are bounded operators. This implies that $\phi(G)$ is infinitesimal w.r.to $H_0$, which shows that $H$ is self-adjoint on $D(H_0)$ and bounded from below. All further assumptions are listed below and will be cited upon use.

(H2) *Exponential decay:* There exists an ionization threshold $\Sigma > \inf \sigma(H)$ with the property that

$$\|e^{\alpha|x|}E_\Delta(H)\| < \infty,$$

for some $\alpha > 0$, and for any given closed interval $\Delta$ contained in $(-\infty, \Sigma)$.

(H3) *Fall-off of the form factor:* For arbitrary $\alpha > 0$,

$$\sup_x e^{-\alpha|x|}\|(1+y^2)G_x\| < \infty.$$

(H4) *Dispersion of the form-factor:* If $h \in C_0^\infty(\mathbb{R}^d \backslash \{0\})$ then, for arbitrary $\alpha > 0$,

$$\int \left( \sup_x e^{-\alpha|x|} |(G_x, h_t)| \right) dt < \infty.$$

Here $h_t = e^{-i\omega t}h$.

(H5) *Short range condition:* There exists a $\mu > 1$ such that

$$\sup_x e^{-\alpha|x|}\|\chi(|y| \geq R)y^n G_x\| \leq C_\alpha R^{-\mu+n} \quad \text{for all} \quad R \geq 0$$

for all $\alpha > 0$ and $n \in \{0, 1, 2\}$.

Further hypotheses (IR), (H6), and (H7) are introduced in Section 10.2.

*Remarks:* (i) These hypotheses are satisfied and, with the exception of (H2), easily verified for many concrete systems (see the next section and Sect. 10.3). (ii) Hypothesis (H2) says that the particle system is localized near the origin for small energies. This is the central physical assumption in this paper. Since (H2) can often be derived from (H0) (see [BFS98, GLL00]) it is legitimate to impose (H2). Assumption (H2) makes (H0) obsolete in all of this paper with the exception of Section 10. (iii) In (H5), $\chi(|y| \geq R)$ stands for multiplication with the characteristic function of the set $\{y : |y| \geq R\}$. More generally we use $\chi(A < c)$ and $\chi(A \leq c)$ to denote the spectral projections $E_{(-\infty,c)}(A)$ and $E_{(-\infty,c]}(A)$ of a given self-adjoint operator $A$. (iv) Hypothesis (H3) follows from (H5).

We conclude this section by defining an extended Hilbert space $\tilde{\mathcal{H}} = \mathcal{H} \otimes \mathcal{F}$ and an extended Hamilton operator

$$\tilde{H} = H \otimes 1 + 1 \otimes d\Gamma(\omega), \tag{32}$$

which describes a time evolution where the photons in the auxiliary Fock space do not interact with the particles. With the help of (32) and the scattering identification $I$ the time evolution (3) will be described as a unitary time evolution on the extended Hilbert space in Section 5. Moreover the extended Hamilton operator is invaluable in the spectral analysis of $H$ (see Section 4). In analogy to $\tilde{H}$ we define $\tilde{H}_0 = H_0 \otimes 1 + 1 \otimes d\Gamma(\omega)$.

## 3.4 Examples of Concrete Physical Systems

In this section some concrete systems are discussed for which the hypotheses (H0) through (H5) are all satisfied. See also Section 10.3 where we explain how the standard model of non-relativistic QED fits into a slightly expanded version of the general framework introduced above.

The example discussed in the introduction, where $K$ is an arbitrary atomic Schrödinger operator, that is,

$$K = -\Delta + V \qquad \text{on} \qquad L^2(\mathbb{R}^{3N}),$$

satisfying hypothesis (H0), $\omega(k) = \sqrt{k^2 + m^2}$, and

$$G_x(k) = g \sum_{j=1}^{N} e^{-ik \cdot x_j} \kappa(k), \qquad \kappa \in C_0^\infty(\mathbb{R}^3),$$

satisfies hypotheses (H1) through (H5), for $g$ sufficiently small. In fact, (H2) is proven in [BFS98], with $\Sigma = \inf \sigma_{\mathrm{ess}}(H) - g \sup_x \int dk \, |G_x(k)|^2/\omega(k)$, (H3) follows from $y^2 = -\Delta_k$, (H4) from $|(G_x, h_t)| \leq Ct^{-3/2}$, see e.g. [RS79], and (H5) even holds in the strong form

$$\sup_x e^{-\alpha|x|} \|\chi(|y| \geq R) y^n G_x\| \leq C_{n,\mu} R^{-\mu} \tag{33}$$

for arbitrary $n, \mu \in \mathbb{N}$. To see this, let $f(y) = (1 + |y|)^\mu \hat{\kappa}(y)$ and put $N = 1$ for simplicity (this means that we consider the case with only one electron). Then

$$e^{-2\alpha|x|} \|\chi(|y| \geq R) y^n G_x\|^2 = e^{-2\alpha|x|} \int_{|y| \geq R} |y^n \hat{\kappa}(x - y)|^2 \, dy$$

$$= e^{-2\alpha|x|} \int_{|y| \geq R} (1 + |x - y|)^{-2\mu} |y^n f(x - y)|^2 \, dy.$$

This decays exponentially as $R \to \infty$ for $|x| \geq R/2$, and if $|x| < R/2$ it is bounded by $(1 + R/2)^{-2n} \sup_x e^{-2\alpha|x|} \|\chi(|y| \geq R) y^n f(x - \cdot)\|^2$ which is of order $O(R^{-2n})$, because $\hat{\kappa} \in C_0^\infty(\mathbb{R}^3)$ by assumption and thus $f$ is rapidly decreasing.

The spin-boson models, where $K$ is a hermitian $n \times n$ matrix on $\mathbb{C}^n = L^2(X)$, with $X = \{1, \ldots, n\}$, also fits into the our general framework. Suppose $\omega$ is as above and $G_x \in \mathcal{S}(\mathbb{R}^3)$ for all $x \in X$. Then hypotheses (H0) through (H5) are satisfied with the convention that $|x| \equiv 0$. Hypotheses (H0) and (H2) are trivial, (H4) is seen as in the first example above, and (H3) and (H5) follow from the fact that $G_x \in \mathcal{S}(\mathbb{R}^3)$, for $x = 1, \ldots, n$.

# 4 The Spectrum of Pauli-Fierz Hamiltonians

## 4.1 Essential Spectrum and Existence of a Ground State

**Theorem 2.** *Assume (H1) and (H2) (Sect. 3), and let $E = \inf(H)$. Then*

$$\inf \sigma_{\text{ess}}(H) \geq \min\{\Sigma, E + m\}.$$

*In particular, $\inf \sigma(H)$ is an isolated eigenvalue of $H$.*

*Proof.* Given $\lambda \in \sigma_{\text{ess}}(H)$ with $\lambda < \Sigma$ we need to show that $\lambda \geq E + m$. Let $\Delta$ be an open interval in $\mathbb{R}$ containing $\lambda$ with $\text{supp}\,\Delta < \Sigma$ and let $(\varphi_n)_{n \geq 0} \subset E_\Delta(H)\mathcal{H}$ with $\|(H - \lambda)\varphi_n\| \to 0$, $\|\varphi_n\| = 1$ and $\varphi_n \rightharpoonup 0$. Then

$$\lambda = \lim_{n \to \infty} \langle \varphi_n, H\varphi_n \rangle$$

and we estimate the r.h.s from below. Let $j_{0,R}, j_{\infty,R} \in C^\infty(\mathbb{R}^3)$ be a partition of unity defined as in Lemma 32 with $j_{0,R}^2 + j_{\infty,R}^2 = 1$. Pick $\alpha > 0$ according to (H2) such that $e^{\alpha|x|}E_\Delta(H)$ is bounded. Then $\sup_n \|e^{\alpha|x|}\varphi_n\| < \infty$ by assumption on $\varphi_n$ and hence by Lemma 32 and Lemma 31

$$
\begin{aligned}
\langle \varphi_n, H\varphi_n \rangle &= \langle \varphi_n, \breve{\Gamma}(j_R)^* \tilde{\Gamma}(j_R) H\varphi_n \rangle \\
&= \langle \varphi_n, \breve{\Gamma}(j_R)^* \tilde{H} \breve{\Gamma}(j_R)\varphi_n \rangle + o(R^0).
\end{aligned}
\tag{34}
$$

uniformly in $n$. From $H \geq E$ and $\mathrm{d}\Gamma(\omega) \geq m - m\chi(N = 0)$, it is clear that

$$\tilde{H} \geq (E + m) - m\chi(N_\infty = 0). \tag{35}$$

Since $\breve{\Gamma}(j_R)^*\chi(N_\infty = 0)\breve{\Gamma}(j_R) = \Gamma(j_{0,R}^2)$ and since $E_\Delta(H)\Gamma(j_{0,R}^2)E_\Delta(H) = (E_\Delta(H)e^{\alpha|x|}) \times (e^{-\alpha|x|}\Gamma(j_{0,R}^2)E_\Delta(H))$ is compact by Lemma 34 in Appendix E, the equation (34) combined with (35) implies that

$$\lambda = \lim_{n \to \infty} \langle \varphi_n, H\varphi_n \rangle \geq E + m + o(R^0).$$

Letting $R \to \infty$ this proves the theorem. □

## 4.2 The Mourre Estimate

Next we establish a type of Mourre theorem with conjugate operator $A = \mathrm{d}\Gamma(a)$ and

$$a = [i\omega, y^2/2] = \frac{1}{2}(\nabla\omega \cdot y + y \cdot \nabla\omega).$$

That is we prove positivity of $i[H, A]$ on spectral subspaces of $H$ away from thresholds and eigenvalues, and, as in $N$-body quantum theory, we obtain important spectral properties of $H$ as a byproduct. Here the thresholds are the elements of $\tau := \sigma_{pp}(H) + \mathbb{N}m$, $(0 \notin \mathbb{N})$. The Mourre inequality will allow us to show that

$$\langle \mathrm{d}\Gamma(y^2) \rangle_{\psi_t} \geq ct^2, \qquad \text{as } |t| \to \infty, \tag{36}$$

where $c > 0$ for states separated in energy from thresholds and eigenvalues. This together with the above mentioned spectral properties suffices to derive AC. As (36) can only be true if the particles are spatially confined, our Mourre estimate only holds for energies below $\Sigma$.

On a suitable dense subspace,

$$i[H, A] = \mathrm{d}\Gamma(|\nabla\omega|^2) - \phi(iaG).$$

We use this equation to *define* the operator $i[H, A]$ on $\cup_{\mu < \Sigma} E_\mu(H)\mathcal{H}$. Note that $\phi(-iaG)\varphi$ makes sense for $\varphi \in \operatorname{Ran} E_\mu(H)$, $\mu < \Sigma$, thanks to the exponential decay and the boundedness of $e^{-\alpha|x|}\phi(-iaG)(N+1)^{-1/2}$. The following virial theorem is an important ingredient in the proof of Theorem 4. Furthermore, in the case of massless bosons and IR-cutoff interaction (see Section 10) it will allow us to prove absence of eigenvalues close to, but different from the ground state energy.

**Lemma 3 (Virial Theorem).** *Assume hypotheses (H1), (H2), and (H3). If $H\varphi = E\varphi$ and $E < \Sigma$ then*

$$\langle\varphi, i[H, A]\varphi\rangle = 0.$$

The proof of this lemma is deferred to Appendix E.

The following theorem is the main result of this section.

**Theorem 4.** *Assume (H1), (H2) and (H5). Then*

(i) *For each $\lambda \in (-\infty, \Sigma)\backslash\tau$ there exists an open interval $\Delta \ni \lambda$, a positive constant $C_\lambda$, and a compact operator $\mathcal{E}$ such that*

$$E_\Delta(H)[iH, A]E_\Delta(H) \geq C_\lambda E_\Delta(H) + \mathcal{E}.$$

(ii) *Non-threshold eigenvalues in $(-\infty, \Sigma)$ have finite multiplicity and can accumulate only at threshold. Furthermore $\tau \cap (-\infty, \Sigma]$ is closed and countable.*

(iii) *If $\lambda \in (-\infty, \Sigma)\backslash\tau$ is not an eigenvalue, then there exists an open interval $\Delta \ni \lambda$ and a positive constant $C_\lambda$ such that*

$$E_\Delta(H)[iH, A]E_\Delta(H) \geq C_\lambda E_\Delta(H).$$

The proof of this theorem follows the lines of the proof in [DG99] with only minor modifications due to the presence of continuous spectrum in the particle Hamiltonian $K$. For the sake of completeness we have included a proof of Theorem (4) in this paper, but it is deferred to Appendix E.

# 5   The Wave Operator

Recall from Section 3 that $\tilde{H} = H \otimes 1 + 1 \otimes d\Gamma(\omega)$ on $\tilde{\mathcal{H}} = \mathcal{H} \otimes \mathcal{F}$, and let $P_B(H)$ in (37) denote the orthogonal projector onto $\mathcal{H}_{pp}(H)$, the closure of the space spanned by all eigenvectors of $H$. The purpose of this section is to establish existence of the wave operator

$$\Omega_+ = s - \lim_{t \to \infty} e^{iHt} I e^{-i\tilde{H}t} P_B(H) \otimes 1 \tag{37}$$

on spectral subspaces of $\tilde{H}$ corresponding to compact intervals $\Delta \subset (-\infty, \Sigma)$. Furthermore we will see that $\Omega_+$ is isometric if restricted to vectors in $\mathcal{H}_{pp}(H) \otimes \mathcal{F}$. The existence of (37) will essentially follow from the existence of asymptotic field operators

$$a_+^\sharp(h)\varphi = \lim_{t \to \infty} e^{iHt} a^\sharp(h_t) e^{-iHt} \varphi \tag{38}$$

and the existence of products of such operators, which is established in the next theorem.

**Theorem 5.** *Assume hypotheses (H1), (H2) and (H4) are satisfied, and let $f, h \in L^2(\mathbb{R}^d)$.*

*i) If $\varphi = E_\eta(H)\varphi$ for some $\eta < \Sigma$ then the limit*

$$a_+^\sharp(h)\varphi = \lim_{t \to \infty} e^{iHt} a^\sharp(h_t) e^{-iHt} \varphi$$

*exists. Here $h_t = e^{-i\omega t}h$.*

*ii) The canonical commutation relations*

$$[a_+(g), a_+^*(h)] = (g, h) \qquad and \qquad [a_+^\sharp(h), a_+^\sharp(g)] = 0,$$

*hold true, in form–sense, on $\chi(H < \eta)\mathcal{H}$ for all $\eta < \Sigma$.*

*iii) Let $m = \inf\{\omega(k) : h(k) \neq 0\}$ and $M = \sup\{\omega(k) : h(k) \neq 0\}$. Then*

$$a_+^*(h) \operatorname{Ran} \chi(H \leq E) \subset \operatorname{Ran} \chi(H \leq E + M)$$
$$a_+(h) \operatorname{Ran} \chi(H \leq E) \subset \operatorname{Ran} \chi(H \leq E - m).$$

*iv) Suppose $\varphi = E_\eta(H)\varphi$, $h_i \in L^2(\mathbb{R}^d; \mathbb{C})$ for $i = 1, \ldots, n$ and let $M_i = \sup\{\omega(k) : h_i(k) \neq 0\}$. Then*

$$a_+^*(h_1) \ldots a_+^*(h_n)\varphi = \lim_{t \to \infty} e^{iHt} a^*(h_{1,t}) \ldots a^*(h_{n,t}) e^{-iHt} \varphi$$

*provided that $\eta + \sum_{i=1}^n M_i < \Sigma$.*

*v) Suppose $\eta < \Sigma$ and $\varphi \in E_\eta(H)\mathcal{H}_{pp}(H)$. Then*

$$a_+(h)\varphi = 0 \quad for \ all \quad h \in L^2(\mathbb{R}^d).$$

*Remark.* For relativistic massive bosons, that is for $\omega(k) = \sqrt{k^2 + m^2}$ with $m > 0$, as well as in the case of relativistic electrons and massless bosons, the asymptotic field operators actually exist on a dense subspace of $\mathcal{H}$ irrespective of $\Sigma$ (see [FGS00]).

*Proof.* i) Assume first that $h \in C_0^\infty(\mathbb{R}^d \backslash \{0\})$. By Cook's argument it suffices to show that

$$\int_1^\infty \|(G_x, h_t)\varphi_t\| dt < \infty. \tag{39}$$

This follows from the assumptions (H2) and (H4).

For the proofs of ii), iii) and iv) we refer to [FGS00].

v) It suffices to show that $a_+(h)\varphi = 0$ if $H\varphi = E\varphi$. Statement v) then follows from the boundedness of $a_+(h)E_\eta(H)$. Since $h_t \rightharpoonup 0$ weakly as $t \to \infty$ we have $s - \lim_{t \to \infty} a(h_t)(H + i)^{-1/2} = 0$. Hence

$$a_+(h)\varphi = \lim_{t \to \infty} e^{iHt} e^{-iEt} (E + i)^{1/2} a(h_t)(H + i)^{-1/2}\varphi = 0. \tag{40}$$

$\square$

Next we prove existence of the *extended wave operator* $\tilde{\Omega}_+ := s - \lim_{t \to \infty} e^{iHt} I e^{-i\tilde{H}t}$ on a suitable spectral subspace of $\tilde{H}$. Since $\tilde{\Omega}_+$ agrees with $\Omega_+$ for vectors in $\mathcal{H}_{pp}(H) \otimes \mathcal{F}$, this will immediately imply existence of $\Omega_+$.

**Lemma 6.** *Assume the hypotheses of the theorem above are satisfied.*

a) *Suppose $\psi = \varphi \otimes a^*(h_1) \ldots a^*(h_n)\Omega$ where $\varphi = E_\lambda(H)\varphi$ and $\lambda + \sum_{i=1}^n M_i < \Sigma$, where $M_i$ is defined as in Theorem 5, iv). Then $\tilde{\Omega}_+\psi := \lim_{t \to \infty} e^{iHt} I e^{-i\tilde{H}t}\psi$ exists and*

$$\tilde{\Omega}_+\psi = a_+^*(h_1) \ldots a_+^*(h_n)\varphi. \tag{41}$$

b) *$\tilde{\Omega}_+$ exists on $E_\mu(\tilde{H})\tilde{\mathcal{H}}$ for all $\mu < \Sigma$.*

*Proof.* Statement a) follows from

$$e^{-i\tilde{H}t}\varphi \otimes a^*(h_1) \ldots a^*(h_n)\Omega = e^{-iHt}\varphi \otimes a^*(h_{1,t}) \ldots a^*(h_{n,t})\Omega \tag{42}$$

the definition of $I$ and Theorem 5, iv).

(b) Since $e^{iHt} I e^{-i\tilde{H}t} E_\mu(\tilde{H})$ is bounded uniformly in $t$, it suffices to prove existence of $\tilde{\Omega}_+$ on a dense subspace of $E_\mu(\tilde{H})\tilde{\mathcal{H}}$. By Lemma 30 finite linear combinations of vectors of the form described in part a) span such a subspace and hence b) follows from a). $\square$

The following theorem is the main result of this section.

**Theorem 7.** *Assume (H1), (H2) and (H4). Then*

$$\Omega_+ \varphi = s - \lim_{t \to \infty} e^{iHt} I e^{-i\tilde{H}t} P_B(H) \otimes 1 \tag{43}$$

*exists on $\cup_{\mu \leq \Sigma} E_\mu(\tilde{H})\tilde{\mathcal{H}}$, $\|\Omega_+\| = 1$ and hence $\Omega_+$ has a unique extension, also denoted by $\Omega_+$, to $E_\Sigma(\tilde{H})\tilde{\mathcal{H}}$. $\Omega_+$ is isometric on $(P_B(H) \otimes 1)E_\Sigma(\tilde{H})\tilde{\mathcal{H}}$ and hence $\operatorname{Ran}\Omega_+$ is closed. Furthermore*

$$e^{iHt}\Omega_+ = \Omega_+ e^{i\tilde{H}t}. \tag{44}$$

*Proof.* The existence of $\Omega_+$ on $\cup_{\mu < \Sigma} E_\mu(\tilde{H})\tilde{\mathcal{H}}$ follows from Lemma 6 b). By Lemma 6 a), the CCR in Theorem 5 ii), and by part v) of that theorem, $\Omega_+$ is isometric on $(P_B(H) \otimes 1) \cup_{\mu < \Sigma} E_\mu(\tilde{H})\tilde{\mathcal{H}}$. So $\Omega_+$ is a partial isometry and hence $\|\Omega_+\| = 1$. All these properties carry over to the closure of $\Omega_+$. $\qquad\square$

# 6  Propagation Estimates

This section establishes the propagation estimates needed later on to construct the asymptotic observable $W$ (see Eq. (8)) and the Deift Simon wave operator $W_+$. To begin with, we define a smooth convex function $S(y, t)$ which modifies $y^2/2t$ for $y$ in a neighborhood of size $t^\delta$, $\delta \in (0, 1)$, of the origin. The operator $W$ will be constructed using $S(y, t)$ in place of $y^2/2t$. This will not affect $W$ but allows us to prove its existence.

Here and henceforth we use the following notation for the various Heisenberg derivatives. Suppose $A$ is an operator in $\mathcal{H}$. Then we define

$$DA := i[H, A] + \frac{\partial A}{\partial t} \tag{45}$$

$$D_0 A := i[H_0, A] + \frac{\partial A}{\partial t} \tag{46}$$

Similarly, the Heisenberg derivative $d$ of an operator $a$ on the one boson sector $\mathfrak{h}$ is defined by

$$da = i[\omega, a] + \frac{\partial a}{\partial t}. \tag{47}$$

Furthermore, if $A(R)$ is a family of operators on $\mathcal{H}$, and $R \in \mathbb{R}_+$, we write

$$A(R) = O(R^m) N^p \qquad \text{if} \qquad \|A(R) N^{-p}\| = O(R^m).$$

## 6.1  Construction of $S(y, t)$

Pick $m \in C_0^\infty(\mathbb{R}_+)$ with $m \geq 0$, $\operatorname{supp}(m) \subset [1, 2]$ and $\int d\sigma\, m(\sigma) = 1$. Set

$$S_0(y) = \int d\sigma\, m(\sigma) \chi(y^2/2 > \sigma)\, (y^2/2 - \sigma). \tag{48}$$

Then $S_0$ is smooth and

$$S_0(y) = \begin{cases} 0 & y^2/2 \leq 1 \\ y^2/2 + b & y^2/2 \geq 2 \end{cases} \tag{49}$$

where $b = -\int d\sigma\, m(\sigma)\sigma$. It follows that

$$S_0(y) = y^2/2 + a(y) + b \tag{50}$$

where $a \in C_0^\infty(\mathbb{R}^3)$. This formula will allow us to apply Lemma 27 in Appendix A to $[i\omega, S_0]$. It is important that $S_0$ is convex, which is easy to see from the definition. Next we define a scaled version of $S_0$ by

$$S(y, t) = t^{-1+2\delta} S_0(y/t^\delta) \tag{51}$$

where $0 < \delta < 1$.

**Lemma 8.** *For all integers $n \geq 0$ and all $\alpha$*

$$\mathcal{D}^n \partial^\alpha \left( S - \frac{y^2}{2t} \right) = O\left( t^{-(n+1)+\delta(2-|\alpha|)} \right).$$

*Here $\mathcal{D} = \partial/\partial t$ or $\mathcal{D} = \partial/\partial t + (y/t) \cdot \nabla$ and $\partial^\alpha$ is any spacial derivative of order $|\alpha|$. In particular*

$$\nabla S = \frac{y}{t} + O(t^{-1+\delta}), \qquad \frac{\partial S}{\partial t} = -\frac{y^2}{2t^2} + O(t^{-2+2\delta}). \tag{52}$$

*These estimates hold uniformly in $y \in \mathbb{R}^d$.*

*Proof.* By definition of $S(y,t)$ and (50), $S(y,t) - y^2/2t = t^{-1+2\delta}a(yt^{-\delta}) + t^{-1+2\delta}b$. The second term clearly enjoys the desired property. For the first term we have

$$\partial^\alpha[t^{-1+2\delta}a(yt^{-\delta})] = t^{-1+\delta(2-|\alpha|)}(\partial^\alpha a)(yt^{-\delta}) \tag{53}$$

where the right hand side is of the form $t^c h(yt^{-\delta})$ with $h \in C_0^\infty(\mathbb{R}^3)$. Both $\partial/\partial t(t^c h)$ and $\mathcal{D}(t^c h)$ are again of this form with $c$ replaced by $c-1$. This proves the lemma. ☐

**Lemma 9.** *Suppose $\omega \in C^\infty(\mathbb{R}^d)$ and $\partial^\alpha \omega$ is bounded for all $\alpha \neq 0$. Then*

$$dS = \frac{1}{2}(\nabla\omega \cdot \nabla S + \nabla S \cdot \nabla\omega) + \frac{\partial S}{\partial t} + O(t^{-1}),$$

*and for any smooth vector field $v(y,t)$ and $\mathcal{D}_v = \partial/\partial t + v \cdot \nabla$,*

$$d(dS) = (\nabla\omega - v) \cdot S''(\nabla\omega - v) \tag{54}$$
$$+ (\mathcal{D}_v \nabla S) \cdot (\nabla\omega - v) + (\nabla\omega - v) \cdot (\mathcal{D}_v \nabla S) \tag{55}$$
$$+ \mathcal{D}_v^2 S - (\mathcal{D}_v v) \cdot \nabla S + O(t^{-1-\delta}). \tag{56}$$

*Proof.* The first part follows from (50) and Lemma 27. By definition of the Heisenberg derivative, $d(dS)$ is given by

$$d(dS) = [i\omega, [i\omega, S]] + 2[i\omega, \partial S/\partial t] + \frac{\partial^2 S}{\partial t^2}.$$

By Lemma 8, (50), and Lemma 27

$$2[i\omega, \partial S/\partial t] = \nabla\omega \cdot \nabla \frac{\partial S}{\partial t} + \nabla \frac{\partial S}{\partial t} \cdot \nabla\omega + O(t^{-2})$$

$$[i\omega, [i\omega, S]] = \nabla\omega S'' \nabla\omega + O(t^{-1-\delta}).$$

To prove the second equation an explicit formula for $[i\omega, S] - \nabla\omega \cdot \nabla S$ is also needed (see the proof of Lemma 27). Since for every smooth vector field $v(y,t)$

$$\nabla\omega \cdot S'' \nabla\omega + \nabla\omega \cdot \nabla \frac{\partial S}{\partial t} + \nabla \frac{\partial S}{\partial t} \cdot \nabla\omega + \frac{\partial^2 S}{\partial t^2}$$
$$= (\nabla\omega - v) \cdot S''(\nabla\omega - v) + (\mathcal{D}_v \nabla S) \cdot (\nabla\omega - v) + (\nabla\omega - v) \cdot (\mathcal{D}_v \nabla S)$$
$$+ \mathcal{D}_v^2 S - (\mathcal{D}_v v) \cdot \nabla S \tag{57}$$

the lemma follows. ☐

## 6.2 Propagation Estimates

We are now ready to state and prove our propagation estimates. Note that these are basic propagation estimates which are well known in other contexts (see [GS97]).

**Proposition 10.** *Assume (H1) and (H3), let $\chi = \bar{\chi} \in C_0^\infty(\mathbb{R})$, and suppose $e^{\alpha|x|}\chi(H)$ is a bounded operator for some $\alpha > 0$. Let $\chi = \chi(H)$ and $v = y/t$. Then, if $\lambda > 0$ is large enough, there exists a constant $C$ such that*

$$\int_1^\infty dt\, \frac{1}{t} \langle \chi\psi_t, \chi(\lambda^2 \leq \mathrm{d}\Gamma(v^2) \leq 2\lambda^2)\chi\psi_t\rangle \leq C\|\psi\|^2$$

*for all $\psi \in \mathcal{H}$. Here $\psi_t = e^{-iHt}\psi$.*

*Remark:* This propagation estimate equally holds on $\tilde{\mathcal{H}}$ and with $H$ and $\mathrm{d}\Gamma(v^2)$ replaced by $\tilde{H}$ and $\mathrm{d}\Gamma(v^2) \otimes 1 + 1 \otimes \mathrm{d}\Gamma(v^2)$. This is needed for the proof of the remark to Proposition 11.

*Proof.* Pick $h \in C_0^\infty(1/2, 3)$ with $h(r) = 1$ on $[1, 2]$, $0 \leq h \leq 1$ and set $\tilde{h}(s) = \int_0^s ds' h^2(s')$. Note that $\tilde{h}(s) = \tilde{h}(3)$ for $s \geq 3$. Hence $g(s) = \tilde{h}(s) - \tilde{h}(3)$ for $s \geq 0$ and $g(-s) = g(s)$ define a $C_0^\infty$-function $g$ on $\mathbb{R}$. The operator $B = -\chi\tilde{h}(\mathrm{d}\Gamma(v^2/\lambda^2))\chi$ is our propagation observable. Since $B$ is bounded the theorem will follow if we show that

$$DB := [iH, B] + \frac{\partial B}{\partial t} \geq \frac{C}{t}\chi(H)\,\chi(1 \leq \mathrm{d}\Gamma(v^2/\lambda^2) \leq 2)\,\chi(H) + \text{integrable terms}$$

for some $C > 0$ if $\lambda$ is large enough. Henceforth we use the abbreviations $h$, $\tilde{h}$ and $g$ to denote the operators $h(\mathrm{d}\Gamma(v^2/\lambda^2))$, $\tilde{h}(\mathrm{d}\Gamma(v^2/\lambda^2))$ and $g(\mathrm{d}\Gamma(v^2/\lambda^2))$, respectively. Clearly

$$\frac{\partial B}{\partial t} = \chi h^2 \mathrm{d}\Gamma(v^2/\lambda^2)\frac{2}{t}\chi \geq \chi h^2 \frac{1}{t}\chi. \tag{58}$$

Next

$$-[iH, B] = \chi[iH, \tilde{h}]\chi = \chi[i\mathrm{d}\Gamma(\omega), \tilde{h}]\chi + \chi[i\phi(G), \tilde{h}]\chi.$$

Consider first the second term on the right hand side. By $[i\phi(G), \tilde{h}] = [i\phi(G), g]$, the Helffer–Sjöstrand functional calculus (see Appendix A.2), and by (H3)

$$\|e^{-\alpha|x|}[i\phi(G), \tilde{h}](N+1)^{-1/2}\| \leq C\|e^{-\alpha|x|}\phi(-iv^2G)(N+1)^{-1/2}\|$$
$$\leq \frac{C}{\lambda^2 t^2}\sup_x e^{-\alpha|x|}\|\phi(iy^2 G_x)(N+1)^{-1/2}\| = O(t^{-2}). \tag{59}$$

Hence $\chi[i\phi(G), \tilde{h}]\chi$ is integrable. By (28)

$$[i\mathrm{d}\Gamma(\omega), \tilde{h}] = [i\mathrm{d}\Gamma(\omega), g]$$
$$= \frac{1}{\lambda t}g'\mathrm{d}\Gamma(\nabla\omega \cdot v/\lambda + v/\lambda \cdot \nabla\omega) + O(t^{-2})N$$
$$= \frac{1}{\lambda t}h\mathrm{d}\Gamma(\nabla\omega \cdot v/\lambda + v/\lambda \cdot \nabla\omega)h + O(t^{-2})N$$
$$\leq \frac{C}{\lambda t}h(N+1)h + O(t^{-2})N$$

Since $[\chi, h] = O(t^{-1})$ it follows that

$$\chi[i d\Gamma(\omega), \tilde{h}]\chi \leq \frac{C}{\lambda t}h^2 + O(t^{-2})N.$$

For $C/\lambda < 1$ this in conjunction with (58) proves the proposition. $\qquad\square$

**Proposition 11.** *Assume (H1) and (H5), let $\chi \in C_0^\infty(\mathbb{R})$, and suppose $e^{\alpha|x|}\chi(H)$ is a bounded operator for some $\alpha > 0$. Let $f \in C_0^\infty(\mathbb{R})$, with $0 \leq f \leq 1$, $f(x) = 0$ for $x \geq 2$, and $f(x) = 1$ for $0 \leq x \leq 1$. Denote $\chi = \chi(H)$ and $f = f(d\Gamma(v^2/\lambda^2))$, where $v = y/t$ and $\lambda \in \mathbb{R}$. Then, for $\lambda$ large enough*

$$\int_1^\infty dt \, \langle \psi_t, \chi f d\Gamma\left((\nabla\omega - v)\cdot S''(\nabla\omega - v)\right) f\chi\psi_t\rangle \leq C\|\psi\|^2.$$

*Remark:* This proposition equally holds on $\tilde{\mathcal{H}}$ and with $H$, $d\Gamma(v^2)$ and $d\Gamma(P)$ (here $P = (\nabla\omega - v)\cdot S''(\nabla\omega - v)$), replaced by $\tilde{H}$, $d\Gamma(v^2) \otimes 1 + 1 \otimes d\Gamma(v^2)$ and $d\Gamma(P) \otimes 1 + 1 \otimes d\Gamma(P)$. This is needed for the proof of Theorem 15.

*Proof.* Let $\gamma(t) = \langle \psi_t, \chi f d\Gamma(dS) f\chi\psi_t\rangle$. From $\pm dS \leq \mathrm{const}(v^2 + 1)$ it follows that $d\Gamma(dS)f(N+1)^{-1}$ is a bounded operator and thus that $\sup_{|t|\geq 1}|\gamma(t)| < \infty$ because of the cut-off $\chi$. Next we show that

$$\gamma'(t) \geq \langle \psi_t, \chi f d\Gamma((\nabla\omega - v)\cdot S''(\nabla\omega - v))f\chi\psi_t\rangle \tag{60}$$
$$+ \text{(integrable w.r. to } t) \times \|\psi\|^2.$$

By the Leibnitz rule

$$\gamma'(t) = \langle \psi_t, \chi(Df)d\Gamma(dS)f\chi\psi_t\rangle + h.c. + \langle \psi_t, \chi f(Dd\Gamma(dS))f\chi\psi_t\rangle \tag{61}$$

Only the last term will contribute to (60).

$$Dd\Gamma(dS) = D_0 d\Gamma(dS) + [i\phi(G), d\Gamma(dS)]$$
$$= d\Gamma(d(dS)) + \phi(-idS\,G)$$

where $d(dS)$ is given by Lemma 9. Since $v = y/t$ the terms in (56) are of order $O(t^{-1-\varepsilon})$ where $\varepsilon = \min(\delta, 2 - 2\delta)$. For the terms of (55) we have

$$\pm\left[(\mathcal{D}_v\nabla S)\cdot(\nabla\omega - v) + (\nabla\omega - v)\cdot(\mathcal{D}_v\nabla S)\right]$$
$$\leq t^{2-\delta}(\mathcal{D}_v\nabla S)^2 + t^{-2+\delta}(\nabla\omega - v)^2 = O(t^{-2+\delta})(1 + v^2)$$

which, thanks to the cutoffs $f$ and $\chi$, gives an integrable contribution to $\gamma'(t)$. This shows that

$$\langle \psi_t, \chi f d\Gamma(d^2 S)f\chi\psi_t\rangle \geq \langle \psi_t, \chi f d\Gamma((\nabla\omega - v)\cdot S''(\nabla\omega - v))f\chi\psi_t\rangle$$
$$+ O(t^{-1-\varepsilon})\|\psi\|^2$$

To estimate the contribution due to $\phi(-idS\,G)$ use

$$\|\chi f\phi(-idSG)f\chi\| \leq \mathrm{const}\|e^{-\alpha|x|}\phi(-idSG)(N+1)^{-1/2}\| \tag{62}$$

and

$$dS = \nabla S \cdot \nabla\omega - i\sum_{r,s}(\partial_{rs}^2 S)(\partial_{rs}^2\omega) + \frac{\partial S}{\partial t} + O(t^{-1-\delta}).$$

This shows that $dS = \chi(|y| \geq t^\delta)dS + O(t^{-\delta-1})$ and in conjunction with Lemma 8 and (H5) it follows that (62) is integrable. To estimate the contribution in (61) due to $Df$ note that

$$Df = D_0 f + [i\phi(G), f].$$

The second term gives a contribution of order $O(t^{-2})$. This is seen in the same way as (59). Next choose $g \in C_0^\infty(\mathbb{R})$ with $\mathrm{supp}(g) \subset (1,2)$ and $gf' = f'$ and denote $g = g(\mathrm{d}\Gamma(v^2/\lambda^2))$. Then

$$\chi(D_0 f)\mathrm{d}\Gamma(dS)f\chi = \chi g(D_0 f)\mathrm{d}\Gamma(dS)gf\chi + O(t^{-2}) \tag{63}$$

and hence

$$|\langle\psi_t, \chi(D_0 f)\mathrm{d}\Gamma(dS)f\chi\psi_t\rangle| \leq \frac{c}{t}\|g(N+1)\chi\psi_t\|^2 + O(t^{-2}) \tag{64}$$

which is integrable by Proposition 10. $\qquad\square$

## 7  The Asymptotic Observable

In this section existence of the asymptotic observable $W$ is proved. An auxiliary version $W_\lambda$ of $W$ will involve a space cutoff at $|y| = \lambda$ in the bosonic configuration space. $W$ is then obtained in the limit $\lambda \to \infty$.

To define the space cutoff we pick, once and for all, a function $f \in C_0^\infty(\mathbb{R})$ with $0 \leq f \leq 1$, $f(x) = 0$ for $x \geq 2$ and $f(x) = 1$ for $0 \leq x \leq 1$. The space cutoff is the operator $f[\mathrm{d}\Gamma(v^2/\lambda^2)]$ or $1 \otimes f[\mathrm{d}\Gamma(v^2/\lambda^2)]$ on $\mathcal{F}$ or $\mathcal{H}$ respectively. Here $v = y/t$ and $\lambda \in \mathbb{R}$. For brevity these operators will also be denoted by $f$ if there is no danger of confusion.

Since $(dS)^2 \leq \mathrm{const}(v^4 + 1)$, which follows from Lemma 9 and Lemma 27, and since $f \in C_0^\infty(\mathbb{R})$, the operator

$$\mathrm{d}\Gamma(dS)f(N+1)^{-1}$$

is bounded.

**Theorem 12.** *Assume Hypotheses (H1), (H2), and (H5). If $\chi \in C_0^\infty(\mathbb{R})$ with $\mathrm{supp}\,\chi \subset (-\infty, \Sigma)$, then*

$$W_\lambda = s - \lim_{t\to\infty} e^{iHt}\chi f\mathrm{d}\Gamma(dS)f\chi e^{-iHt}$$

*exists, is self-adjoint, and commutes with $H$.*

*Proof.* To prove existence of $W_\lambda \varphi$ we use Cook's argument, i.e. we show that

$$\int_1^\infty dt \left| \frac{d}{dt} \langle \psi_t, \chi f \mathrm{d}\Gamma(dS) f \chi \varphi_t \rangle \right| \leq C \|\psi\|.$$

As in the proof of Proposition 11 one shows that

$$\begin{aligned}
\frac{d}{dt} \langle \psi_t, \chi f \mathrm{d}\Gamma(dS) f \chi \varphi_t \rangle &= \langle \psi_t, \chi f \mathrm{d}\Gamma[(\nabla \omega - v) \cdot S''(\nabla \omega - v)] f \chi \varphi_t \rangle \\
&\quad + O(t^{-1-\varepsilon}) \|\psi\| \|\varphi\|.
\end{aligned} \tag{65}$$

Since $S'' \geq 0$ the first term on the right side defines a non-negative sesquilinear form in $\psi$ and $\varphi$ and hence we can apply Schwarz

$$\begin{aligned}
|\langle \psi_t, \chi f \mathrm{d}\Gamma[(\nabla \omega - v) \cdot S''(\nabla \omega - v)] f \chi \varphi_t \rangle| & \\
\leq |\langle \psi_t, \chi f \mathrm{d}\Gamma[(\nabla \omega - v) \cdot S''(\nabla \omega - v)] f \chi \psi_t \rangle|^{1/2} & \\
\times |\langle \varphi_t, \chi f \mathrm{d}\Gamma[(\nabla \omega - v) \cdot S''(\nabla \omega - v)] f \chi \varphi_t \rangle|^{1/2}. &
\end{aligned}$$

This, together with Proposition 11 after an application of Hölder's inequality, shows that also the first term in (65) is integrable. This proves existence of $W_\lambda \varphi$.

Clearly $W_\lambda$ is bounded and symmetric. To prove that $W_\lambda$ commutes with $H$ it suffices to show that $e^{-iHt} W_\lambda e^{iHt} = W_\lambda$ for all $t \in \mathbb{R}$. This follows from

$$e^{-iHs} W_\lambda e^{iHs} \varphi - W_\lambda \varphi = \lim_{t \to \infty} e^{iHt} \chi [f \mathrm{d}\Gamma(dS) f]_t^{t+s} \chi e^{-iHt} \varphi$$

because $\chi [f \mathrm{d}\Gamma(dS) f]_t^{t+s} \chi = O(t^{-1})$. Indeed

$$\frac{\partial}{\partial t}(f \mathrm{d}\Gamma(dS) f) = f \frac{\partial}{\partial t} \mathrm{d}\Gamma(dS) f + \frac{\partial f}{\partial t} \mathrm{d}\Gamma(dS) f + f \mathrm{d}\Gamma(dS) \frac{\partial f}{\partial t}$$

where $\partial f / \partial t = O(t^{-1})$ and $\chi f \mathrm{d}\Gamma(d(\partial S/\partial t)) f \chi = O(t^{-1})$. The latter follows from (50) and Lemma 27. $\qquad \Box$

In the next step we remove the space cutoff $f$. This will allow us to prove positivity of $W = \lim_{\lambda \to \infty} W_\lambda$ away from thresholds and eigenvalues.

**Proposition 13.** *Under the assumptions of Theorem 12, the limit $W = \lim_{\lambda \to \infty} W_\lambda$ exists in operator norm sense, and $W$ is given by*

$$\langle \varphi, W \psi \rangle = \lim_{t \to \infty} \langle \varphi, e^{iHt} \chi \mathrm{d}\Gamma(dS) \chi e^{-iHt} \psi \rangle$$

*for all $\varphi, \psi \in D(\mathrm{d}\Gamma(y^2)) \cap D(N)$. $W$ commutes with $H$.*

*Proof.* We pick $\varphi, \psi \in D(\mathrm{d}\Gamma(y^2)) \cap D(N)$ and we consider the difference

$$\begin{aligned}
|\langle \varphi_t, \chi \mathrm{d}\Gamma(dS) \chi \psi_t \rangle - \langle \varphi_t, \chi f \mathrm{d}\Gamma(dS) f \chi \psi_t \rangle| & \\
\leq |\langle \varphi_t, \chi (1-f) \mathrm{d}\Gamma(dS) f \chi \psi_t \rangle| + |\langle \varphi_t, \chi \mathrm{d}\Gamma(dS)(1-f) \chi \psi_t \rangle| & \tag{66} \\
\leq \|(1-f) \chi \varphi_t\| \|\mathrm{d}\Gamma(dS) f \chi \psi_t\| + \|(1-f) \chi \psi_t\| \|\mathrm{d}\Gamma(dS) \chi \varphi_t\|. &
\end{aligned}$$

Since $(1 - f(s))^2 \leq s^2$, we have

$$\|(1 - f)\chi\varphi_t\| \leq \frac{1}{\lambda^2}\|d\Gamma(y^2/t^2)\chi\varphi_t\| \leq \frac{C}{\lambda^2}(1/t^2\|d\Gamma(y^2 + 1)\varphi\| + \|\varphi\|), \qquad (67)$$

and analogously for $\varphi$ replaced by $\psi$. The second inequality in (67) follows by Lemma 36 in Appendix F. To handle the factor $\|d\Gamma(dS)\chi\varphi_t\|$ on the r.h.s. of (66) use that, by Lemma 8 and Lemma 9,

$$dS = \frac{1}{2t}(\nabla\omega \cdot y + y \cdot \nabla\omega) - \frac{y^2}{2t^2} + O(t^{-1+\delta}). \qquad (68)$$

Part (iv) of Lemma 37 (with $y$ replaced by $y/t$) and Lemma 36 in Appendix F lead then to

$$\|d\Gamma(dS)\chi\varphi_t\| \leq C\,(1/t^2\|d\Gamma(y^2 + 1)\varphi\| + \|\varphi\|). \qquad (69)$$

Insering (67) and (69) into (66) and using $\|d\Gamma(dS)f\chi\psi_t\| \leq C\|\psi\|$ we find

$$|\langle\varphi_t, \chi d\Gamma(dS)\chi\psi_t\rangle - \langle\varphi_t, \chi f d\Gamma(dS)f\chi\psi_t\rangle| \leq \frac{C}{\lambda^2}(1/t^2\|d\Gamma(y^2 + 1)\varphi\| + \|\varphi\|)$$
$$\times (1/t^2\,\|d\Gamma(y^2 + 1)\psi\| + \|\psi\|),$$

for arbitrary $\varphi, \psi \in D(d\Gamma(y^2)) \cap D(N)$, and for all $t \geq 1$. This shows that $W$ exists as a weak limit and that

$$|\langle\varphi, (W - W_\lambda)\psi\rangle| \leq \frac{C}{\lambda}\|\varphi\|\,\|\psi\|,$$

which proves that $W$ also exists as a norm limit. Finally, that $W$ commutes with $H$ follows from the fact that $W_\lambda$ commutes with $H$ for each $\lambda$. $\qquad\square$

Using the Mourre inequality from Theorem 4, we next prove positivity of $W$ away from thresholds and eigenvalues. Note that this is the only place where the Mourre inequality is used.

**Proposition 14.** *Assume hypotheses (H1), (H2) and (H5) are satisfied. Assume, moreover, that the energy cutoff $\chi$ in the definition of $W$ satisfies* $\mathrm{supp}\,\chi \subset (-\infty, \Sigma)\backslash\mathcal{S}$*, where $\mathcal{S} = \sigma_{pp}(H) + m \cdot (\mathbb{N} \cup \{0\})$ is the set of all eigenvalues and thresholds of $H$. Then*

$$W \geq d\,\chi^2,$$

*for some $d > 0$. In particular, if $\Delta \subset (-\infty, \Sigma)\backslash\mathcal{S}$ and $\chi{\restriction}\Delta = 1$, then $W \geq d$ on* $\mathrm{Ran}E_\Delta(H)$*.*

*Proof.* By the compactness of $\overline{\sigma(H) \cap (-\infty, \Sigma)\backslash\mathcal{S}}$, and since $W$ commutes with $H$, it is enough if we prove that, for each $x \in (-\infty, \Sigma)\backslash\mathcal{S}$,

$$W{\restriction}\mathrm{Ran}E_{U_x}(H) = E_{U_x}(H)WE_{U_x}(H) \geq d_x E_{U_x}(H)\chi^2,$$

where $U_x$ is an arbitrarily small neighborhood of $x$, and $d_x > 0$. Let $\tilde{\chi} = E_{U_x}(H)$ be the spectral projection of $H$ on $U_x$. If we choose $U_x$ to be sufficiently small, then, by the Mourre estimate (Theorem 4), we have

$$\tilde{\chi}[iH, A]\tilde{\chi} \geq 2d_x \tilde{\chi}(H), \tag{70}$$

for some $d_x > 0$. Here $A = \mathrm{d}\Gamma(a) = \mathrm{d}\Gamma([i\omega, y^2/2])$. Choose now $\psi \in D(\mathrm{d}\Gamma(y^2)) \cap D(N)$ and let $\psi_t = e^{-iHt}\psi$. Then, by Proposition 13, we have

$$
\begin{aligned}
\langle \psi, \tilde{\chi} W \tilde{\chi} \psi \rangle &= \lim_{t\to\infty} \langle \psi_t, \chi \, \tilde{\chi} \mathrm{d}\Gamma(dS) \tilde{\chi} \, \chi \psi_t \rangle \\
&= \lim_{t\to\infty} \frac{d}{dt} \langle \psi_t, \chi \, \tilde{\chi} \mathrm{d}\Gamma(S) \tilde{\chi} \, \chi \psi_t \rangle,
\end{aligned}
\tag{71}
$$

where the second equality holds because, by Hypothesis (H5), $\tilde{\chi}[i\phi(G), \mathrm{d}\Gamma(S)]\tilde{\chi} = -\tilde{\chi}\phi(iSG)\tilde{\chi} = o(t^{-1+\delta})$. From (71) it now follows that

$$
\begin{aligned}
\langle \psi, \tilde{\chi} W \tilde{\chi} \psi \rangle &= \lim_{t\to\infty} \frac{1}{t} \langle \psi_t, \chi \, \tilde{\chi} \mathrm{d}\Gamma(S) \tilde{\chi} \, \chi \psi_t \rangle \\
&= \lim_{t\to\infty} \frac{1}{t^2} \langle \psi_t, \chi \, \tilde{\chi} \mathrm{d}\Gamma(y^2/2) \tilde{\chi} \, \chi \psi_t \rangle,
\end{aligned}
\tag{72}
$$

where, in the second equality, we used the definition of the function $S(y)$. Now we have

$$
\begin{aligned}
e^{iHt}\tilde{\chi}\mathrm{d}\Gamma(y^2/2)\tilde{\chi}e^{-iHt} &= \tilde{\chi}\mathrm{d}\Gamma(y^2/2)\tilde{\chi} + \int_0^t ds \, e^{iHs}\tilde{\chi}[iH, \mathrm{d}\Gamma(y^2/2)]\tilde{\chi}e^{-iHs} \\
&= \tilde{\chi}\mathrm{d}\Gamma(y^2/2)\tilde{\chi} - \int_0^t ds \, e^{iHs}\tilde{\chi}\phi(i\frac{y^2}{2}G)\tilde{\chi}e^{-iHs} + \int_0^t ds \, e^{iHs}\tilde{\chi}A\tilde{\chi}e^{-iHs} \\
&= \tilde{\chi}\mathrm{d}\Gamma(y^2/2)\tilde{\chi} - \int_0^t ds \, e^{iHs}\tilde{\chi}\phi(i\frac{y^2}{2}G)\tilde{\chi}e^{-iHs} \\
&\quad + t\tilde{\chi}A\tilde{\chi} + \int_0^t ds \int_0^s dr \, e^{iHr}\tilde{\chi}[iH, A]\tilde{\chi}e^{-iHr}.
\end{aligned}
\tag{73}
$$

Note that the operator $\phi(iy^2G)\tilde{\chi}$ is bounded. Moreover the expectation values of $\tilde{\chi}\mathrm{d}\Gamma(y^2)\tilde{\chi}$ and of $\tilde{\chi}A\tilde{\chi}$ in the state $\chi\psi$ are finite, because $\psi \in D(\mathrm{d}\Gamma(y^2)) \cap D(N)$, $\pm A \leq C \, \mathrm{d}\Gamma(y^2 + 1)$ and because of Lemma 35 (see Appendix F). Thus, after division by $t^2$, only the last term in (73) gives a non-vanishing contribution to (72) in the limit $t \to \infty$. By (70)

$$\langle \psi, \tilde{\chi} W \tilde{\chi} \psi \rangle = \lim_{t\to\infty} \frac{1}{t^2} \int_0^t ds \int_0^s dr \, \langle \psi_r, \chi \, \tilde{\chi}[iH, A]\tilde{\chi} \, \chi \psi_r \rangle \geq d_x \, \langle \psi, \chi^2 \tilde{\chi} \psi \rangle,$$

which proves the proposition because $D(\mathrm{d}\Gamma(y^2)) \cap D(N)$ is dense in $\mathcal{H}$. $\qquad \square$

## 8 Inverting the Wave Operator

The Deift-Simon wave operator $W_+$, to be constructed in this section, inverts the extended wave operator $\tilde{\Omega}_+$ with respect to $W_\lambda$ in the sense that

$$W_\lambda = \tilde{\Omega}_+ W_+ = \lim_{t \to \infty} e^{iHt} I e^{-i\tilde{H}t} W_+. \tag{74}$$

On spectral subspaces where $W_\lambda$ is positive and thus invertible, $W_+ W_\lambda^{-1}$ is then in fact a right inverse of $\tilde{\Omega}_+$. Formally, and when space and energy cutoffs are ignored, then $W_+$ is given by

$$W_+ = s - \lim_{t \to \infty} e^{i\tilde{H}t} \big[ \mathrm{d}\check{\Gamma}(j_t, dj_t)\mathrm{d}\Gamma(S) + \check{\Gamma}(j_t)\mathrm{d}\Gamma(dS) \big] e^{-iHt} \tag{75}$$

where $j_t = (j_{0,t}, j_{\infty,t})$ and $j_{0,t} + j_{\infty,t} = 1$. By the last identity, $I\check{\Gamma}(j_t) = 1$ and $I\mathrm{d}\check{\Gamma}(j_t, dj_t) = D_0[I\check{\Gamma}(j_t)] = 0$. Hence (74) is obvious at least on this formal level. The functions $j_{0,t}$ and $j_{\infty,t}$ are constructed as follows. Let $j_0$, $j_\infty \in C^\infty(\mathbb{R}^d)$ where $j_0(y) = 1$ for $|y| < 1$, $j_0(y) = 0$ for $|y| > 2$, and let $j_\infty = 1 - j_0$. Next set $j_{\sharp,t}(y) = j_\sharp(y/ut)$ where $u > 0$ is a fixed parameter. By construction of $j_t$ and $W_+$, Eq. (75), photons with velocity $u$ or larger are mapped to the second Fock space where their interaction with the electrons is turned off.

First we prove existence of $W_+$ in Theorem 15 and then we prove (74). Theorem 15 together with the Mourre estimate, Theorem 4, is the heart of our proof of AC.

Recall from Section 7 that $DA$, $D_0A$ and $da$ denote Heisenberg derivatives of operators $A$ and $a$ on $\mathcal{H}$ and $\mathfrak{h}$ respectively. If $B$ is an operator on the extended Hilbert space $\tilde{\mathcal{H}}$ and if $C$ maps $\mathcal{H}$ to $\tilde{\mathcal{H}}$, then we set

$$DB := i[\tilde{H}, B] + \frac{\partial B}{\partial t}$$

$$\tilde{D}C := i\tilde{H}C - CiH + \frac{\partial C}{\partial t}.$$

The derivatives $D_0$, and $\tilde{D}_0$ are defined in a similar way using $H_0$ and $\tilde{H}_0$ rather than $H$ and $\tilde{H}$. Finally the Heisenberg derivative $db$ of an operator $b$ mapping the one-boson sector $\mathfrak{h}$ to $\mathfrak{h} \oplus \mathfrak{h}$ (that is, $bh = (b_0h, b_\infty h)$ with $b_{0,\infty}$ being operators on $\mathfrak{h}$) is defined by

$$db = i \begin{pmatrix} \omega & 0 \\ 0 & \omega \end{pmatrix} b - b\, i\omega + \frac{\partial b}{\partial t} = \begin{pmatrix} db_0 \\ db_\infty \end{pmatrix}.$$

The next theorem is the main result of this section.

**Theorem 15.** *Assume Hypotheses (H1), (H2), and (H5). If $\chi \in C_0^\infty(\mathbb{R})$ with* $\mathrm{supp}\,\chi \subset (-\infty, \Sigma)$, *then*

$$W_+ = s - \lim_{t \to \infty} e^{i\tilde{H}t} \tilde{\chi} \tilde{f} \tilde{D}_0 \left[ \check{\Gamma}(j_t)\mathrm{d}\Gamma(S) \right] f\chi e^{-iHt}$$

*exists. Here $\chi = \chi(H)$ and $\tilde{\chi} = \chi(\tilde{H})$. Furthermore*

$$e^{-i\tilde{H}t} W_+ = W_+ e^{-iHt}.$$

*Proof.* By Cook's argument we need to show that there exists a constant $C$ such that

$$\int_1^\infty dt \left| \frac{d}{dt} \langle \psi, W_+(t)\varphi \rangle \right| \leq C \|\psi\| \tag{76}$$

for all $\psi \in \mathcal{H}$, where

$$W_+(t) = e^{i\tilde{H}t}\tilde{\chi}\tilde{f}Qf\chi e^{-iHt}$$

$$Q = \tilde{D}_0 \left[ \breve{\Gamma}(j_t)\mathrm{d}\Gamma(S) \right] = \mathrm{d}\breve{\Gamma}(j_t, dj_t)\mathrm{d}\Gamma(S) + \breve{\Gamma}(j_t)\mathrm{d}\Gamma(dS).$$

In form sense

$$\frac{d}{dt}W_+(t) = e^{i\tilde{H}t}\tilde{\chi}\tilde{D}[\tilde{f}Qf]\chi e^{-iHt}$$

$$\tilde{D}[\tilde{f}Qf] = (D\tilde{f})Qf + \tilde{f}(\tilde{D}Q)f + \tilde{f}Q(Df).$$

The contributions due to $D\tilde{f}$ and $Df$ are dealt with as in Proposition 11 and are integrable due to Proposition 10. The operator $\tilde{D}Q$ is the sum

$$\tilde{D}Q = \tilde{D}_0 Q + i(\phi(G) \otimes 1)Q - Qi\phi(G) \tag{77}$$

where the last two terms give a contribution of order $t^{-\mu}$ because of Hypothesis (H5). To show this write

$$Q = i\left[\mathrm{d}\Gamma(\omega) \otimes 1 + 1 \otimes \mathrm{d}\Gamma(\omega)\right] \breve{\Gamma}(j_t)\mathrm{d}\Gamma(S) - \breve{\Gamma}(j_t)\mathrm{d}\Gamma(S)i\mathrm{d}\Gamma(\omega)$$
$$- \frac{\partial}{\partial t}\breve{\Gamma}(j_t)\mathrm{d}\Gamma(S) \tag{78}$$

and commute $i\phi(G)$ from the right through the terms in (78) using (17) and (24) (before differentiating with respect to $t$). This is a lengthy computation which leads to

$$i(\phi(G) \otimes 1)Q - Qi\phi(G) = i\left[\phi(j_\infty G) \otimes 1 - 1 \otimes \phi(j_\infty G)\right] \tilde{D}_0 \left(\breve{\Gamma}(j_t)\mathrm{d}\Gamma(S)\right)$$
$$- i\left[\phi(d(j_0 S)iG) \otimes 1 + 1 \otimes \phi(d(j_\infty S)iG)\right] \breve{\Gamma}(j_t)$$
$$- i\left[\phi(j_0 SiG) \otimes 1 + 1 \otimes \phi(j_\infty SiG)\right] \tilde{D}_0 \left(\breve{\Gamma}(j_t)\right)$$
$$- \left[\phi(d(j_0)G) \otimes 1 + 1 \otimes \phi(d(j_\infty)G)\right] \breve{\Gamma}(j_t)\mathrm{d}\Gamma(S).$$

Using Hypothesis (H5), Eq. (50) and Lemma 27 one shows that each of these terms is of order $t^{-\mu}$. We demonstrate this for the last term.

By Lemma 27

$$t(dj_\infty)G = t\left[\nabla\omega \cdot \nabla j_\infty - i\sum_{r,s}(\partial_{rs}\omega)(\partial_{rs}j_\infty) + \partial_t j_\infty\right] G + O(t^{-2})G. \tag{79}$$

Since derivatives of $\omega$ are bounded, and since derivatives of $j_\infty$ are of order $t^{-1}$ and live on $\{|y| \geq ut\}$, Equation (79) in conjunction with Hypothesis (H5) implies

$$\left\| e^{-\alpha|x|}\phi(t(dj_\infty)G)(N+1)^{-1} \right\| = O(t^{-\mu}). \tag{80}$$

On the other hand $\mathrm{d}\Gamma(S/t)f(N+1)^{-1}$ is bounded. Hence

$$
\begin{aligned}
\tilde{\chi}\tilde{f}\,1 & \otimes \phi((dj_\infty)G)\breve{\Gamma}(j_t)\mathrm{d}\Gamma(S)f\chi \\
&= \tilde{\chi}e^{\alpha|x|}\tilde{f}\left[1\otimes e^{-\alpha|x|}\phi(t(dj_\infty)G)(N_\infty+1)^{-1}\right](N_\infty+1)(N_0+N_\infty+1)^{-1} \\
&\quad \times \breve{\Gamma}(j_t)\mathrm{d}\Gamma(S/t)f(N+1)^{-1}(N+1)^2\chi
\end{aligned}
$$

is of order $t^{-\mu}$.

To handle the contribution due to $\tilde{D}_0Q$ we work in the representation $\mathcal{F}\otimes\mathcal{F} = \oplus_{n\geq 0}\oplus_{k=0}^n \mathcal{F}_{n-k}\otimes\mathcal{F}_k$ where $\mathcal{F}_k$ is the $k$-boson subspace of $\mathcal{F}$. Since $\tilde{D}_0Q$ maps $\mathcal{F}_n$ to $\oplus_{k=0}^n\mathcal{F}_{n-k}\otimes\mathcal{F}_k$ one has

$$
\langle\psi_t,\tilde{\chi}\tilde{f}\tilde{D}_0Qf\chi\varphi_t\rangle = \sum_{n=0}^\infty\sum_{k=0}^n \langle\alpha_{t,nk},P_{n-k}\otimes P_k\tilde{D}_0Q\beta_{t,n}\rangle \tag{81}
$$

where $\alpha_{t,nk} = P_{n-k}\otimes P_k\tilde{f}\tilde{\chi}\psi_t$, $\beta_{t,n} = P_nf\chi\varphi_t$ and $P_n$ is the orthogonal projection $\mathcal{F}\to\mathcal{F}_n$. Note that $\tilde{\chi} = \tilde{\chi}\chi(N_\infty \leq n_\infty)$ for some $n_\infty$ large enough, and hence $\alpha_{t,nk} = 0$ for $k > n_\infty$. Next we estimate $|\langle\alpha_{t,nk},P_{n-k}\otimes P_k\tilde{D}_0Q\beta_{t,n}\rangle|$ for each given $n$ and $k$ separately. To this end we identify $\mathcal{F}_{n-k}\otimes\mathcal{F}_k$ with a subspace of $L(\mathbb{R}^{dn})$. Then, on $\mathcal{F}_n$, the operator

$$
P_{n-k}\otimes P_k\breve{\Gamma}(j_t)\mathrm{d}\Gamma(S) = J_{nk}S_n =: S_{nk}
$$

acts by multiplication with a function $J_{nk}S_n$ where $S_n(y,t) = \sum_{i=1}^n S(y_i,t)$ and by (26)

$$
J_{nk} = \binom{n}{k}^{1/2}\underbrace{j_0\otimes\ldots\otimes j_0}_{n-k\text{ factors}}\otimes\underbrace{j_\infty\otimes\ldots\otimes j_\infty}_{k\text{ factors}}. \tag{82}
$$

In terms of $S_{nk}$ the operator $P_{n-k}\otimes P_k\tilde{D}_0Q$ is given by

$$
P_{n-k}\otimes P_k\tilde{D}_0Q = D_0^2\left(P_{n-k}\otimes P_k\breve{\Gamma}(j_t)\mathrm{d}\Gamma(S)\right) = D_0^2S_{nk} \tag{83}
$$

where $D_0S_{nk} = [i\Omega,S_{nk}]+\partial S_{nk}/\partial t$ and $\Omega(k_1,\ldots,k_n) = \sum_{i=1}^n\omega(k_i)$. Let $V = (y_1,\ldots,y_n)/t \in \mathbb{R}^{dn}$ and let $\mathcal{D}_V = \partial/\partial t+V\cdot\nabla$ denote the material derivative w.r. to $V$. In the appendix we show that

$$
\begin{aligned}
D_0^2S_{nk} =&(\nabla\Omega - V)\cdot S_{nk}''(\nabla\Omega - V) \\
&+ (\mathcal{D}_V\nabla S_{nk})\cdot(\nabla\Omega - V) + (\nabla\Omega - V)\cdot(\mathcal{D}_V\nabla S_{nk}) \\
&+ \mathcal{D}_V^2S_{nk} + n^2\binom{n}{k}^{1/2}O(t^{-1-\delta})
\end{aligned} \tag{84}
$$

for $|y|/t \leq 2\lambda$ (Lemma 40) and that

$$
\mathcal{D}_V^m\partial^\alpha\left(S_{nk} - J_{nk}\sum_{i=1}^n y_i^2/2t\right) = n^{m+1}\binom{n}{k}^{1/2}O(t^{-(1+m)+2(\delta-|\alpha|)}) \tag{85}
$$

(Lemma 39). These are analogs of Lemma 9 and Lemma 8. The binomial factor in (84) and (85) stems from (82) and will be estimated by $n^k$. From Eq. (85) we get for the last term of (84)

$$\mathcal{D}_V^2(J_{nk}S_n) = J_{nk}\mathcal{D}_V^2\left(S_n - \sum_{i=1}^n \frac{y_i^2}{2t}\right) = n^{k/2+3}O(t^{-3+2\delta}). \tag{86}$$

where we used that $\mathcal{D}_V f = 0$ for any function $f(y,t)$ which only depends on $V$ and $D_V^2(y^2/2t) = 0$. For the second and third term in Eq. (84), Eq. (85) shows that

$$\pm\left[(\mathcal{D}_V\nabla S_{nk})\cdot(\nabla\Omega - V) + (\nabla\Omega - V)\cdot(\mathcal{D}_V\nabla S_{nk})\right]$$
$$\leq t^{2-\delta}(\mathcal{D}_V\nabla S_{nk})^2 + t^{-2+\delta}(\nabla\Omega - V)^2 = O(t^{-2+\delta})(n+V^2)n^{k+4} \tag{87}$$

Combining (81) and (83) with (84), (86) and (87) we get

$$\left|\langle\psi_t, \tilde{\chi}\tilde{f}\tilde{D}_0 Qf\varphi_t\rangle\right| \leq \sum_{n=0}^\infty \sum_{k=0}^n |\langle\alpha_{t,nk}, (\nabla\Omega - V)\cdot S_{nk}''(\nabla\Omega - V)\beta_{t,n}\rangle|$$
$$+ \sum_{n=0}^\infty \sum_{k=0}^n \|\alpha_{t,nk}\|\,\|\beta_{t,n}\|n^{k+4}O(t^{-1-\varepsilon}) \tag{88}$$

where $\varepsilon = \min(1-\delta, \delta) > 0$. Here $n^{k+1} \leq n^{n_\infty+1}$ because $\alpha_{t,nk} = 0$ for $k > n_\infty$. Hölder's inequality in the form

$$\sum_{n=0}^\infty \sum_{k=0}^n A_{nk}B_n \leq \left(\sum_{n=0}^\infty \sum_{k=0}^n A_{nk}^2\right)^{1/2}\left(\sum_{n=0}^\infty (n+1)B_n^2\right)^{1/2} \tag{89}$$

where $A_{nk},\ B_n \geq 0$, will be used frequently in the following. It shows that the second term in (88) is bounded from above by $\|\tilde{f}\tilde{\chi}\psi_t\|\,\|(N+1)^{n_\infty+5/2}f\chi\varphi_t\|O(t^{-1-\varepsilon})$ which is integrable.

To deal with the first term in (88) we use that

$$\pm S_{nk}''(y,t) \leq \text{const}\times n^2\binom{n}{k}^{1/2}S_n''(y,t),$$

for $|y| \leq 2\lambda t$ and $ut^{\delta-1} \geq 2$ by Lemma 38. This allows us to estimate the contribution due to $S_{nk}''\chi(|V| \leq 2\lambda)$. The contribution due to $S_{nk}''\chi(|V| > 2\lambda)$ is bounded by $O(t^{-2})(n+1)^4\|\alpha_{t,nk}\|\,\|\beta_{t,n}\|$ thanks to the space cutoff $f$. Together with (88), the Schwarz inequality, and $\|\alpha_{t,nk}\| = 0$ for $k > n_\infty$, this implies that

$$|\langle\alpha_{t,nk}, (\nabla\Omega - V)\cdot S_{nk}''(\nabla\Omega - V)\beta_{t,n}\rangle| \leq |\langle\alpha_{t,nk}, (\nabla\Omega - V)\cdot S_n''(\nabla\Omega - V)\alpha_{t,nk}\rangle|^{1/2}$$
$$\times |\langle\beta_{t,n}, (\nabla\Omega - V)\cdot S_n''(\nabla\Omega - V)\beta_{t,n}\rangle|^{1/2}\,n^{n_\infty/2+2}$$
$$+ O(t^{-2})(n+1)^4\|\alpha_{t,nk}\|\,\|\beta_{t,n}\|$$

for $ut^{\delta-1} \geq 2$. Now insert this bound into (88), use that $(\nabla\Omega - V) \cdot S''_{nk}(\nabla\Omega - V) = \sum_{i=1}^{n} (\nabla\omega(k_i) - y_i/t) \cdot S''(\nabla\omega(k_i) - y_i/t)$, sum over $k$ and $n$, and apply (89) to see that

$$\sum_{n=0}^{\infty} \sum_{k=0}^{n} |\langle \alpha_{t,nk}, (\nabla\Omega - V) \cdot S''_{nk}(\nabla\Omega - V)\beta_{t,n}\rangle|$$

$$\leq \left| \langle \tilde{f}\tilde{\chi}\psi_t, \mathrm{d}\Gamma\left[(\nabla\omega - v) \cdot S''(\nabla\omega - v)\right] \otimes 1 + 1 \otimes \mathrm{d}\Gamma\left[(\nabla\omega - v) \cdot S''(\nabla\omega - v)\right]\tilde{f}\tilde{\chi}\psi_t\rangle \right|^{1/2}$$

$$\times \left| \langle f\chi\varphi_t, N^{(n_\infty+5)/2}\mathrm{d}\Gamma\left[(\nabla\omega - v) \cdot S''(\nabla\omega - v)\right]N^{(n_\infty+5)/2}f\chi\varphi_t\rangle \right|^{1/2}$$

$$+ O(t^{-2})\,\|\tilde{f}\tilde{\chi}\psi_t\|\,\|(N+1)^{9/2}f\,\chi\varphi_t\|.$$

This is integrable w. r. to $t$ by Proposition 11.

To prove the last statement it suffices to show that $e^{-i\tilde{H}s}W_+e^{iHs}\varphi = W_+\varphi$ for all $s \in \mathbb{R}$. This follows from

$$e^{-i\tilde{H}s}W_+e^{iHs}\varphi - W_+\varphi = \lim_{t\to\infty} e^{i\tilde{H}t}\tilde{\chi}[\tilde{f}Qf]_t^{t+s}\chi e^{-iHt}\varphi$$

if we prove that $\tilde{\chi}\partial/\partial t[\tilde{f}Qf]\chi = O(t^{-1})$. The contributions due to $\partial_t \tilde{f}$ and $\partial_t f$ are easily seen to be of order $t^{-1}$. As for $\partial_t Q$ note that, by Lemma 39,

$$P_{n-k} \otimes P_k \partial_t Q = D_0 \partial_t S_{nk} = \nabla\Omega \cdot \nabla(\partial_t S_{nk}) + \frac{\partial^2 S_{nk}}{\partial t^2} + n^2\binom{n}{k}^{1/2} O(t^{-2})$$

$$= n^2\binom{n}{k}^{1/2} O(t^{-1})$$

for $|y/t| \leq 2\lambda$. Use this, (89), and that $k \leq n_\infty$ thanks to the energy cutoff $\tilde{\chi}$. $\qquad\square$

By construction of $W_+$, $W_\lambda = \tilde{\Omega}_+W_+$ as we show in the next lemma. Some minor technical difficulties in its proof are due to the presence of the cutoffs and due to the unboundedness of $I$.

**Lemma 16.** *Suppose $W_\lambda$ and $W_+$ are defined as in Theorem 12 and Theorem 15. Then, under the assumptions of these theorems,*

$$W_\lambda = s - \lim_{t\to\infty} e^{iHt}Ie^{-i\tilde{H}t}W_+ = \tilde{\Omega}_+W_+.$$

*Proof.* By definition of $W_+$ we have, for all $\varphi \in \mathcal{H}$,

$$Ie^{-i\tilde{H}t}W_+\varphi = I\tilde{\chi}\tilde{f}\tilde{D}_0[\check{\Gamma}(j_t)\mathrm{d}\Gamma(S)]f\chi e^{-iHt}\varphi + o(1), \quad \text{for } t \to \infty. \tag{90}$$

Note here that $e^{-i\tilde{H}t}W_+\varphi$ is in the domain of $I$, for all $t \in \mathbb{R}$, because $W_+\varphi = \chi'(\tilde{H})W_+\varphi$, for any $\chi' \in C_0^\infty(\mathbb{R})$, with $\chi' = 1$ on supp $\chi$. From (90) it follows now, if we expand the free Heisenberg derivative, that

$$Ie^{-i\tilde{H}t}W_+\varphi = I\tilde{\chi}\tilde{f}\mathrm{d}\check{\Gamma}(j_t, dj_t)\mathrm{d}\Gamma(S)f\chi e^{-iHt}\varphi + I\tilde{\chi}\tilde{f}\check{\Gamma}(j_t)\mathrm{d}\Gamma(dS)f\chi e^{-iHt}\varphi + o(1)$$

$$= f\,I\tilde{\chi}\mathrm{d}\check{\Gamma}(j_t, dj_t)\mathrm{d}\Gamma(S)f\chi e^{-iHt}\varphi + I\tilde{\chi}\check{\Gamma}(j_t)f\mathrm{d}\Gamma(dS)f\chi e^{-iHt}\varphi + o(1). \tag{91}$$

To prove the second equality commute $\tilde{f}$ to the left in the first term, using that, by Lemma 28, $[\tilde{\chi}, \tilde{f}] = O(t^{-1})$, and to the right in the second term, using that $\tilde{f}\breve{\Gamma}(j_t) = \breve{\Gamma}(j_t)f$. Choose now $\chi_1 \in C_0^\infty(\mathbb{R})$, with $\chi_1 = 1$ on supp $\chi$, and supp $\chi_1 \subset (-\infty, \Sigma)$, and set $\chi_1 = \chi_1(H)$. Then $\chi_1\chi = \chi$, and thus, by (91),

$$Ie^{-i\tilde{H}t}W_+\varphi = f\, I\tilde{\chi}\, \mathrm{d}\breve{\Gamma}(j_t, dj_t)\mathrm{d}\Gamma(S)f\chi_1\,\chi e^{-iHt}\varphi + I\tilde{\chi}\,\breve{\Gamma}(j_t)f\mathrm{d}\Gamma(dS)f\chi_1\,\chi e^{-iHt}\varphi + o(1)$$
$$= f\, I\tilde{\chi}\, \mathrm{d}\breve{\Gamma}(j_t, dj_t)\chi_1\mathrm{d}\Gamma(S)f\chi e^{-iHt}\varphi + I\tilde{\chi}\,\breve{\Gamma}(j_t)\chi_1 f\mathrm{d}\Gamma(dS)f\chi e^{-iHt}\varphi + o(1),$$
$$(92)$$

where, in the second equality, we used that $[f\mathrm{d}\Gamma(S)f, \chi_1]\chi = O(t^{-1})$, for $t \to \infty$, which easily follows, expanding $\chi_1$ in a Hellfer-Sjöstrand integral, by Hypothesis (H3), by Lemma 28 and because, by assumption, $e^{\alpha|x|}\chi$ is a bounded operator, for some $\alpha > 0$. Below we will show that

$$I\tilde{\chi}\,\breve{\Gamma}(j_t)\,\chi_1 = \chi + o(1) \quad \text{and} \tag{93}$$

$$I\tilde{\chi}\,\mathrm{d}\breve{\Gamma}(j_t, dj_t)\,\chi_1 = o(t^{-1}). \tag{94}$$

Inserting these two equations in (92) it follows, since $\mathrm{d}\Gamma(S/t)f\chi$ is uniformly bounded in $t$,

$$Ie^{-i\tilde{H}t}W_+\varphi = \chi f\mathrm{d}\Gamma(dS)f\chi e^{-iHt}\varphi + o(1) = e^{-iHt}W_\lambda\varphi + o(1),$$

which proves the lemma. It only remains to prove (93) and (94). We begin proving (93). To this end we note that $I\breve{\Gamma}(j_t) = 1_\mathcal{H}$ (because, by construction of $j_t$, $j_{0,t} + j_{\infty,t} = 1$) and thus, for any $n \in \mathbb{N}$,

$$\chi = I\breve{\Gamma}(j_t)\chi = IE_{[0,n]}(N_\infty)\breve{\Gamma}(j_t)\chi + IE_{(n,\infty)}(N_\infty)\breve{\Gamma}(j_t)\chi, \tag{95}$$

where $E_\Delta$ denotes the characteristic function of the set $\Delta$. Now we claim that the norm of the second term on the r.h.s. of the last equation can be made arbitrarily small by choosing $n$ sufficiently large. This follows because

$$IE_{(n,\infty)}(N_\infty)\breve{\Gamma}(j_t)\chi = IE_{(n,\infty)}(N_\infty)\breve{\Gamma}(j_t)E_{(n,\infty)}(N)\chi,$$

which implies, since $\|IE_{(n,\infty)}(N_\infty)\breve{\Gamma}(j_t)\| \le 1$ for all $n \in \mathbb{N}$, that

$$\|IE_{(n,\infty)}(N_\infty)\breve{\Gamma}(j_t)\chi\| \le \|IE_{(n,\infty)}(N_\infty)\breve{\Gamma}(j_t)\|\, \|E_{(n,\infty)}(N)(N+1)^{-1}\|\, \|(N+1)\chi\|$$
$$\le \frac{C}{n+2}. \tag{96}$$

On the other hand the first term on the r.h.s. of (95) can be written as

$$IE_{[0,n]}(N_\infty)\breve{\Gamma}(j_t)\chi = I(N_0+1)^{-n}E_{[0,n]}(N_\infty)(N_0+1)^n\breve{\Gamma}(j_t)\chi\chi_1$$
$$= I(N_0+1)^{-n}E_{[0,n]}(N_\infty)(N_0+1)^n\tilde{\chi}\breve{\Gamma}(j_t)\chi_1 + o(1), \tag{97}$$

where the second equality follows because of Lemma 32, and because $I(N_0+1)^{-n}E_{[0,n]}(N_\infty)$ is a bounded operator (see Lemma 1). Now, for $n \in \mathbb{N}$ sufficiently large, $E_{[0,n]}(N_\infty)\tilde{\chi} = \tilde{\chi}$. This remark, together with (95), (96) and (97) shows that, for any $\varepsilon > 0$, $\|\chi - I\tilde{\chi}\breve{\Gamma}(j_t)\chi_1\| < \varepsilon$, for $t$ sufficiently large. This proves (93). Eq.(94) follows in a very similar way by $I\mathrm{d}\breve{\Gamma}(j_t, dj_t) = D_0(I\breve{\Gamma}(j_t)) = D_0(1_\mathcal{H}) = 0$, using that $\|IE_{(n,\infty)}(N_\infty)\mathrm{d}\breve{\Gamma}(j_t, dj_t)(N+1)^{-1}\| \le$ const and applying Lemma 33 (see Appendix D). $\qquad\square$

So far the positive parameter $u$ in the construction of $W_+$ was arbitrary. The next lemma now shows that, by choosing $u$ small, we can neglect the possibility to find zero "escaping photons" in states in the range of $W_+$. This will be important in the proof of asymptotic completeness, Theorem 19.

**Lemma 17.** *Assume Hypotheses (H1), (H2) and (H5) hold, and let the Deift-Simon wave operator $W_+$ be defined as in Theorem 15. Then*

$$\|(1 \otimes E_{\{0\}}(N))W_+\| \leq 2\,u^2 \|(N+1)^{1/2}\chi(H)\|^2.$$

*Proof.* For $\varphi \in \mathcal{H}$ and $\psi \in \mathcal{H} \otimes \mathcal{F}$ we define $\varphi_t = e^{-iHt}\varphi$ respectively $\psi_t = e^{-i\tilde{H}t}\psi$. Then by definition of $W_+$, we have

$$\begin{aligned}
\langle \psi, W_+\varphi \rangle &= \lim_{t\to\infty} \langle \psi, e^{i\tilde{H}t}\tilde{\chi}\tilde{f}\tilde{D}_0 \left\{ \breve{\Gamma}(j_t)\mathrm{d}\Gamma(S) \right\} f e^{-iHt}\chi\varphi \rangle \\
&= \lim_{t\to\infty} \langle \psi_t, \tilde{\chi}\tilde{f}\tilde{D} \left\{ \breve{\Gamma}(j_t)\mathrm{d}\Gamma(S) \right\} f\chi\varphi_t \rangle.
\end{aligned} \tag{98}$$

The last equality follows because, by assumption, the operator $e^{\alpha|x|}\chi$ is bounded for some $\alpha > 0$ and because the norm of the operator

$$e^{-\alpha|x|} \left\{ (i\phi(G) \otimes 1)\breve{\Gamma}(j_t)\mathrm{d}\Gamma(S) - \breve{\Gamma}(j_t)\mathrm{d}\Gamma(S)i\phi(G) \right\} f(N+1)^{-2} \tag{99}$$

tends to 0, as $t \to \infty$. To see this write the operator in (99) as

$$\begin{aligned}
e^{-\alpha|x|} &\left\{ (i\phi(G) \otimes 1)\breve{\Gamma}(j_t) - \breve{\Gamma}(j_t)i\phi(G) \right\} \mathrm{d}\Gamma(S)f(N+1)^{-2} \\
&+ \breve{\Gamma}(j_t)e^{-\alpha|x|}[i\phi(G), \mathrm{d}\Gamma(S)]f(N+1)^{-2}.
\end{aligned} \tag{100}$$

Now the operator in the first line equals, by (24),

$$-ie^{-\alpha|x|} \left\{ \phi((j_{0,t}-1)G) \otimes 1 + 1 \otimes \phi(j_{\infty,t}G) \right\} (N_0 + N_\infty + 1)^{-1}\breve{\Gamma}(j_t)\,\mathrm{d}\Gamma(S)f(N+1)^{-1},$$

and tends to 0 as $t \to \infty$. This follows because $\mathrm{d}\Gamma(S)f(N+1)^{-1} = O(t)$, while, by Hypothesis (H5), the factor on the left of $\breve{\Gamma}(j_t)$ is of order $O(t^{-\mu})$, for some $\mu > 1$.

To handle the operator in the second line of (100) use that $[i\phi(G), \mathrm{d}\Gamma(S)] = \phi(-iSG)$, and that, by (H5),

$$\sup_x e^{-\alpha|x|} \|\phi(-iSG_x)(N+1)^{-1}\| = \sup_x e^{-\alpha|x|} \|\phi(\chi(|y| \geq t^\delta)iSG_x)(N+1)^{-1}\| = O(t^{-1+\delta}).$$

This implies that also the term in the second line of (100) tends to zero as $t \to \infty$.

From (98) it follows now, because terms involving the Heisenberg derivative of $f$, or $\tilde{f}$ give, by Lemma 28, a vanishing contribution in the limit $t \to \infty$, that

$$\begin{aligned}
\langle \psi, W_+\varphi \rangle &= \lim_{t\to\infty} \frac{d}{dt} \langle \psi_t, \tilde{\chi}\tilde{f}\breve{\Gamma}(j_t)\mathrm{d}\Gamma(S)f\chi\varphi_t \rangle = \lim_{t\to\infty} \frac{1}{t} \langle \psi_t, \tilde{\chi}\tilde{f}\breve{\Gamma}(j_t)\mathrm{d}\Gamma(S)f\chi\varphi_t \rangle \\
&= \lim_{t\to\infty} \frac{1}{t} \langle \breve{\Gamma}(j_t)^*\tilde{\chi}\psi_t, f\mathrm{d}\Gamma(y^2/2t)f\chi\varphi_t \rangle,
\end{aligned} \tag{101}$$

where, in the last equality, we used Lemma 8 to replace $S$ by $y^2/2t$. Consider in particular the case $\psi = (1 \otimes E_{\{0\}}(N))\psi$; then there exists some $\alpha \in \mathcal{H}$ with $\psi = \alpha \otimes \Omega$, and

$$\breve{\Gamma}(j_t)^* \tilde{\chi} \psi_t = I \left( \Gamma(j_{0,t}) \otimes \Gamma(j_{\infty,t}) \right) \left( (\chi \alpha_t) \otimes \Omega \right) = \Gamma(j_{0,t}) \chi \alpha_t.$$

For such $\psi = \alpha \otimes \Omega$, (101) reads

$$\langle \psi, W_+ \varphi \rangle = \lim_{t \to \infty} \frac{1}{2t^2} \langle \alpha_t, \chi f \Gamma(j_{0,t}) \mathrm{d}\Gamma(y^2) f \chi \varphi_t \rangle. \tag{102}$$

Now we note that

$$
\begin{aligned}
|\langle \alpha_t, \chi f \Gamma(j_{0,t}) \mathrm{d}\Gamma(y^2) f \chi \varphi_t \rangle| &= |\langle \alpha_t, \chi f \mathrm{d}\Gamma(j_{0,t}, j_{0,t} y^2) f \chi \varphi_t \rangle| \\
&\leq \|(N+1)^{1/2} \chi \alpha_t\| \, \|(N+1)^{1/2} \chi \varphi_t\| \, \|j_{0,t} y^2\| \\
&\leq 4u^2 t^2 \|\alpha\| \, \|\varphi\| \, \|(N+1)^{1/2} \chi\|^2,
\end{aligned}
\tag{103}
$$

where, in the last step, we used that $\|j_{0,t} y^2\| \leq \sup_{|y| \leq 2ut} y^2 \leq 4t^2 u^2$. Since $\|\alpha\| = \|\psi\|$, it follows from (103) and (102) that

$$|\langle \psi, (1 \otimes E_{\{0\}}(N)) W_+ \varphi \rangle| \leq 2u^2 \|(N+1)^{1/2} \chi(H)\|^2 \|\psi\| \, \|\varphi\|.$$

$\square$

# 9 Asymptotic Completeness

As explained in the introduction we prove asymptotic completeness by induction in the energy measured in units of $m$. The first step is the following, essentially trivial lemma. The idea is that AC on $\operatorname{Ran} E_\eta(H)$ as explained in the introduction in Eq. 3, implies the same property for $I e^{-i\tilde{H}t}$ on $\operatorname{Ran} E_\eta(H) \otimes \mathcal{F}$, the photons from $\mathcal{F}$ merely contributing to the asymptotically free radiation.

**Lemma 18.** *Assume that hypotheses (H1), (H2) and (H4) are satisfied, and let the wave operators $\tilde{\Omega}_+$ and $\Omega_+$ be defined as in Lemma 6 and in Theorem 7, respectively. Suppose $\operatorname{Ran}\Omega_+ \supset E_\eta(H)\mathcal{H}$ and $\mu < \Sigma$. Then for every $\varphi \in \operatorname{Ran} E_\mu(\tilde{H})$ there exists a vector $\psi \in \operatorname{Ran} E_\mu(\tilde{H})$ such that*

$$\tilde{\Omega}_+(E_\eta(H) \otimes 1)\varphi = \Omega_+\psi.$$

*If $\varphi \in E_\Delta(\tilde{H})\tilde{\mathcal{H}}$ then $\psi \in E_\Delta(\tilde{H})\tilde{\mathcal{H}}$.*

*Proof.* By Lemma 30 every given $\varphi \in E_\mu(\tilde{H})\tilde{\mathcal{H}}$ can be approximated by a sequence of vectors $\varphi_n \in E_\mu(\tilde{H})\tilde{\mathcal{H}}$ which are finite linear combinations of vectors of the from

$$\gamma = \alpha \otimes a^*(h_1)\ldots a^*(h_n)\Omega, \qquad \lambda + \sum_{i=1}^n M_i < \mu \tag{104}$$

where $\alpha = E_\lambda(H)\alpha$ and $M_i = \{|k| : h_i(k) \neq 0\}$. Let $\gamma \in \tilde{\mathcal{H}}$ be of the form (104). Then

$$\begin{aligned}
I e^{-i\tilde{H}t}(E_\eta(H) \otimes 1)\,\gamma &= I e^{-i\tilde{H}t} E_\eta(H)\alpha \otimes a^*(h_1)\ldots a^*(h_n)\Omega \\
&= a^*(h_{1,t})\ldots a^*(h_{n,t})\,e^{-iHt}\,E_\eta(H)\alpha.
\end{aligned} \tag{105}$$

By assumption $E_\eta(H)\alpha = \Omega_+\beta$ for some $\beta \in \tilde{\mathcal{H}}$ and we may assume $\beta = E_\eta(\tilde{H})\beta$ by the intertwining relation for $\Omega_+$. From (105) it follows that

$$\begin{aligned}
I e^{-i\tilde{H}t}(E_\eta(H) \otimes 1)\,\gamma &= a^*(h_{1,t})\ldots a^*(h_{n,t})e^{-iHt}\Omega_+\beta \\
&= a^*(h_{1,t})\ldots a^*(h_{n,t})I e^{-i\tilde{H}t}(P_B \otimes 1)\,\beta \\
&\quad + a^*(h_{1,t})\ldots a^*(h_{n,t})\left\{e^{-iHt}\,\Omega_+\beta - I e^{-i\tilde{H}t}\,(P_B \otimes 1)\,\beta\right\},
\end{aligned} \tag{106}$$

where $P_B$ denotes the orthogonal projector onto $\mathcal{H}_{pp}(H)$. After inserting a factor $(N+1)^{-n/2}(N+1)^{n/2}$ in the second factor on the r.h.s. of last equation, just in front of the braces, we get

$$\begin{aligned}
\|I e^{-i\tilde{H}t}&(E_\eta(H) \otimes 1)\,\gamma - I e^{-i\tilde{H}t}(P_B \otimes 1)\,(1 \otimes a^*(h_1)\ldots a^*(h_n))\beta\| \\
&\leq \|a^*(h_{1,t})\ldots a^*(h_{n,t})(N+1)^{-n/2}\|\,\|(N+1)^{n/2}\left\{e^{-iHt}\Omega_+\beta - I e^{-i\tilde{H}t}(P_B \otimes 1)\beta\right\}\|.
\end{aligned} \tag{107}$$

The first factor on the r.h.s. is bounded by a finite constant, uniformly in $t$. The second factor converges to zero as $t \to \infty$. To see this use that it stays bounded for *all* integers $n$, a fact which follows from the boundedness of $(H+i)^{n/2}\,\Omega_+ E_\mu(\tilde{H})$ and of $(N+1)^{n/2}\,I\,E_\mu(\tilde{H})$. Since $(1 \otimes a^*(h_1)\ldots a^*(h_n))\beta \in E_\mu(\tilde{H})\tilde{\mathcal{H}}$ it follows that

$$\tilde{\Omega}_+(E_\eta(H) \otimes 1)\gamma = \Omega_+(1 \otimes a^*(h_1)\ldots a^*(h_n))\beta.$$

Hence to each $\varphi_n$, as defined at the very beginning, there exists a vector $\psi_n \in E_\mu(H)\tilde{\mathcal{H}}$ such that $\tilde{\Omega}_+(E_\eta(H) \otimes 1)\varphi_n = \Omega_+\psi_n$. The left side converges to $\tilde{\Omega}_+(E_\eta(H) \otimes 1)\varphi$ as $n \to \infty$, and hence the right side converges as well. Since $\Omega_+$ is isometric on $\mathcal{H}_{pp}(H) \otimes \mathcal{F}$ it follows that $(P_B \otimes 1)\psi_n$ is Cauchy and hence has a limit $\psi \in E_\mu(H)\mathcal{H}$. Thus $\tilde{\Omega}_+(E_\eta(H) \otimes 1)\varphi = \Omega_+\psi$ which proves the first part of the lemma. The second part follows from the intertwining relations for $\tilde{\Omega}_+$ and $\Omega_+$. $\qquad\square$

**Theorem 19 (Asymptotic Completeness).** *Assume hypotheses (H1) through (H5) are satisfied. Then*

$$\operatorname{Ran}\Omega_+ \supset E_{(-\infty,\Sigma)}(H)\mathcal{H}.$$

*Proof.* The proof is by induction. We show that

$$\operatorname{Ran}(\Omega_+) \supset E_{(-\infty,km)}(H)E_{(-\infty,\Sigma)}(H)\mathcal{H} \tag{108}$$

holds for all integers $k$ with $(k-1)m \leq \Sigma$. For $k < 0$ and $|k|$ large enough (108) is trivially correct because $H$ is bounded from below. Hence we may assume (108) holds for $k = n-1$. To prove it for $k = n$ it suffices to show that

$$\operatorname{Ran}\Omega_+ \supset E_\Delta(H)\mathcal{H} \tag{109}$$

for compact intervals $\Delta \subset (-\infty, nm) \cap (-\infty, \Sigma)\backslash\mathcal{S}$ because, by Theorem 4, the union of such subspaces is dense in $E_{(-\infty,nm)}(H)E_{(-\infty,\Sigma)}(H)\mathcal{H}$ and because $\operatorname{Ran}\Omega_+$ is closed. Now choose $\chi \in C_0^\infty(\mathbb{R})$ such that $\chi = 1$ on $\Delta$ and $\operatorname{supp}\chi \subset (-\infty, \Sigma)\backslash S$. Let $\lambda,\ u > 0$ and define $W_\lambda$ and $W_+$ in terms of $\chi$, $\lambda$ and $u$ as in Theorem 12 and 15. Define moreover the asymptotic observable $W = \lim_{\lambda \to \infty} W_\lambda$ as in Proposition 13. By Proposition 14, the operator $W : E_\Delta(H)\mathcal{H} \to E_\Delta(H)\mathcal{H}$ is onto and hence for every given $\varphi = E_\Delta(H)\varphi$ there exists a $\psi \in \operatorname{Ran} E_\Delta(H)$ such that $\varphi = W\psi$. Given $\varepsilon > 0$ we pick $u, \lambda$ small, respectively large enough so that

$$\|\tilde{\Omega}_+\chi(N_\infty = 0)W_+\psi\| < \varepsilon, \qquad \|(W - W_\lambda)\psi\| < \varepsilon, \tag{110}$$

by using Lemma 17, Proposition 13, and the boundedness of $\tilde{\Omega}_+ E_\Delta(\tilde{H})$. Then by Lemma 16

$$W_\lambda\psi = \tilde{\Omega}_+ W_+\psi = \tilde{\Omega}_+\chi(N_\infty > 0)W_+\psi + \tilde{\Omega}_+\chi(N_\infty = 0)W_+\psi. \tag{111}$$

The vector $\chi(N_\infty > 0)W_+\psi$ has at least one boson in the outer Fock space and thus an energy of at most $(n-1)m$ in the inner one. More precisely

$$\chi(N_\infty > 0)W_+\psi = \chi(H < (n-1)m) \otimes \chi(N > 0)W_+\psi.$$

Hence by induction hypothesis and Lemma 18 there exist $\gamma \in E_\Delta(\mathcal{H})\tilde{\mathcal{H}}$ such that

$$\tilde{\Omega}_+\chi(N_\infty > 0)W_+\psi = \Omega_+\gamma.$$

This equality together with (110) and (111) shows that

$$\|\varphi - \Omega_+\gamma\| = \|W\psi - \tilde{\Omega}_+\chi(N_\infty > 0)W_+\varphi\| < 2\varepsilon$$

which proves the theorem. $\qquad\qquad\square$

# 10    Massless Photons

This section is devoted to the case where the bosons are massless photons, but the soft modes do not interact with the particles (electrons). That is

$$H = K \otimes 1 + 1 \otimes d\Gamma(|k|) + \phi(G) = H_0 + \phi(G) \tag{112}$$

and

(IR)                    $$G_x(k) = 0 \quad \text{if} \quad |k| < m$$

for some $m > 0$. As before, we assume that $G_x \in \mathfrak{h}$ for each $x \in X$ and that $\sup_x \|G_x\| < \infty$. Then $\phi(G)(H_0 + 1)^{-1/2}$ is again bounded and hence $H$ is self-adjoint on $D(H_0)$. The key idea is to compare $H$ with the modified Hamiltonian

$$H_{\text{mod}} = K \otimes 1 + 1 \otimes d\Gamma(\omega) + \phi(G) \tag{113}$$

where $K$ and $G$ are as above but the dispersion $\omega(k) = |k|$ is modified for $|k| < m$. We choose $\omega$ in such a way that (H1) is satisfied (with $m/2$ instead of $m$) and $\omega(k) = |k|$ for $|k| \geq m$.

## 10.1    Asymptotic Completeness

Asymptotic completeness for $H$ is essentially a corollary of Theorem 19. Let $\tilde{H} = H \otimes 1 + 1 \otimes d\Gamma(|k|)$ and let $\Omega_+$ and $\Omega_{+,\text{mod}}$ be defined in terms of $H$, $H_{\text{mod}}$, and $\tilde{H}_{\text{mod}} = H_{\text{mod}} \otimes 1 + 1 \otimes d\Gamma(\omega)$.

**Theorem 20 (Asymptotic Completeness).** *Assume (IR), (H2), and (H5) for the system defined by (112). Then the wave operator $\Omega_+$ exists on $E_{(-\infty,\Sigma)}(\tilde{H})\tilde{\mathcal{H}}$ and*

$$\text{Ran}(\Omega_+) \supset E_{(-\infty,\Sigma)}(H)\mathcal{H}.$$

*Proof.* We split the Fock space into two Fock spaces, one with interacting photons the other one with non-interacting photons. Henceforth the subindices $i$ and $s$ refer to *interacting* and *soft* respectively. Let

$$K_i = \{k \in \mathbb{R}^d : |k| \geq m\} \qquad \mathfrak{h}_i = L^2(K_i) \tag{114}$$

$$K_s = \{k \in \mathbb{R}^d : |k| < m\} \qquad \mathfrak{h}_s = L^2(K_s). \tag{115}$$

Then $\mathfrak{h} = \mathfrak{h}_i \oplus \mathfrak{h}_s$ and correspondingly $\mathcal{F}(\mathfrak{h})$ is isomorphic to $\mathcal{F}(\mathfrak{h}_i) \otimes \mathcal{F}(\mathfrak{h}_s)$ with an isomorphism $U$ as given in Section 2.5. By assumption on $G$ and $\omega$ and by (21)

$$U H_{\text{mod}} U^* = H_i \otimes 1 + 1 \otimes d\Gamma(\omega_s)$$
$$U H U^* = H_i \otimes 1 + 1 \otimes d\Gamma(|k|_s)$$

with respect to the factorization $\mathcal{H} = (L^2(X) \otimes \mathcal{F}(\mathfrak{h}_i)) \otimes \mathcal{F}(\mathfrak{h}_s)$. Here $H_i = K \otimes 1 + 1 \otimes d\Gamma(|k|) + \phi(G)$ on $\mathcal{H}_i := L^2(X) \otimes \mathcal{F}(\mathfrak{h}_i)$. It follows that $H_{\text{mod}}$ and $H$ have the same

eigenvectors, they are of the form $U^*(\varphi_i \otimes \Omega_s)$ where $\varphi_i$ is an eigenvector of $H_i$. Hence $P_B(H) = P_B(H_{\text{mod}})$. Furthermore

$$e^{iHt} I e^{-i\tilde{H}t} = e^{iH_{\text{mod}}t} e^{-id\Gamma(\omega - |k|)t} I e^{i[d\Gamma(\omega - |k|) \otimes 1 + 1 \otimes d\Gamma(\omega - |k|)]t} e^{-i\tilde{H}_{\text{mod}}t}$$

$$= e^{iH_{\text{mod}}t} I e^{-i\tilde{H}_{\text{mod}}t}$$

because $H_{\text{mod}}$ and $d\Gamma(\omega - |k|)$ commute. This shows that $\Omega_+ = \Omega_{+,\text{mod}}$ and hence that $\Omega_+$ exists on $\chi(\tilde{H}_{\text{mod}} \leq \mu)\mathcal{H}$ and that $\text{Ran}\,\Omega_+ \supset \chi(H_{\text{mod}} \leq \mu)\mathcal{H}$ for all $\mu < \Sigma$, by Theorem 19.

To reformulate these results in terms of $H_i$ we consider $U \otimes U$ as a map from $\mathcal{F} \otimes \mathcal{F}$ to $\mathcal{F}_i \otimes \mathcal{F}_i \otimes \mathcal{F}_s \otimes F_s$. Then $U\Omega_+(U^* \otimes U^*)$ exists on $(U \otimes U)\chi(\tilde{H}_{\text{mod}} \leq \mu)\mathcal{H} \supset \chi(\tilde{H}_i \leq \mu)\tilde{\mathcal{H}}_i \otimes \Omega_s \otimes \Omega_s$, where $\tilde{H}_i = H_i \otimes 1 + 1 \otimes d\Gamma(|k|)$ on $\tilde{\mathcal{H}}_i = \mathcal{H}_i \otimes \mathcal{F}_i$, and $\text{Ran}\,U\Omega_+(U^* \otimes U^*) \supset U\chi(H_{\text{mod}} \leq \mu)\mathcal{H} \supset \chi(H_i \leq \mu)\mathcal{H}_i \otimes \Omega_s$. Furthermore

$$U\Omega_+(U^* \otimes U^*) = \Omega_i \otimes (I_s(\chi(N_s = 0) \otimes 1_s))$$

where $\Omega_i = s - \lim_{t \to \infty} e^{iH_i t} I_i e^{-i\tilde{H}_i t}(P_B(H_i) \otimes 1)$. In fact $UI(U^* \otimes U^*) = I_i \otimes I_s$ and $UP_B(H)U^* = P_B(H_i) \otimes \chi(N_s = 0)$. It follows that $U\Omega_+(U^* \otimes U^*)$ exists on $\chi(\tilde{H}_i \leq \mu)\tilde{\mathcal{H}}_i \otimes \mathcal{F}_s \otimes \mathcal{F}_s \supset (U \otimes U)\chi(\tilde{H} \leq \mu)\tilde{\mathcal{H}}$ and that its range contains $\chi(H_i \leq \mu)\mathcal{H}_i \otimes \mathcal{F}_s \supset U\chi(H \leq \mu)\mathcal{H}$. Hence $\Omega_+$ exists on $\chi(\tilde{H} \leq \mu)\tilde{\mathcal{H}}$ and by the intertwining relation for $\Omega_+$, the range of $\Omega_+ \restriction \chi(\tilde{H} \leq \mu)\tilde{\mathcal{H}}$ contains $\chi(H \leq \mu)\mathcal{H}$. The theorem now follows because $\mu < \Sigma$ was arbitrary and $\|\Omega_+\| = 1$. $\qquad\square$

## 10.2 Relaxation to the Ground State

With the help of AC established in the last section we next show that states below the ionization threshold relax to the ground state in the sense (7) under the dynamics generated by the Hamilton operator (112).

We begin by summarizing results due to Bach et al. [BFSS99], [BFS98] on the point spectrum of $H$ that are needed in this section. No infrared cutoff is assumed in the following discussion.

Consider the Hamilton operator

$$H_g = K \otimes 1 + 1 \otimes d\Gamma(|k|) + g\phi(G) \tag{116}$$

where $K = -\Delta + V$ on $L^2(\mathbb{R}^n)$ and $V$ is operator-bounded w.r.t. $-\Delta$ with relative bound zero. This assumption allows for typical $N$-body Schrödinger operators [HS00]. We assume that

$$\sup_x \int |G_x(k)|^2 \left(\frac{1}{|k|} + 1\right) dk < \infty \tag{117}$$

to ensure self adjointness of $H_g$ on $D(H_{g=0})$. Following [BFSS99] we furthermore assume that

$$\text{(H6)} \qquad \sup_x (1 + |x|^2)^{-M/2} \int dk \, \frac{|(k \cdot \nabla_k)^2 G_x(k)|^2}{|k|} < \infty$$

for some $M \geq 0$. All exited bound states of $H_g$ will be unstable if their life time as given by Fermi's Golden Rule is finite. To state this condition we have to introduce some notation. Suppose $E_0 < E_1 < \cdots < \inf \sigma_{ess}(K)$ are the isolated eigenvalues of $K$ with finite multiplicity. Let $m_j$ be the multiplicity of the eigenvalue $E_j$, and let $\varphi_{j,l} \in L^2(\mathbb{R}^n)$, for $l = 1, \ldots m_j$ be an orthonormal base of the eigenspace of $K$ to the eigenvalue $E_j$. Then for each $0 \leq i < j$ and for each $k \in \mathbb{R}^d$ we define the $m_i \times m_j$ matrix $A_{ij}(k) := \langle \varphi_{i,r}, G_x(k)\varphi_{j,s} \rangle$. Now, for each $j \geq 0$ we define the $m_j \times m_j$ matrix

$$\Gamma_j = \sum_{i:i<j} \int A_{ij}^*(k) A_{ij}(k) \delta(\omega(k) - E_j + E_i) dk \tag{118}$$

The eigenvalues of this matrix are then the resonance widths in second order perturbation theory corresponding to the eigenvalues $E_j$. To show that $H_g$ has no eigenvalues in neighborhoods of the eigenvalues of the unperturbed Hamiltonian $H_0$, we therefore need the following assumption:

(H7) *Fermi Golden Rule.* For each $j \geq 1$ we have $\Gamma_j > 0$.

The following theorem summarizes results from [BFSS99] and [BFS98].

**Theorem 21.** *i) (Exponential decay, [BFS98]) Suppose $\mu < \inf \sigma_{ess}(K)$ and $\varepsilon > 0$. Then there exists a constant $M = M(\varepsilon)$ such that*

$$\|e^{\alpha|x|}\chi(H_g < \mu)\| < M \tag{119}$$

*for all $\alpha, g$ with $\inf \sigma_{ess}(K) - \mu - \alpha^2 - g^2 \sup_x \int dk \, |G_x(k)|^2/|k| > \varepsilon$.*

*ii) (Existence and uniqueness of the ground state, [BFS98, GLL00]) If $\inf \sigma(K) < \inf \sigma_{ess}(K) - g^2 \sup_x \int dk \, |G_x(k)|^2/|k|$ and $G_x(-k) = \overline{G_x(k)}$ then $E_g := \inf \sigma(H_g)$ is a non-degenerate eigenvalue of $H_g$. Moreover if $\psi_g$ is an eigenvector of $H_g$ corresponding to the eigenvalue $E_g$, then*

$$\|P_{\psi_g} - P_{\varphi_0 \otimes \Omega}\| \to 0, \quad as \quad g \to 0, \tag{120}$$

*where $P_{\psi_g}$ and $P_{\varphi_0 \otimes \Omega}$ denote the orthogonal projections onto the spaces spanned by the ground states $\psi_g$ and $\varphi_0 \otimes \Omega$, respectively.*

*iii) (Absence of exited eigenstates, [BFSS99]) Assume (H6) and (H7). Set $\Delta = [E_0 + \varepsilon, \mu]$, for fixed $\varepsilon > 0$ and $\mu < \inf \sigma_{ess}(K)$. Then*

$$\sigma_{pp}(H_g) \cap \Delta = \emptyset \tag{121}$$

*for $g > 0$ sufficiently small.*

If the infrared cutoff (IR) is imposed, then assumption (117) simplifies to $\sup_x \|G_x\| < \infty$ and all results of the above theorem then hold for $g$ sufficiently small. Note that $m$ must be small in order for (H7) to hold, because transitions between energy levels $E_j$ with separation less than $m$ are suppressed by the infrared cutoff.

Next we prove absence of eigenvalues above and close to $E_g$ for $g$ small, and assuming (IR). As in [BFSS99] we argue by contradiction and prove a virial theorem as well as the positivity of $[iH, A]$ on a spectral interval $(E_g, E_1 - \varepsilon]$ and for a suitable conjugate operator $A$.

**Lemma 22 (Virial Theorem).** *Assume (IR) and (H3). If $H\varphi = E\varphi$ and $E < \inf \sigma_{ess}(K) - g^2 \sup_x \int dk \, |G_x(k)|^2/|k|$ then $\varphi \in D(N)$ and*

$$\langle \varphi, (N - g\phi(iaG)) \, \varphi \rangle = 0.$$

*Proof.* With the notation of the proof of Theorem 20 the eigenvector $\varphi$ is of the form $U^*(\varphi_i \otimes \Omega_s)$. It follows that $\varphi$ is an eigenvector of $H_{\text{mod}}$ and thus in $D(H_{\text{mod}}) \subset D(N)$. By Theorem 21, part i), $e^{\alpha|x|}\varphi \in \mathcal{H}$ for some $\alpha > 0$ and hence Lemma 3 applies to $\varphi$ and $\mathcal{H}_{\text{mod}}$. This shows that

$$\langle \varphi, [d\Gamma(|\nabla\omega|^2) - g\phi(iaG)]\varphi \rangle = 0$$

which proves the theorem because, by the form of $\varphi$, $d\Gamma(|\nabla\omega|^2)\varphi = N\varphi$. $\qquad\square$

**Theorem 23 (Positive commutator).** *Assume (IR), (H0), and (H3). Set $\Delta = (E_g, \, E_1 - \varepsilon]$, for some fixed $\varepsilon > 0$ (here $E_1$ is the first point in the spectrum of $K$ above $\inf \sigma(K)$). Then there is a constant $C > 0$ such that*

$$E_\Delta(H_g)\,(N - g\,\phi(iaG))\,E_\Delta(H_g) \geq C\,E_\Delta(H_g),$$

*for all $g > 0$ sufficiently small. In particular, by Lemma 22, it follows that $H_g$ has no eigenvalue in $\Delta$, if $g > 0$ is small enough.*

*Proof.* Using $N \geq 1 - 1 \otimes P_\Omega$ we get

$$\begin{aligned}
E_\Delta(H_g)\,(N - g\,\phi(iaG))\,E_\Delta(H_g) &\geq E_\Delta(H_g)\,(1 - 1 \otimes P_\Omega - g\,\phi(iaG))\,E_\Delta(H_g) \\
&= E_\Delta(H_g) - E_\Delta(H_g)\,(1 \otimes P_\Omega)\,E_\Delta(H_g) - g E_\Delta(H_g)\phi(iaG)E_\Delta(H_g) \\
&\geq E_\Delta(H_g)\left(1 - g\left(\sup_x e^{-\alpha|x|}\|iaG_x\|\right)\|E_\Delta(H_g)e^{\alpha|x|}\|\,\|(N_i+1)^{1/2}E_\Delta(H_g)\|\right) \\
&\quad - E_\Delta(H_g)\,(1 \otimes P_\Omega)\,E_\Delta(H_g)
\end{aligned} \tag{122}$$

where $N_i = \int_{|k|>m} dk\, a^*(k)a(k)$ is the operator counting the number of interacting bosons (which is bounded w.r.t. $H_g$), and where $P_\Omega$ is the orthogonal projector onto $\Omega$. By Hypothesis (H3), and because $\|aG_x\| \leq \text{const} \, (\|yG_x\| + \|G_x\|)$, the number in the parenthesis in the first term on the r.h.s. of the last equation is larger than $C$, for any $C < 1$, if $g > 0$ is small enough. It remains to show that the last term in (122) converges to 0 as $g \to 0$. To do this we split it into two parts, according to

$$\begin{aligned}
E_\Delta(H_g)\,(1 \otimes P_\Omega)\,E_\Delta(H_g) &= E_\Delta(H_g)\,\big(E_{\{E_0\}}(K) \otimes P_\Omega\big)\,E_\Delta(H_g) \\
&\quad + E_\Delta(H_g)\,\big(E_{[E_1,\infty)}(K) \otimes P_\Omega\big)\,E_\Delta(H_g),
\end{aligned} \tag{123}$$

where $K$ is the particle Hamiltonian, and $E_0$ and $E_1$ are its ground state energy and its first exited eigenvalue. Since $P_{\psi_g}E_\Delta(H_g) = 0$ the first term in Eq. (123) can be written as

$$E_\Delta(H_g)\,\big(E_{\{E_0\}}(K) \otimes P_\Omega\big)\,E_\Delta(H_g) = E_\Delta(H_g)\,\big(P_{\varphi_0 \otimes \Omega} - P_{\psi_g}\big)\,E_\Delta(H_g), \tag{124}$$

which converges to zero by Theorem 21, part ii). Consider now the second term on the r.h.s. of (123). Choose $\chi \in \mathbb{C}_0^\infty(\mathbb{R})$, such that $0 \leq \chi \leq 1$, $\chi(s) = 0$ if $s > E_1 - \varepsilon/2$, and $\chi(s) = 1$ if $s \in \Delta$ (this is a smooth version of the characteristic function $E_\Delta$). Then we have on the one hand $\chi(H_g) E_\Delta(H_g) = E_\Delta(H_g)$ and on the other hand $\left(E_{[E_1,\infty)}(K) \otimes P_\Omega\right) \chi(H_0) = 0$. Thus we get

$$\left(E_{[E_1,\infty)}(K) \otimes P_\Omega\right) E_\Delta(H_g) = \left(E_{[E_1,\infty)}(K) \otimes P_\Omega\right) \left(\chi(H_g) - \chi(H_0)\right) E_\Delta(H_g). \quad (125)$$

Now if $\tilde{\chi}$ is an almost analytic extension of $\chi$, in the sense of the Helffer–Sjöstrand functional calculus (see Appendix A.2), then we have

$$\chi(H_g) - \chi(H_0) = -\frac{g}{\pi} \int dx dy \, \partial_{\bar{z}} \tilde{\chi}(z - H_g)^{-1} \phi(G)(z - H_0)^{-1}. \quad (126)$$

This implies, since $\chi$ has a compact support, that $\|\chi(H_g) - \chi(H_0)\| \leq Cg$, for some constant $C > 0$, and thus, by (125), that $\|\left(E_{[E_1,\infty)}(K) \otimes P_\Omega\right) E_\Delta(H_g)\| \to 0$ as $g \to 0$. This completes the proof of the theorem. $\qquad\square$

The last Theorem, together with Theorem 21, proves the following corollary.

**Corollary 24.** *Assume Hypotheses (IR), (H3), (H6), and (H7). If $\mu < \inf \sigma_{ess}(K)$, then*

$$\sigma_{pp}(H_g) \cap (E_g, \mu) = \emptyset,$$

*for all $g > 0$ sufficiently small.*

With the help of this corollary and Theorem 20 we next prove *relaxation to the groundstate* in the sense of the following theorem. To define the algebra of observables let $\mathcal{A}$ denote the $C^*$ algebra generated by all Weyl operators $W(h) = \exp(i\phi(h))$, with $h \in L^2(\mathbb{R}^d, dk)$. By taking tensor products of operators in $\mathcal{A}$ with bounded operators acting on the Hilbert space $\mathcal{H}_{el} = L^2(\mathbb{R}^n, dx)$ of the electrons one obtains a $C^*$ algebra, which we denote by $\tilde{\mathcal{A}}$.

**Theorem 25 (Relaxation to the ground state).** *Assume Hypotheses (IR), (H0), and (H3) through (H7). Choose $\mu < \inf \sigma_{ess}(K)$. Then, for sufficiently small values of the coupling constant $g > 0$, the Hamiltonian $H_g$ exhibits the property of* relaxation to the ground state *for states with energy less than $\mu$. This means that, if $g > 0$ is sufficiently small, then, for all $A \in \tilde{\mathcal{A}}$ and for all $\psi \in \text{Ran}(H \leq \mu)$, we have*

$$\lim_{t\to\infty} \langle \psi_t, A\psi_t \rangle = \langle \psi_g, A\psi_g \rangle \langle \psi, \psi \rangle, \quad (127)$$

*where $\psi_t = e^{-iHt}\psi$ and $\psi_g$ denotes the groundstate of $H_g$.*

*Proof.* Since the $C^*$ algebra $\mathcal{A}$ is generated by the Weyl-operators $W(h) = e^{i\phi(h)}$, and because products of Weyl-operators are again Weyl-operator (up to some unimportant phase) it is enough if we prove (127) for $A = B \otimes W(h)$, where $B$ is a bounded operator on $\mathcal{H}_{el}$ and $h \in \mathcal{S}(\mathbb{R}^d)$.

By Theorem 20 we know that the system we are considering is asymptotically complete. On the other hand we know, from Corollary 24, that the ground state $\psi_g$ is the only eigenstate of $H_g$, which lies in the range of the spectral projection $\chi(H_g \leq \mu)$. These two results imply that each $\psi \in \text{Ran}\chi(H_g \leq \mu)$ can be written as limit of a sequence of finite linear combinations of states like $a_+^*(f_1)\ldots a_+^*(f_m)\psi_g$, with $f_i \in \mathcal{S}(\mathbb{R}^d)$. Since we are dealing only with bounded operators, it follows that it is enough to prove (127) for $A = B \otimes W(h)$ and for

$$\psi = \sum_{i=1}^{N} c_i \, a_+^*(f_1^i) a_+^*(f_2^i) \ldots a_+^*(f_{n_i}^i) \psi_g. \tag{128}$$

In this case we have

$$\lim_{t \to \infty} \langle e^{-iHt}\psi, A e^{-iHt}\psi \rangle = \sum_{i,j=1}^{N} \bar{c}_i \, c_j \, \lim_{t \to \infty} \langle e^{-iHt} \prod_{l=1}^{n_i} a_+^*(f_l^i)\psi_g, A e^{-iHt} \prod_{m=1}^{n_j} a_+^*(f_m^j)\psi_g \rangle$$

$$= \sum_{i,j=1}^{N} \bar{c}_i \, c_j \, \lim_{t \to \infty} \langle \prod_{l=1}^{n_i} a^*(f_{l,t}^i)\psi_g, A \prod_{m=1}^{n_j} a^*(f_{m,t}^j)\psi_g \rangle, \tag{129}$$

where we used the definition of the asymptotic fields $a_+^*(h)$. Notice now that the ground state $\psi_g$ of $H_g$ is in the domain of arbitrary powers of the field-Hamiltonian $d\Gamma(|k|)$. Moreover we know that the Weyl operators leave $D(d\Gamma(|k|)^n)$ invariant. This follows by the commutation relations $[W(h), d\Gamma(|k|)] = -\phi(i|k|h)W(h) + 1/2 \operatorname{Re}(|k|h, h) W(h)$. These remarks imply that we can rewrite the limit in the r.h.s. of the last equation as

$$\lim_{t \to \infty} \langle \prod_{l=1}^{n_i} a^*(f_{l,t}^i)\psi_g, A \prod_{m=1}^{n_j} a^*(f_{m,t}^j)\psi_g \rangle$$

$$= \lim_{t \to \infty} \langle \psi_g, \prod_{l=1}^{n_i} a(f_{l,t}^i) \, (B \otimes W(h)) \prod_{m=1}^{n_j} a^*(f_{m,t}^j)\psi_g \rangle$$

$$= \lim_{t \to \infty} \langle A^* \psi_g, \prod_{l=1}^{n_i} a(f_{l,t}^i) \prod_{m=1}^{n_j} a^*(f_{m,t}^j)\psi_g \rangle \tag{130}$$

$$+ \lim_{t \to \infty} \langle \psi_g, B \otimes \left[ \prod_{l=1}^{n_i} a(f_{l,t}^i), W(h) \right] \prod_{m=1}^{n_j} a^*(f_{m,t}^j)\psi_g \rangle.$$

If we expand the commutator in the last equation, we get a sum of $n_i$ terms. Each of these terms contains a contraction $(f_{l,t}^i, h)_{L^2}$. Now, since we have assumed that $f_l^i, h_r \in \mathcal{S}(\mathbb{R}^d)$, we have

$$\left| (f_{l,t}^i, h_r) \right| = \left| \int dk \, \overline{f_l^i(k)} h_r(k) e^{i|k|t} \right| \to 0 \quad \text{as} \quad t \to \infty,$$

and thus the second term on the r.h.s. of (130) vanishes. To handle the first term on the r.h.s. of (130) we use that, by Lemma 26, $\lim_{t \to \infty} \prod_{l=1}^{n} a(f_{l,t})\psi_g = 0$. Assuming $n_i \geq n_j$, this implies that the first term on the r.h.s. of (130) vanishes if $n_i > n_j$, and that

$$\lim_{t \to \infty} \prod_{l=1}^{n_i} a(f_{l,t}^i) \prod_{m=1}^{n_j} a^*(f_{m,t}^j)\psi_g = \lim_{t \to \infty} \psi_g \, \langle \psi_g, \prod_{l=1}^{n_i} a(f_{l,t}^i) \prod_{m=1}^{n_j} a^*(f_{m,t}^j)\psi_g \rangle$$

$$= \psi_g \langle \psi_g, \prod_{l=1}^{n_i} a_+(f_l^i) \prod_{m=1}^{n_j} a_+^*(f_m^j)\psi_g \rangle$$

$$= \psi_g \langle \prod_{l=1}^{n_i} a_+^*(f_l^i)\psi_g, \prod_{m=1}^{n_j} a_+^*(f_m^j)\psi_g \rangle,$$

for all $n_i \geq n_j$. Using the antisymmetry of the inner product it is analogously proven that the r.h.s. of (130) vanishes if $n_i < n_j$. Thus, for arbitrary $n_i, n_j$,

$$\lim_{t \to \infty} \langle \prod_{l=1}^{n_i} a^*(f_{l,t}^i)\psi_g, A \prod_{m=1}^{n_j} a^*(f_{m,t}^j)\psi_g \rangle = \langle A^* \psi_g, \psi_g \rangle \langle \prod_{l=1}^{n_i} a_+^*(f_l^i)\psi_g, \prod_{m=1}^{n_j} a_+^*(f_m^j)\psi_g \rangle.$$

The theorem now follows if we insert this result into (129) and compute the sum over $i, j$. $\qquad \square$

**Lemma 26.** *Suppose* $\mathbb{N} \ni n > 0$ *and* $\varphi \in D((H_g + i)^{n/2})$. *Then, if* $h_1, \ldots, h_n \in L^2(\mathbb{R}^d, dk)$ *with* $\|h_i\|_\omega^2 = \int dk \, |h_i(k)|^2 \, (1 + 1/|k|) < \infty$, *we have*

$$\lim_{t \to \infty} \prod_{j=1}^n a(h_{j,t})\varphi = 0. \tag{131}$$

*Proof.* Notice that $\| \prod_{i=1}^n a^*(h_i)(d\Gamma(|k|) + 1)^{-n/2}\| \leq C \prod_{i=1}^n \|h_i\|_\omega$. This implies that

$$\begin{aligned}
\| \prod_{j=1}^n a(h_{j,t})\varphi\| &\leq \|(\prod_{j=1}^n a(h_{j,t}) - \prod_{j=1}^n a(\tilde{h}_{j,t}))(H + i)^{-n/2}\|\|(H + i)^{n/2}\varphi\| + \|\prod_{j=1}^n a(\tilde{h}_{j,t})\varphi\| \\
&\leq \sum_{j=1}^n \| \left( a(h_{1,t}) \ldots a(h_{j,t} - \tilde{h}_{j,t}) \ldots a(\tilde{h}_{n,t}) \right) (H + i)^{-n/2}\|\|(H + i)^{n/2}\varphi\| \\
&\quad + \|\prod_{j=1}^n a(\tilde{h}_{j,t})\varphi\| \\
&\leq C \sum_{j=1}^n \|h_1\|_\omega \ldots \|h_j - \tilde{h}_j\|_\omega \ldots \|h_n\|_\omega + \|\prod_{j=1}^n a(\tilde{h}_{j,t})\varphi\|,
\end{aligned} \tag{132}$$

where we used that $\|(d\Gamma(|k|) + 1)^{n/2}(H_g + i)^{-n/2}\| < \infty$ (see [FGS00]). Because of the last equation it is enough to prove (131) when $h_j \in C_0^\infty(\mathbb{R}^d \backslash \{0\})$. In this case we have $M = \min_j \inf\{|k| : h_j(k) \neq 0\} > 0$, and there exists $f \in C^\infty(\mathbb{R}^d)$ with $f(k) = 0$ if $|k| < M/2$, and $f(k) = 1$ if $|k| \geq M$. Then, on the one hand, $d\Gamma(f)$ is bounded w.r.t. $H_g$ (and higher powers of $d\Gamma(f)$ are bounded w.r.t. corresponding powers of $H_g$). This implies that $\varphi \in D((d\Gamma(f) + 1)^{n/2})$. On the other hand $\prod_{j=1}^n a(h_{j,t})$ is bounded w.r.t. $(d\Gamma(f) + 1)^{-n/2}$. Thus, with $\psi = (d\Gamma(f) + 1)^{+n/2}\varphi$, we have

$$\prod_{j=1}^n a(h_{j,t})\varphi = \prod_{j=1}^n a(h_{j,t})(d\Gamma(f) + 1)^{-n/2}\psi. \tag{133}$$

Since $\prod_{j=1}^n a(h_{j,t})(d\Gamma(f) + 1)^{-n/2}$ is uniformly bounded in $t$, it is enough if we show that the r.h.s. of the last equation converges to 0, as $t \to \infty$ for $\psi = \alpha \otimes a^*(f_1) \ldots a^*(f_m)\Omega$.

To this end we write

$$\prod_{j=1}^{n} a(h_{j,t})(\mathrm{d}\Gamma(f)+1)^{-n/2} \prod_{i=1}^{m} a^*(f_i)\Omega = (\mathrm{d}\Gamma(f)+1+n)^{-n/2} \prod_{j=1}^{n} a(h_{j,t}) \prod_{i=1}^{m} a^*(f_i)\Omega$$

$$= (\mathrm{d}\Gamma(f)+1+n)^{-n/2} \left[ \prod_{j=1}^{n} a(h_{j,t}) , \prod_{i=1}^{m} a^*(f_i) \right] \Omega.$$
(134)

Expanding the commutator, we find a sum of terms containing $n$ contractions $(h_{j,t}, f_i)_{L^2}$. The lemma now follows, because all these contractions converge to 0 as $t \to \infty$. □

## 10.3 QED in Dipole Approximation

As mentioned in the introduction, our methods can be extended to prove AC, as well as our further main results, for atoms described by "non-relativistic QED" in the dipole approximation. In this section we briefly explain how this is accomplished. For an introduction to the standard model of non-relativistic QED and for the justification of the dipole approximation we refer to [BFS98]. Here we merely show how this model fits into our general framework.

We consider a non-relativistic electron interacting with the quantized radiation field. (The generalization to $N$ electrons is straightforward.) States of this system are described by vectors in the Hilbert space $\mathcal{H} = \mathcal{H}_{\mathrm{at}} \otimes \mathcal{F}$, where $\mathcal{H}_{\mathrm{at}} = L^2(\mathbb{R}^3, dx)$, and $\mathcal{F}$ is the bosonic Fock space over $\mathfrak{h} = L^2(\mathbb{R}^3, \mathbb{C}^2)$. The Hamilton operator is

$$H = K + \mathrm{d}\Gamma(|k|) + \phi(G),$$
(135)

where $K = -\Delta + V$ is assumed to satisfy Hypothesis (H0). To describe QED in the dipole approximation, we set

$$G_x(\lambda, k) = \kappa(k)g(x)x \cdot \varepsilon_\lambda(k)$$

where $x \in \mathbb{R}^3$ is the position of the electron, and $\varepsilon_\lambda(k)$, $\lambda = 1, 2$, are the polarization vectors orthogonal to $k$. As above $\kappa \in C_0^\infty(\mathbb{R}^3)$, and $\kappa(k) = 0$ if $|k| < m$, for some $m > 0$. The factor $g \in C_0^\infty(\mathbb{R}^3)$ is a space cutoff necessary to make $G_x$ bounded as a function of $x$. From a physical point of view this simplification is legitimate, because the electrons are exponentially localized near the origin (see also [BFS98]).

The Hamilton operator (135) clearly satisfies assumptions (H1), (H2), and (H4). The problem is that the polarization vectors $\varepsilon_\lambda(k)$, as functions of $k$, cannot be chosen in such a way that they are twice differentiable on the unit sphere (an easy application of a famous theorem due to H. Hopf), and hence hypotheses (H3) and (H5) cannot be satisfied. In order to circumvent this problem, we introduce two systems of polarization vectors, the "north"-system $\varepsilon_\lambda^N(k)$ and the "south"-system $\varepsilon_\lambda^S(k)$. The north-system $\varepsilon_\lambda^N(k)$ depends smoothly on $k$, for $k \in \mathbb{R}^3 \backslash Z_S$, where $Z_S$ is an open neighborhood of the negative z-axis $\{k \in \mathbb{R}^3 : k_1 = k_2 = 0, \text{ and } k_3 \leq 0\}$, whereas the south-system $\varepsilon_\lambda^S(k)$ depends smoothly on $k$, for $k \in \mathbb{R}^3 \backslash Z_N$, where $Z_N$ is an open neighborhood of the positive z-axis $\{k \in \mathbb{R}^3 : k_1 = k_2 = 0, \text{ and } k_3 \geq 0\}$. The sets $Z_N$ and $Z_S$ are chosen such that $Z_N \cap Z_S \subset B_{m/2}(0) = \{k \in \mathbb{R}^3 : |k| < m/2\}$. For each $k \neq 0$, $\{\varepsilon_\lambda^N(k)\}_{\lambda=1,2}$ and $\{\varepsilon_\lambda^S(k)\}_{\lambda=1,2}$ are orthogonal bases in the plane perpendicular to $k$. Hence there exists a matrix $R(k) \in O(2)$ such that

$$\varepsilon_\lambda^N(k) = \sum_{\mu=1,2} \varepsilon_\mu^S(k) R_{\mu\lambda}(k).$$

Correspondingly, if $f \in L^2(\mathbb{R}^3, \mathbb{C}^2)$ is a wave function describing a photon with respect to the north base, then the *same* photon is described by

$$(Rf)(k, \lambda) = \sum_{\mu=1,2} R_{\lambda\mu}(k) f(k, \mu),$$

with respect to the south base $\varepsilon_\lambda^S(k)$.

The Hamilton operator (135) depends on the choice of polarization vectors, but two Hamilton operators corresponding to two different choices are unitarily equivalent, and hence we may make a choice by convenience. We choose the north system and denote the corresponding Hamilton operator by $H_N$. As a reminder of this choice we also attach a subindex $N$ to the Hilbert space $\mathcal{H}_N = \mathcal{H}_{at} \otimes \mathcal{F}_N$ and to its Fock space $\mathcal{F}_N$.

The idea is now to split each photon into two parts, a part supported in the northern half space and a part supported in the southern half space. The parts in the *south* will then be mapped into an auxiliary Fock space, where they are described with respect to the *south* base. The Hamilton operator describing the time evolution in this split representation then turns out to involve only interactions with smooth coupling functions.

In order to separate the north-photons from the south-photons, we introduce two functions $j_N, j_S \in C^\infty(\mathbb{R}^3)$, with $j_N^2(k) + j_S^2(k) = 1$, if $|k| > m$, and with $\text{supp} j_N \subset \mathbb{R}^3 \backslash Z_S$, and $\text{supp} j_S \subset \mathbb{R}^3 \backslash Z_N$ (such functions exist because of the assumption $Z_N \cap Z_S \subset B_{m/2}(0)$). Next we define an isometry $u : \mathfrak{h} \to \mathfrak{h} \oplus \mathfrak{h}$ by $f \mapsto (j_N f, Rj_S f)$. This induces another isometry $\breve{\Gamma}(u) : \mathcal{F}_N \to \mathcal{F} \otimes \mathcal{F}$, which is characterized by

$$\breve{\Gamma}(u)\Omega = \Omega \otimes \Omega \quad \text{and} \quad \breve{\Gamma}(u)a^\sharp(f) = (a^\sharp(j_N f) \otimes 1 + 1 \otimes a^\sharp(Rj_S f))\breve{\Gamma}(u).$$

States of the total system are now described by vectors in the new Hilbert space $\mathcal{H}_{\text{new}} = \mathcal{H}_{at} \otimes \mathcal{F} \otimes \mathcal{F}$; however, only vectors in the subspace $\mathcal{H}_u := \mathcal{H}_{at} \otimes \text{Ran} \breve{\Gamma}(u)$ correspond to physical states. On $\mathcal{H}_{\text{new}}$ we define the new Hamiltonian

$$H_{\text{new}} = K \otimes 1 \otimes 1 + 1 \otimes (d\Gamma(|k|) \otimes 1 + 1 \otimes d\Gamma(|k|)) + \phi(j_N G_N) \otimes 1 + 1 \otimes \phi(Rj_S G_N). \tag{136}$$

The operators $\phi(j_N G_N) \otimes 1$ and $1 \otimes \phi(Rj_S G_N)$ couple the north- and the south photons, respectively, to the electron. The new Hamiltonian $H_{\text{new}}$ leaves the subspace $\mathcal{H}_u$ of physical states invariant, and its restriction to $\mathcal{H}_u$, denoted by $H_u := H_{\text{new}}{\restriction}\mathcal{H}_u$, is unitarily equivalent to the Hamiltonian $H_N$, acting on the Hilbert space $\mathcal{H}_N$. In fact $H_u = \breve{\Gamma}(u)H_N\breve{\Gamma}(u)^*$. Most importantly, both form factors

$$(j_N G_N)_x(k, \lambda) = j_N(k)\kappa(k)g(x)x \cdot \varepsilon_\lambda^N(k) \quad \text{and}$$
$$(Rj_S G_N)_x(k, \lambda) = (j_S G_S)_x(k, \lambda) = j_S(k)\kappa(k)g(x)x \cdot \varepsilon_\lambda^S(k)$$

in (136) are smooth on the entire k-space, and hence they satisfy Hypotheses (H3) and (H5).

In order to describe asymptotically free photons, we introduce the new extended Hilbert space $\tilde{\mathcal{H}}_{\text{new}} := \mathcal{H}_{\text{at}} \otimes \mathcal{F} \otimes \mathcal{F} \otimes \mathcal{F} \otimes \mathcal{F}$. Vectors in the first two copies of $\mathcal{F}$ describe interacting photons, while photons in the third and fourth copies of $\mathcal{F}$ are asymptotically free and live in the north and in the south of $k$-space, respectively. Physical states with asymptotically free photons are contained in the subspace $\tilde{\mathcal{H}}_u = \mathcal{H}_{\text{at}} \otimes \text{Ran} \, \breve{\Gamma}(u) \otimes \text{Ran} \, \breve{\Gamma}(\underline{u})$ of $\tilde{\mathcal{H}}_{\text{new}}$. The Hamiltonian generating the dynamics on the extended Hilbert space $\tilde{\mathcal{H}}_{\text{new}}$ is given by $\tilde{H}_{\text{new}} = H_{\text{new}} \otimes 1_{\mathcal{F} \otimes \mathcal{F}} + 1_{\mathcal{H}_{\text{new}}} \otimes (\text{d}\Gamma(|k|) \otimes 1 + 1 \otimes \text{d}\Gamma(|k|))$. This operator leaves the subspace $\tilde{\mathcal{H}}_u$ invariant, and $\tilde{H}_u := \tilde{H}_{\text{new}} {\restriction} \tilde{\mathcal{H}}_u$ is unitary equivalent to the extended Hamiltonian $\tilde{H}_N = H_N \otimes 1 + 1 \otimes \text{d}\Gamma(|k|)$, acting on the extended Hilbert space $\tilde{\mathcal{H}}_N = \mathcal{H}_{\text{at}} \otimes \mathcal{F}_N \otimes \mathcal{F}_N$. On the spectral subspace of $\tilde{\mathcal{H}}_{\text{new}}$ where $\tilde{H}_{\text{new}} < \Sigma$ we define the new wave operator

$$\Omega_+^{\text{new}} : \chi(\tilde{H}_{\text{new}} < \Sigma)\tilde{\mathcal{H}}_{\text{new}} \to \chi(H_{\text{new}} < \Sigma)\mathcal{H}_{\text{new}}$$

$$\Omega_+^{\text{new}} := s - \lim_{t \to \infty} e^{iH_{\text{new}}t} I_{\text{new}} e^{-i\tilde{H}_{\text{new}}t} (P_B(H_{\text{new}}) \otimes 1_{\mathcal{F} \otimes \mathcal{F}}),$$

where $I_{\text{new}}$ takes all bosons from the third and the fourth Fock space and puts them into the first and into the second Fock space, respectively. The restriction of $\Omega_+^{\text{new}}$ to $\chi(\tilde{H}_u < \Sigma)\tilde{\mathcal{H}}_u$, denoted by $\Omega_+^u$, maps $\tilde{\mathcal{H}}_u$ to $\mathcal{H}_u$ and is unitary equivalent to $\Omega_+^N : \tilde{\mathcal{H}}_N \to \mathcal{H}_N$, the usual wave operator defined in terms of $H_N$ and $\tilde{H}_N$.

In this new description, asymptotic completeness means that $\text{Ran}\Omega_+^u = \text{Ran}\chi(H_u < \Sigma)$. But this follows if we show that

$$\text{Ran}\Omega_+^{\text{new}} = \text{Ran}\chi(H_{\text{new}} < \Sigma), \tag{137}$$

because $\Omega_+^{\text{new}}$ maps $\tilde{\mathcal{H}}_u$ into $\mathcal{H}_u$ and its orthogonal complement $\tilde{\mathcal{H}}_u^\perp \subset \tilde{\mathcal{H}}_{\text{new}}$ into $\mathcal{H}_U^\perp$. Eq. (137) can be proved by extending, in an obvious manner, the methods of Sections 6 to 9 to $H_{\text{new}}$ on $\mathcal{H}_{\text{new}}$, to the new extended Hilbert space $\tilde{\mathcal{H}}_{\text{new}}$, and to the new wave operator $\Omega_+^{\text{new}}$.

# A    Pseudo Differential Calculus and Functional Calculus

This appendix collects our main tools for computing commutators.

## A.1    Pseudo Differential Calculus on Fock Space

**Lemma 27.** *Suppose* $f \in \mathcal{S}(\mathbb{R}^d)$, $g \in C^n(\mathbb{R}^d)$ *and* $\sup_{|\alpha|=n} \|\partial^\alpha g\|_\infty < \infty$. *Let* $p = -i\nabla$. *Then*

$$i[g(p), f(x)] = i \sum_{1 \le |\alpha| \le n-1} \frac{(-i)^{|\alpha|}}{\alpha!} (\partial^\alpha f)(x)(\partial^\alpha g)(p) + R_{1,n}$$

$$= (-i) \sum_{1 \le |\alpha| \le n-1} \frac{i^{|\alpha|}}{\alpha!} (\partial^\alpha g)(p)(\partial^\alpha f)(x) + R_{2,n}$$

*where*

$$\|R_{j,n}\| \le C_n \sup_{|\alpha|=n} \|\partial^\alpha g\|_\infty \int dk \, |k|^n |\hat{f}(k)|.$$

*In particular, and most importantly, if $n = 2$ in the limit $\varepsilon \to 0$*

$$i[g(p), f(\varepsilon x)] = \varepsilon \nabla g(p) \cdot \nabla f(\varepsilon x) + O(\varepsilon^2)$$
$$= \varepsilon \nabla f(\varepsilon x) \cdot \nabla g(p) + O(\varepsilon^2).$$

*Proof.* Let $f(x) = \int dk \, e^{ikx} \hat{f}(k)$. The first equation follows from

$$g(p)e^{ikx} - e^{ikx}g(p) = e^{ikx}\left[e^{-ikx}g(p)e^{ikx} - g(p)\right]$$
$$= e^{ikx}[g(p+k) - g(p)] \tag{138}$$

and Taylor's formula

$$g(p+k) - g(p) = \sum_{1 \le |\alpha| \le n-1} (\partial^\alpha g)(p)\frac{k^\alpha}{\alpha!}$$
$$+ n \int_0^1 dt \, (1-t)^{n-1} \sum_{|\alpha|=n} (\partial^\alpha g)(p+tk)k^\alpha/\alpha!.$$

To obtain the second equation write $g(p)e^{ikx} - e^{ikx}g(p) = -[g(p-k) - g(p)]e^{ikx}$ instead of (138). $\qquad\square$

**Lemma 28.** *Suppose $\omega$ is in $\in C^\infty(\mathbb{R}^d)$ and has bounded derivatives. If $f \in C_0^\infty(\mathbb{R})$, then*

$$[i\mathrm{d}\Gamma(\omega), f] = \frac{1}{t} f' \, \mathrm{d}\Gamma(\nabla\omega \cdot y/t + y/t \cdot \nabla\omega) + O(t^{-2})N$$

*where $f = f(\mathrm{d}\Gamma(v^2))$, $v = y/t$ and $f' = f'(\mathrm{d}\Gamma(v^2))$.*

*Proof.* The operators $\mathrm{d}\Gamma(\omega)$ and $f$ commute with $N$, hence it suffices to prove the equation on $\otimes_s^n \mathfrak{h}$. On this subspace

$$i[\mathrm{d}\Gamma(\omega), f] = \sum_{j=1}^n \left[\omega(k_j), f(\sum_{i=1}^n v_i^2)\right].$$

To evaluate $[\omega(k_j), f(\sum_{i=1}^n v_i^2)]$ for given fixed $j \in \{1, \ldots, n\}$ consider $f(\sum v_i^2)$ as a function of $y_j$ only, and apply Lemma 27. It follows that

$$\left[\omega(k_j), f(\sum_{i=1}^n v_i^2)\right] = \frac{1}{t^2} f' \, 2y_j \cdot \nabla\omega(k_j) + R_{j,t}$$

where

$$\|R_{j,t}\| \leq t^{-2} \sup_{|\alpha|=2} \|\partial^\alpha \omega\|_\infty \int |k|^2 |\hat{f}_j(k)| dk$$

and

$$\hat{f}_j(k) = (2\pi)^{-d} \int e^{-ik \cdot y_j} f\left(y_j^2 + \sum_{l=1, l \neq j}^n y_l^2\right) dk.$$

It is easy to see that

$$|\hat{f}_j(k)| \leq \frac{C_p}{(1+|k|)^p} \qquad \text{for all} \qquad p \in \mathbb{N}$$

where $C_p$ only depends on $f$ but not on $j$ or $\sum_{l=1, l \neq j}^n y_l^2$. It follows that $\|R_{j,t}\| \leq C t^{-2}$ and hence that on $\otimes_s^n \mathfrak{h}$

$$\|[i\mathrm{d}\Gamma(\omega), f] - \frac{1}{t} f' 2 \sum_{j=1}^n y_j/t \cdot \nabla \omega(k_j)\| \leq \frac{Cn}{t^2}. \tag{139}$$

The same equation holds with $\nabla \omega(k_j) \cdot y_j/t$ instead of $y_j/t \cdot \nabla \omega(k_j)$, as follows from $[\nabla \omega(k_j), y_j/t] = \Delta \omega(k_j)/t = O(t^{-1})$. In conjunction with (139) this proves the lemma. $\qquad \square$

## A.2   Helffer-Sjöstrand Functional Calculus

Suppose $f \in C_0^\infty(\mathbb{R}; \mathbb{C})$ and $A$ is a self-adjoint operator. A convenient representation for $f(A)$, which is often used in this paper, is then given by

$$f(A) = -\frac{1}{\pi} \int dx dy \, \frac{\partial \tilde{f}}{\partial \bar{z}}(z) (z - A)^{-1}, \qquad z = x + iy,$$

which holds for any extension $\tilde{f} \in C_0^\infty(\mathbb{R}^2; \mathbb{C})$ of $f$ with $|\partial_{\bar{z}} \tilde{f}| \leq C|y|$,

$$\tilde{f}(z) = f(z) \qquad \text{and} \qquad \frac{\partial \tilde{f}}{\partial \bar{z}}(z) = \frac{1}{2}\left(\frac{\partial f}{\partial x} + i \frac{\partial f}{\partial y}\right)(z) = 0 \qquad \text{for all} \qquad z \in \mathbb{R}. \tag{140}$$

Such a function $\tilde{f}$ is called an *almost analytic extension* of $f$. A simple example is given by $\tilde{f}(z) = (f(x) + iyf'(x)) \chi(z)$ where $\chi \in C_0^\infty(\mathbb{R}^2)$ and $\chi = 1$ on some complex neighborhood of supp $f$. Sometimes we need faster decay of $|\partial_{\bar{z}} \tilde{f}|$ as $|y| \to 0$ in the form $|\partial_{\bar{z}} \tilde{f}| \leq C|y|^n$. In that case we work with the almost analytic extension

$$\tilde{f}(z) = \left(\sum_{k=0}^n f^{(k)}(x) \frac{(iy)^k}{k!}\right) \chi(z)$$

where $\chi$ is as above. We call this an almost analytic extension *of order $n$*. For more details and extensions of this functional calculus the reader is referred to [HS00] or [Dav95].

# B   Representation of States in $\chi(\tilde{H} < c)\tilde{\mathcal{H}}$

The representation of states in Ran $\chi(\tilde{H} < c)$ proved in this section is used in Lemma 18 and Theorem 20.

**Lemma 29.** *Suppose $\omega(k) = |k|$ or $\omega$ satisfies (H1), and let $c > 0$. Then the space of linear combinations of vectors of the form $a^*(h_1) \ldots a^*(h_n)\Omega$ with $h_i \in L^2(\mathbb{R}^d)$ and $\sum_{i=1}^n \sup\{\omega(k) : k \in \mathrm{supp}(h_i)\} < c$ is dense in $\chi(\mathrm{d}\Gamma(\omega) < c)\mathcal{F}$.*

*Proof.* Let $\mathcal{D}_c$ denote the space specified in the lemma. Clearly $\mathcal{D}_c \subset \chi(\mathrm{d}\Gamma(\omega) < c)\mathcal{F}$. Since the span of vectors of the form $\chi(\mathrm{d}\Gamma(\omega) < c)a^*(h_1) \ldots a^*(h_n)\Omega$ with $h_i$ and $n$ arbitrary is dense in $\chi(\mathrm{d}\Gamma(\omega) < c)\mathcal{F}$, it suffices to show that such vectors can be approximated by vectors in $\mathcal{D}_c$. Let $n \in \mathbb{N}$ and $h_1, \ldots, h_n \in L^2(\mathbb{R}^d)$ be fixed and let $\psi = \chi(\mathrm{d}\Gamma(\omega) < c)a^*(h_1) \ldots a^*(h_n)\Omega$. Then

$$\psi(k_1, \ldots, k_n) = \chi(\sum \omega(k_i) < c)\frac{1}{\sqrt{n!}}\sum_\sigma h_1(k_{\sigma 1}) \cdots h_n(k_{\sigma n}) \tag{141}$$

where the sum extends over all permutations $\sigma$ of $\{1, \ldots, n\}$. The set $T = \{t \in \overline{\mathbb{R}}_+^n \mid \sum_{i=1}^n t_i < c\}$ can be written as a countable union of cubes $Q_\alpha$ with disjoint interiors and edges parallel to the coordinate axis. That is $T = \cup_\alpha Q_\alpha$, $Q_\alpha^\circ \cap Q_\beta^\circ = \emptyset$ and $Q_\alpha$ is the cartesian product $\prod_{i=1}^n I_{\alpha_i}$ of $n$ intervals $I_{\alpha_i}$ in $\overline{\mathbb{R}}_+$ ($\alpha = (\alpha_1, \ldots, \alpha_n)$). It follows that

$$\begin{aligned}
S &= \left\{(k_1, \ldots, k_n) : \sum_{i=1}^n \omega(k_i) < c\right\} \\
&= \bigcup_\alpha \{(k_1, \ldots, k_n) : (\omega(k_1), \ldots, \omega(k_n)) \in Q_\alpha\} \\
&= \bigcup_\alpha \bigcap_{i=1}^n \{(k_1, \ldots, k_n) : \omega(k_i) \in I_{\alpha_i}\}
\end{aligned}$$

and correspondingly

$$\chi_S(k_1, \ldots, k_n) = \sum_\alpha \prod_{i=1}^n \chi(\omega(k_i) \in I_{\alpha_i}) \qquad a.e. \text{ in } \mathbb{R}^{nd}. \tag{142}$$

This shows that $\chi_S\varphi = \sum_\alpha \prod_{i=1}^n \chi(\omega(k_i) \in I_{\alpha_i})\varphi$ for any function $\varphi \in L^2(\mathbb{R}^{nd})$ where the sum converges in $L^2$-norm by the monotone convergence theorem. Now let $J_{\alpha_i}(k) = \chi(\omega(k) \in I_{\alpha_i})$. By (141) and (142) and the symmetry of $\chi_S(k_1, \ldots, k_n)$ with respect to permutation of the variables we get

$$\begin{aligned}
(\chi_S\psi)(k_1, \ldots, k_n) &= \frac{1}{\sqrt{n!}}\sum_\sigma \chi_S(k_{\sigma 1}, \ldots, k_{\sigma n})h_1(k_{\sigma 1}) \cdots h_n(k_{\sigma n}) \\
&= \sum_\alpha \frac{1}{\sqrt{n!}}\sum_\sigma (J_{\alpha_1}h_1)(k_{\sigma 1}) \cdots (J_{\alpha_n}h_n)(k_{\sigma n}).
\end{aligned}$$

Hence $\psi = \sum_\alpha a^*(J_{\alpha_1}h_1) \cdots a^*(J_{\alpha_n}h_n)\Omega \in \overline{\mathcal{D}}_c$ which proves the lemma. $\qquad\square$

**Lemma 30.** *Suppose $\omega(k) = |k|$ or $\omega$ satisfies (H1), and let $c > 0$. Then the set of all linear combinations of vectors of the form*

$$\varphi \otimes a^*(h_1)\ldots a^*(h_n)\Omega, \qquad \lambda + \sum_{i=1}^{N} M_i < c \qquad (143)$$

*where $\varphi = \chi(H < \lambda)\varphi$ for some $\lambda < c$, $n \in \mathbb{N}$ and $M_i = \sup\{\omega(k) : h_i(k) \neq 0\}$, is dense in $\chi(\tilde{H} < c)\tilde{\mathcal{H}}$.*

*Proof.* Obviously the set of vectors $\chi(\tilde{H} < c)\varphi \otimes a^*(h_1)\ldots a^*(h_n)\Omega$ with $\varphi \in \mathcal{H}$, $h_i \in \mathfrak{h}$ and $n$ arbitrary is dense in $\chi(\tilde{H} < c)\tilde{\mathcal{H}}$. Thus it suffices to approximate such vectors by vectors of the form (143). In the sense of a strong Stieltjes integral

$$\chi(\tilde{H} < c) = \int dE_\lambda(H) \otimes \chi(\mathrm{d}\Gamma(\omega) < c - \lambda).$$

This means that

$$\chi(\tilde{H} < c)\varphi \otimes a^*(h_1)\ldots a^*(h_n)\Omega = \lim_{\sup|\Delta_i| \to 0} \sum_i E_{\Delta_i}(H)\varphi \otimes \chi(\mathrm{d}\Gamma(\omega) < c - \lambda_i)a^*(h_1)\ldots a^*(h_n)\Omega$$

where $\Delta_i = (\lambda_{i-1}, \lambda_i]$. The lemma now follows from Lemma 29 applied to $\chi(\mathrm{d}\Gamma(\omega) < c - \lambda_i)a^*(h_1)\ldots a^*(h_n)\Omega$. $\qquad\square$

# C   Number–Energy Estimates.

Thanks to the positivity of the boson mass, assumption (H1), one has the operator inequality

$$N \leq aH + b, \qquad (144)$$

for some constants $a$ and $b$. The purpose of this Section is to prove that also higher powers of $N$ are bounded with respect to the same powers of $H$. This easily follows from (144) if the commutator $[N, H]$ is zero, that is, for vanishing interaction. Otherwise it follow, as we will see, from the boundedness of $ad_N^k(H)(H+i)^{-1}$ for all $k$.

**Lemma 31.** *Assume the hypotheses (H1) and (H2) and suppose $m \in \mathbb{N} \cup \{0\}$.*

   *i) Then uniformly in $z$, for $z$ in a compact subset of $\mathbb{C}$,*

   $$\|(N+1)^{-m}(z-H)^{-1}(N+1)^{m+1}\| = O(|\operatorname{Im} z|^{-m}).$$

   *ii) $(N+1)^m(H+i)^{-m}$ is a bounded operator.*

   *iii) If $\chi \in C^\infty(\mathbb{R})$ with $\operatorname{supp}\chi \subset (-\infty, \Sigma)$, then $N^m e^{\alpha|x|}\chi(H)$ is a bounded operator, provided $\alpha > 0$ is small enough.*

*Proof.*  i) This is proved by induction in $m$. For $m = 0$ it suffices to show that $N(H+i)^{-1}$ is bounded, because $\|(H+i)(H-z)^{-1}\| = O(|\operatorname{Im}(z)|^{-1})$ for $z$ in a compact subset of $\mathbb{C}$. The operator $N(H_0+i)^{-1}$ is bounded because of the positivity of the boson mass, assumption (H1). Since $\phi(G)$ is infinitesimally small with respect to $H_0$, it follows that $H_0(H+i)^{-1}$ is bounded and hence so is $N(H+i)^{-1}$.

Next assume that the assertion in i) holds for any non–negative integer less than a given $m \in \mathbb{N}$. To prove it for $m$ we commute $(z-H)^{-1}$ with $(N+1)^{m+1}$ and get

$$
\begin{aligned}
(N+1)&^{-m}(z-H)^{-1}(N+1)^{m+1} \\
&= (N+1)(z-H)^{-1} + (N+1)^{-m}(z-H)^{-1}[H,(N+1)^{m+1}](z-H)^{-1} \\
&= (N+1)(z-H)^{-1} \\
&\quad + (N+1)^{-m}(z-H)^{-1}\sum_{l=1}^{m+1}\binom{m+1}{l}(-1)^l(N+1)^{m+1-l}\mathrm{ad}_N^l(H)(z-H)^{-1} \\
&= (N+1)(z-H)^{-1} \\
&\quad + (N+1)^{-1}\sum_{l=1}^{m+1}\binom{m+1}{l}\underbrace{(N+1)^{-(m-1)}(z-H)^{-1}(N+1)^{m+1-l}}_{O(|\operatorname{Im}z|^{-m+1})}i^l\phi(i^lG)(z-H)^{-1}.
\end{aligned}
$$
(145)

By induction hypothesis and because $\phi(i^lG)(z-H)^{-1}$ is of order $O(|\operatorname{Im}z|^{-1})$, for $z$ in a compact set in $\mathbb{C}$, it follows that the r.h.s. of the last equation is of order $O(|\operatorname{Im}z|^{-m})$.

ii) Follows directly from i) if we put $z = -i$ and write

$$
\begin{aligned}
(N+1)^m(H+i)^{-m} &= (N+1)^m(H+i)^{-1}(N+1)^{-m+1} \\
&\quad \times (N+1)^{m-1}(H+i)^{-1}(N+1)^{-m+2}\cdots \\
&\quad \cdots(N+1)(H+1)^{-1}.
\end{aligned}
$$

iii) We begin to write

$$
\begin{aligned}
N^m e^{\alpha|x|}\chi(H) &= N^m e^{\alpha|x|}(H+i)^{-m}e^{-\alpha|x|}e^{+\alpha|x|}\chi(H)(H+i)^m \\
&= N^m(i+e^{+\alpha|x|}He^{-\alpha|x|})^{-m}e^{+\alpha|x|}\chi(H)(H+i)^m \\
&= N^m(i+H+\delta H_\alpha)^{-m}e^{+\alpha|x|}\chi(H)(H+i)^m,
\end{aligned}
$$
(146)

where $\delta H_\alpha = 2i\alpha p\cdot x/|x| - 2\alpha/|x| - \alpha^2$, and $\alpha > 0$ is so small that $\|\delta H_\alpha(H+i)^{-1}\| < 1$ (this ensures that $(i+H+\delta H_\alpha) = (1+\delta H_\alpha(H+i)^{-1})(H+i)$ is invertible, with a bounded inverse given by $(H+i)^{-1}(1+\delta H_\alpha(H+i)^{-1})^{-1})$. The last equation implies iii), since $N^m(i+H+\delta H_\alpha)^{-m}$ is a bounded operator. This can be shown in the same way we showed that $N^m(H+i)^{-m}$ is a bounded operator, using that $N$ commutes with $\delta H_\alpha$. $\qquad\square$

# D   Commutator Estimates

Let $j_0$, $j_\infty \in C^\infty(\mathbb{R}^d)$ be real-valued with $j_0^2 + j_\infty^2 \leq 1$, $j_0(y) = 1$ for $|y| \leq 1$ and $j_0(y) = 0$ for $|y| \geq 2$. Given $R > 0$ set $j_{\#,R}(y) = j_\#(y/R)$ and let $j_R : \mathfrak{h} \to \mathfrak{h} \oplus \mathfrak{h}$ denote the operator defined by $j_R h = (j_{0,R} h, j_{\infty,R} h)$.

**Lemma 32.** *Assume hypotheses (H1), (H2), (H5). Suppose $\alpha > 0$, $m \in \mathbb{N} \cup \{0\}$, and $j_R$ is as above. Suppose also that $\chi, \chi' \in C_0^\infty(\mathbb{R})$, with supp $\chi' \subset (-\infty, \Sigma)$. Then, for $R \to \infty$,*

*i)* $e^{-\alpha|x|}(N_0 + N_\infty + 1)^m \left( \breve{\Gamma}(j_R)H - \tilde{H}\breve{\Gamma}(j_R) \right)(N+1)^{-m-1} = O(R^{-1})$,

*ii)* $(N_0 + N_\infty + 1)^m \left( \chi(\tilde{H})\breve{\Gamma}(j_R) - \breve{\Gamma}(j_R)\chi(H) \right)\chi'(H) = O(R^{-1})$.

*Proof.*    i) From the intertwining relations (24), and (25) we have that

$$\breve{\Gamma}(j_R)H - \tilde{H}\breve{\Gamma}(j_R) = -\,\mathrm{d}\breve{\Gamma}(j_R, \mathrm{ad}_\omega(j_R))$$
$$+ [\phi((j_{0,R}-1)G) \otimes 1 + 1 \otimes \phi(j_{\infty,R}G)]\breve{\Gamma}(j_R).$$

By Lemma 27 $\mathrm{ad}_\omega(j_R) = O(R^{-1})$. Hence

$$(N_0 + N_\infty + 1)^m \mathrm{d}\breve{\Gamma}(j_R, \mathrm{ad}_\omega(j_R))(N+1)^{-m-1} = O(R^{-1}).$$

To see that the other two terms lead to contributions of order $O(R^{-1})$ write

$$(N_0 + N_\infty + 1)^m[\phi((j_{0,R}-1)G) \otimes 1 + 1 \otimes \phi(j_{\infty,R}G)]$$
$$= [\phi((j_{0,R}-1)G) \otimes 1 + 1 \otimes \phi(j_{\infty,R}G)](N_0 + N_\infty + 1)^m$$
$$+ \sum_{l=1}^{m} \binom{m}{l}(-i)^l[\phi(i^l(j_{0,R}-1)G) \otimes 1 + 1 \otimes \phi(i^l j_{\infty,R}G)](N_0 + N_\infty + 1)^{m-l},$$

then use $(N_0 + N_\infty + 1)^j \breve{\Gamma}(j_R) = \breve{\Gamma}(j_R)(N+1)^j$ and use Hypothesis (H5).

ii) Let $\tilde{\chi}$ be an almost analytic extension of $\chi$ of order $m$, as defined in Sect. A.2. Then we have

$$(N_0 + N_\infty + 1)^m(\chi(\tilde{H})\breve{\Gamma}(j_R) - \breve{\Gamma}(j_R)\chi(H))\chi'(H)$$
$$= -\frac{1}{\pi}\int dxdy\,\partial_{\bar{z}}\tilde{\chi}(N_0 + N_\infty + 1)^m(z - \tilde{H})^{-1}e^{-\alpha|x|}(\tilde{H}\breve{\Gamma}(j_R) - \breve{\Gamma}(j_R)H)$$
$$\times e^{\alpha|x|}\chi'(H)(z-H)^{-1}.$$

Then the statement follows by i) because

$$(N_0 + N_\infty + 1)^m(z - \tilde{H})^{-1}(N_0 + N_\infty + 1)^{-m+1} = O(|\operatorname{Im} z|^{-m}), \tag{147}$$

and because $(N+1)^m e^{\alpha|x|}\chi'(H)$ is a bounded operator, provided $\alpha > 0$ is sufficiently small. $\qquad\square$

**Lemma 33.** *Assume hypotheses (H1), (H2), (H5). Suppose $\alpha > 0$, $m \in \mathbb{N} \cup \{0\}$. Let $j_R$ be as above and set $dj_R = [i\omega, j_R] + \partial j_R / \partial R$. Suppose also that $\chi, \chi' \in C_0^\infty(\mathbb{R})$, with* supp $\chi' \subset (-\infty, \Sigma)$. *Then*

   *i)* $e^{-\alpha|x|}(N_0 + N_\infty + 1)^m \left( \mathrm{d}\breve{\Gamma}(j_R, dj_R)H - \tilde{H}\mathrm{d}\breve{\Gamma}(j_R, dj_R) \right)(N+1)^{-m-2} = O(R^{-2})$,

   *ii)* $(N_0 + N_\infty + 1)^m \left( \mathrm{d}\breve{\Gamma}(j_R, dj_R)\chi(H) - \chi(\tilde{H})\mathrm{d}\breve{\Gamma}(j_R, dj_R) \right)\chi'(H) = O(R^{-2})$.

*Proof.* We begin proving part i). To this end note that

$$\mathrm{d}\breve{\Gamma}(j_R, dj_R)H - \tilde{H}\mathrm{d}\breve{\Gamma}(j_R, dj_R) = \mathrm{d}\breve{\Gamma}(j_R, dj_R)\mathrm{d}\Gamma(\omega) - (\mathrm{d}\Gamma(\omega) \otimes 1 + 1 \otimes \mathrm{d}\Gamma(\omega))\mathrm{d}\breve{\Gamma}(j_R, dj_R)$$
$$+ \mathrm{d}\breve{\Gamma}(j_R, dj_R)\phi(G) - (\phi(G) \otimes 1)\mathrm{d}\breve{\Gamma}(j_R, dj_R).$$
$$(148)$$

Consider the term on the first line on the r.h.s. of the last equation. We have

$$\mathrm{d}\breve{\Gamma}(j_R, dj_R)\mathrm{d}\Gamma(\omega) - (\mathrm{d}\Gamma(\omega) \otimes 1 + 1 \otimes \mathrm{d}\Gamma(\omega))\mathrm{d}\breve{\Gamma}(j_R, dj_R) \!\restriction \otimes_s^n \mathfrak{h}$$
$$= U \left[ \sum_{i \neq j} (j_R \otimes \cdots \otimes dj_R \otimes \cdots \otimes [j_R, \omega] \otimes \cdots \otimes j_R) \right.$$
$$\left. + \sum_{i=1}^n (j_R \otimes \cdots \otimes [dj_R, \omega] \otimes \cdots \otimes j_R) \right].$$
$$(149)$$

Each one of the $n^2$ terms on the r.h.s. of the last equation is of order $O(R^{-2})$, by Lemma 27. Hence

$$(N_0 + N_\infty + 1)^m \left( \mathrm{d}\breve{\Gamma}(j_R, dj_R)\mathrm{d}\Gamma(\omega) \right.$$
$$\left. - (\mathrm{d}\Gamma(\omega) \otimes 1 + 1 \otimes \mathrm{d}\Gamma(\omega))\mathrm{d}\breve{\Gamma}(j_R, dj_R) \right)(N+1)^{-m-2} = O(R^{-2}).$$

Consider now the term on the second line on the r.h.s. of (148). This is given by

$$\mathrm{d}\breve{\Gamma}(j_R, dj_R)\phi(G) - (\phi(G) \otimes 1)\mathrm{d}\breve{\Gamma}(j_R, dj_R) = \left( \phi((j_{0,R} - 1)G) \otimes 1 + 1 \otimes \phi(j_{\infty,R}G) \right)\mathrm{d}\breve{\Gamma}(j_R, dj_R)$$
$$+ \left( \phi(dj_{\infty,R}G) \otimes 1 + 1 \otimes \phi(dj_{\infty,R}G) \right)\breve{\Gamma}(j_R).$$

By Hypothesis (H5) it follows, in the same way as in the proof of Lemma 32, that

$$e^{-\alpha|x|}(N_0 + N_\infty + 1)^m \left( \mathrm{d}\breve{\Gamma}(j_R, dj_R)\phi(G) - (\phi(G) \otimes 1)\mathrm{d}\breve{\Gamma}(j_R, dj_R) \right)(N+1)^{-m-2} = O(R^{-\mu-1}).$$

This completes the proof of part i). Part ii) of the Lemma follows in the same way as in Lemma 32. $\qquad\square$

# E   Mourre Estimate

This section contains the proofs of Lemma 3 (the Virial Theorem) and of Theorem 4 (the Mourre Estimate). Recall that $A$ stands for $\mathrm{d}\Gamma(a)$ with $a = 1/2(\nabla\omega \cdot y + y \cdot \nabla\omega)$.

**Proof of Lemma 3, (Virial Theorem).** First we prove the virial theorem for a regularized variant, $A_\varepsilon$, of $A = A_{\varepsilon=0}$ defined on $D(H)$, and then we let $\varepsilon \to 0$. Let $a_\varepsilon = 1/2(\nabla\omega \cdot y_\varepsilon + y_\varepsilon \cdot \nabla\omega)$ where $y_\varepsilon = y(1 + \varepsilon y^2)^{-1}$ and let $A_\varepsilon = \mathrm{d}\Gamma(a_\varepsilon)$. Then

$$i\langle H\varphi, A_\varepsilon\varphi\rangle - i\langle A_\varepsilon\varphi, H\varphi\rangle = \langle\varphi, \left[\mathrm{d}\Gamma(i[\omega, a_\varepsilon]) - \phi(ia_\varepsilon G)\right]\varphi\rangle \tag{150}$$

for all $\varphi \in D(K) \otimes \mathcal{F}_0$, which is a core for $H$. Since all operators in (150) are $H$-bounded this equation extends to $D(H)$. If $\varphi$ is an eigenvector $H$ then the left side of (150) vanishes because $A_\varepsilon \subset A_\varepsilon^*$. Thus it suffices to prove that the right side converges to $\langle\varphi, i[H, A]\varphi\rangle$, as $\varepsilon \to 0$, for $\varphi \in E_\mu(H)\mathcal{H}$, $\mu < \Sigma$. We show that

$$\phi(ia_\varepsilon G)\varphi \to \phi(iaG)\varphi, \tag{151}$$

$$\text{and} \qquad \mathrm{d}\Gamma(i[\omega, a_\varepsilon])\varphi \to \mathrm{d}\Gamma(|\nabla\omega|^2)\varphi \tag{152}$$

as $\varepsilon \to 0$. Eq. (151) follows from $\|a_\varepsilon G - aG\| \to 0$ and $\sup_x e^{-\alpha|x|}\|\phi((a - a_\varepsilon)G)(N + 1)^{-1/2}\| < \infty$ by Lebesgue's dominated convergence theorem if $\alpha > 0$ is chosen in such a way that $e^{\alpha|x|}(N + 1)^{1/2}\varphi$ belongs to $\mathcal{H}$. Here we used (H2) and (H3). To prove (152) note that $i[\omega, y_\varepsilon] \to \nabla\omega$ strongly and hence that $i[\omega, a_\varepsilon] \to |\nabla\omega|^2$ strongly. Since moreover $\sup_\varepsilon \|i[\omega, a_\varepsilon]\| < \infty$ this proves (152) for all $\varphi \in D(N)$. $\qquad\square$

The property proved in the following lemma has a well known analog, called local compactness, in the theory of Schrödinger operators. We used it in the proof of Theorem 2 and will need it again in the subsequent proof of the Mourre theorem, Thm. 4.

**Lemma 34.** *Assume (H1) and suppose $f \in L^\infty(X)$ and $g \in L^\infty(\mathbb{R}^d, dy)$ with $\|g\|_\infty \leq 1$. If $f(x) \to 0$, as $|x| \to \infty$, and $g(y) \to 0$, as $|y| \to \infty$ in $\mathbb{R}^d$, then*

$$f(x) \otimes \Gamma(g)(H + i)^{-1/2} \qquad \text{is compact.}$$

This lemma follows from the fact that $p^2$ and $\mathrm{d}\Gamma(\omega)$ are form bounded w.r. to $H$ and from the positivity of the boson mass.

*Proof.* From $m > 0$ it follows that $\chi(N > n)\Gamma(g)(\mathrm{d}\Gamma(\omega) + 1)^{-1/4} \to 0$ as $n \to \infty$. Furthermore

$$\Gamma(g)(\mathrm{d}\Gamma(\omega) + 1)^{-1/4}\!\upharpoonright \otimes_s^n \mathfrak{h} = g(y_1)\ldots g(y_n)[\omega(k_1) + \ldots + \omega(k_n) + 1]^{-1/4} \tag{153}$$

which is compact. It follows that $\Gamma(g)(\mathrm{d}\Gamma(\omega) + 1)^{-1/4}$ is a compact operator on $\mathcal{F}$. Since $f(x)(p^2 + 1)^{-1/4}$ is compact on $L^2(X)$ and $(p^2 + 1)^{1/4} \otimes (\mathrm{d}\Gamma(\omega) + 1)^{1/4}(H + i)^{-1/2}$ is a bounded operator by ?, the operator $f \otimes \Gamma(g)(H + i)^{-1/2}$ is compact. $\qquad\square$

The proof of Theorem 4 is by induction in the number of energy steps of size $m$, the minimal energy of a free boson. Assuming that the theorem holds for $\lambda < min(\Sigma, (n-1)m)$ we prove it for $\lambda < min(\Sigma, nm)$. To this end we need a partition of unity in the bosonic configuration space. Let $j_0, \ j_\infty \in C^\infty(\mathbb{R}^d)$, with $j_0^2 + j_\infty^2 = 1$, and with $0 \le j_0 \le 1$, $j_0(y) = 1$ if $|y| \le 1$ and $j_0(y) = 0$ if $|y| \ge 2$. Given $R > 0$ we set $j_{\#,R}(y) = j_\#(y/R)$ and $q_R(y) = q(y/R)$. Each boson $h \in \mathfrak{h}$ will be split into the two parts $j_{0,R}h$ and $j_{\infty,R}$ localized near the origin and near infinity respectively. Recall from Section 2.6 that $\check{\Gamma}(j_R)$ does the corresponding localization in Fock space. In Appendix D we saw how to move $\check{\Gamma}(j_R)$ through $H$ or any bounded function of $H$ (see Lemma 32). Since $[iH, A]$ has a structure similar to $H$, with $\omega$ and $G$ replaced by $|\nabla\omega|^2$ and by $iaG$, respectively, it is easy to show that an analogue of Lemma 32 also holds for $[iH, A]$. In particular we use that

$$e^{-\alpha|x|}\left\{\check{\Gamma}(j_R)[iH, A] - ([iH, A] \otimes 1 + 1 \otimes d\Gamma(|\nabla\omega|^2))\check{\Gamma}(j_R)\right\}(N+1)^{-1} = o(R^0), \quad (154)$$

as $R \to \infty$.

**Proof of Theorem 4 (Mourre Estimate).** First of all we introduce the Mourre constants

$$d(\lambda) := \overset{\Omega^\perp}{\underset{\sigma_{pp}(H) + d\Gamma(\omega(k)) = \lambda}{\inf}} \quad d\Gamma(|\nabla\omega(k)|^2) \qquad (155)$$

$$\tilde{d}(\lambda) := \underset{\sigma_{pp}(H) + d\Gamma(\omega(k)) = \lambda}{\inf} \quad d\Gamma(|\nabla\omega(k)|^2), \qquad (156)$$

where the superscript $\Omega^\perp$ in the definition of $d$ means that we exclude the vacuum sector to compute the infimum. Note that $d$ vanishes only on thresholds, while $\tilde{d}$ vanishes on thresholds and on eigenvalues of $H$. We introduce, moreover, the smeared out versions of the functions $d, \tilde{d}$. For $\kappa > 0$, we set $\Delta_\lambda^\kappa = [\lambda - \kappa, \lambda + \kappa]$, and then we define $d^\kappa(\lambda) = \inf_{\mu \in \Delta_\lambda^\kappa} d(\mu)$ and $\tilde{d}^\kappa(\lambda) = \inf_{\mu \in \Delta_\lambda^\kappa} \tilde{d}(\mu)$. By definition of these functions we have the inequality

$$\overset{\Omega^\perp}{\inf}\left(\tilde{d}^\kappa(\lambda - d\Gamma(\omega(k))) + d\Gamma(|\nabla\omega(k)|^2)\right) \ge d^\kappa(\lambda). \qquad (157)$$

For all $n \in \mathbb{N}$ we will show the following statements.

$H_1(n)$: Let $\varepsilon > 0$ and $\lambda \in [E_0, E_0 + nm) \cap (-\infty, \Sigma)$. Then there exists an open interval $\Delta \ni \lambda$ and a compact operator $\mathcal{E}$ such that $E_\Delta(H)[iH, A]E_\Delta(H) \ge (d(\lambda) - \varepsilon)E_\Delta(H) + \mathcal{E}$.

$H_2(n)$: Let $\varepsilon > 0$ and $\lambda \in [E_0, E_0 + nm) \cap (-\infty, \Sigma)$. Then there exists an open interval $\Delta \ni \lambda$ such that $E_\Delta(H)[iH, A]E_\Delta(H) \ge (\tilde{d}(\lambda) - \varepsilon)E_\Delta(H)$.

$H_3(n)$: Let $\kappa, \ \varepsilon_0, \ \varepsilon > 0$. Then there exists $\delta > 0$ such that for all $\lambda \in [E_0, E_0 + nm - \varepsilon_0] \cap (-\infty, \Sigma)$, one has $E_{\Delta_\lambda^\delta}(H)[iH, A]E_{\Delta_\lambda^\delta}(H) \ge (\tilde{d}^\kappa(\lambda) - \varepsilon)E_{\Delta_\lambda^\delta}(H)$.

$S_1(n)$: $\tau$ is a closed and countable set in $[E_0, E_0 + nm) \cap (-\infty, \Sigma)$.

$S_2(n)$: For any closed interval $I \subset [E_0, E_0 + nm) \cap (-\infty, \Sigma)$, with $I \cap \tau = \varnothing$, one has $\dim \text{Ran } E_{I \cap \sigma_{pp}(H)} < \infty$.

Note here that all the claims of the theorem follow from these statements, if we prove them for any $n \in \mathbb{N}$. Actually, the statements give no new information if $n$ is so large that $E_0 + nm > \Sigma$. We will prove these statements by induction in $n$. Since $H_1(1)$ and $S_1(1)$ are obvious, all the statements, for any $n \in \mathbb{N}$, follow if we prove the implications: $H_1(n) \Rightarrow H_2(n)$, $H_2(n) \Rightarrow H_3(n)$, $H_1(n) \Rightarrow S_2(n)$, $S_2(n-1) \Rightarrow S_1(n)$, and $S_1(n) \wedge H_3(n-1) \Rightarrow H_1(n)$. Now the implication $S_2(n-1) \Rightarrow S_1(n)$ is trivial. Moreover the implications $H_1(n) \Rightarrow H_2(n)$, $H_2(n) \Rightarrow H_3(n)$ and $H_1(n) \Rightarrow S_2(n)$ are proved by standard abstract arguments, which are typical in the proof of any Mourre inequality. It only remains to prove that $S_1(n)$ together with $H_3(n-1)$ imply the statement $H_1(n)$. To this end we fix $\lambda \in [E_0, E_0 + nm) \cap (-\infty, \Sigma)$ and $\varepsilon > 0$. Choose now $\chi \in C_0^\infty(\mathbb{R})$ with $\operatorname{supp} \chi \subset [\lambda - \delta, \lambda + \delta] \cap (-\infty, \Sigma)$, where $\delta$ is some positive constant which will be fixed later on. Then we have

$$
\begin{aligned}
\chi(H) & [iH, A] \chi(H) \\
&= \breve{\Gamma}(j_R)^* E_{\{0\}}(N_\infty) \breve{\Gamma}(j_R) \chi(H)[iH, A]\chi(H) + \breve{\Gamma}(j_R)^* E_{[1,\infty)}(N_\infty) \breve{\Gamma}(j_R) \chi(H)[iH, A]\chi(H) \\
&= \Gamma(q_R)\chi(H)[iH, A]\chi(H) \\
&\quad + \breve{\Gamma}(j_R)^* \chi(\tilde{H}) \left\{ [iH, A] \otimes 1 + 1 \otimes d\Gamma(|\nabla\omega|^2) \right\} \chi(\tilde{H}) E_{[1,\infty)}(N_\infty) \breve{\Gamma}(j_R) + o(R^0),
\end{aligned}
\tag{158}
$$

where we used Lemma 32 and Eq. (154) to commute $\breve{\Gamma}(j_R)$ to the right. The first term on the r.h.s. of (158) is compact, by Lemma 34. To handle the second term we want to diagonalize $1 \otimes d\Gamma(\omega)$ and $1 \otimes d\Gamma(|\nabla\omega|^2)$ on the range of $E_{[1,\infty)}(N_\infty)$. Using the induction hypothesis $S_1(n)$ we find $\kappa > 0$ such that $d^\kappa(\lambda) \geq d(\lambda) - \varepsilon/3$. Then, by $H_3(n-1)$, we know that if $\delta > 0$ is small enough we have

$$
\begin{aligned}
E_{\Delta_\lambda^\delta}&(H + d\Gamma(\omega(k))) \left\{ [iH, A] \otimes 1 + 1 \otimes d\Gamma(|\nabla\omega(k)|^2) \right\} E_{\Delta_\lambda^\delta}(H + d\Gamma(\omega(k))) E_{[1,\infty)}(N_\infty) \\
&\geq E_{\Delta_\lambda^\delta}(H + d\Gamma(\omega(k))) \left\{ \tilde{d}^\kappa(\lambda - d\Gamma(\omega(k))) + d\Gamma(|\nabla\omega(k)|^2) - \frac{\varepsilon}{3} \right\} E_{[1,\infty)}(N_\infty) \\
&\geq (d^\kappa(\lambda) - \frac{\varepsilon}{3}) E_{\Delta_\lambda^\delta}(H + d\Gamma(\omega(k))) E_{[1,\infty)}(N_\infty) \\
&\geq (d(\lambda) - \frac{2\epsilon}{3}) E_{\Delta_\lambda^\delta}(H + d\Gamma(\omega(k))) E_{[1,\infty)}(N_\infty),
\end{aligned}
$$

where, in the second inequality we used (157). It follows, since $\operatorname{supp} \chi \subset \Delta_\lambda^\delta$, that

$$
\chi(\tilde{H}) \left\{ [iH, A] \otimes 1 + 1 \otimes d\Gamma(|\nabla\omega|^2) \right\} \chi(\tilde{H}) E_{[1,\infty)} \geq (d(\lambda) - \frac{2\epsilon}{3}) \chi^2(\tilde{H}) E_{[1,\infty)}(N_\infty).
$$

Now we insert this in the second term on the r.h.s. of (158) and then we commute, using again Lemma 32 and Eq. (154), $\breve{\Gamma}(j_R)$ back to the left. Using that $\breve{\Gamma}(j_R)^* E_{[1,\infty)}(N_\infty) \breve{\Gamma}(j_R) = 1 - \Gamma(q_R)$ and that, by Lemma 34, $\Gamma(q_R)\chi(H)$ is a compact operator, we find, from (158),

$$
\chi(H)[iH, A]\chi(H) \geq (d(\lambda) - \frac{2\varepsilon}{3}) \chi^2(H) + \mathcal{E} + o(R^0).
$$

The statement $H_1(n)$ then follows if we choose $\chi$ such that $\chi = 1$ on $\Delta_\lambda^{\delta/2}$, if we multiply (158) from the right and from the left with $E_{\Delta_\lambda^{\delta/2}}(H)$ and if we choose $R$ sufficiently large. $\qquad \square$

# F  Invariance of Domains

In this section the invariance of the domain of $d\Gamma(y^2)$ with respect to action of $\chi(H)$ for smooth functions $\chi$ is proven. Moreover we prove in Lemma 36 that the norm of $d\Gamma(y^2/t^2)\chi(H)e^{-iHt}\varphi$ remains uniformly bounded for all $t \geq 1$, if $\varphi \in D(d\Gamma(y^2 + 1))$. These results are used in Proposition 13 to prove the existence of the asymptotic observable $W$.

**Lemma 35.** *Assume Hypotheses (H1), (H2) and (H3) are satisfied. Suppose moreover that $\varphi \in D(d\Gamma(y^2)) \cap D(N)$ and that $\chi \in C^\infty(\mathbb{R})$ with* supp $\chi \subset (-\infty, \Sigma)$. *Then we have*

$$\|d\Gamma(y^2)\chi(H)\varphi\| \leq C(\|d\Gamma(y^2 + 1)\varphi\| + \|\varphi\|).$$

*Proof.* Find $\chi_1 \in C_0^\infty(\mathbb{R})$ with $\chi\chi_1 = \chi$, and supp $\chi_1 \subset (-\infty, \Sigma)$. Put $\chi = \chi(H)$ and $\chi_1 = \chi_1(H)$. Then we have

$$d\Gamma(y^2)\chi = d\Gamma(y^2)\chi\chi_1 = \chi d\Gamma(y^2)\chi_1 + [d\Gamma(y^2), \chi]\chi_1 \tag{159}$$

Now expand the $\chi$ in the commutator in an integral, according to the Helffer–Sjöstrand functional calculus (see Appendix A.2). We get

$$\begin{aligned}
[d\Gamma(y^2), \chi]\chi_1 = &-\frac{2i}{\pi} \int dxdy\, \partial_{\bar{z}}\tilde{\chi}\,(z - H)^{-1} d\Gamma(a)(z - H)^{-1}\chi_1 \\
&+ \frac{i}{\pi} \int dxdy\, \partial_{\bar{z}}\tilde{\chi}\,(z - H)^{-1}\phi(iy^2 G)\chi_1(z - H)^{-1},
\end{aligned} \tag{160}$$

where $\tilde{\chi}$ is an almost analytic extension of $\chi$, in the sense of the Helffer-Sjöstrand functional calculus, and $a = [i\omega(k), y^2/2] = 1/2(\nabla\omega \cdot y + y \cdot \nabla\omega)$. The second term on the r.h.s. of the last equation is bounded, because, by hypotheses (H2) and (H3), $\|\phi(y^2G)\chi_1\| < \infty$. To handle the first term on the r.h.s. of (160) we commute the factor $d\Gamma(a)$ to the right of the resolvent $(z - H)^{-1}$. Using $[d\Gamma(a), H] = id\Gamma(|\nabla\omega|^2) - i\phi(iaG)$, we see that the contributions arising from this commutator are bounded, because $|\nabla\omega| <$ const and because, using again (H2) and (H3), $\|\phi(iaG)\chi\| < \infty$ is a bounded operator. Thus we have, from (159)

$$d\Gamma(y^2)\chi = \chi d\Gamma(y^2)\chi_1 + C\,\chi'(H)d\Gamma(a)\chi_1 + \text{ bounded}, \tag{161}$$

where $\chi'$ is the first derivative of $\chi$. Now we commute the two operators $d\Gamma(y^2)$ and $d\Gamma(a)$ in the two terms on the r.h.s. of the last equation to the right of $\chi_1$. For example , for the term $\chi d\Gamma(y^2)\chi_1$, we find

$$\begin{aligned}
\chi d\Gamma(y^2)\chi_1 = &\,\chi\chi_1 d\Gamma(y^2) + \chi[d\Gamma(y^2), \chi_1] \\
= &\,\chi d\Gamma(y^2) - \frac{1}{\pi} \int dxdy\, \partial_{\bar{z}}\tilde{\chi}_1\,\chi(z - H)^{-1}[d\Gamma(y^2), H](z - H)^{-1} \\
= &\,\chi d\Gamma(y^2) - \frac{2i}{\pi t^2} \int dxdy\, \partial_{\bar{z}}\tilde{\chi}_1\chi(z - H)^{-1} d\Gamma(a)(z - H)^{-1} \\
&+ \frac{i}{\pi} \int dxdy\, \partial_{\bar{z}}\tilde{\chi}_1(z - H)^{-1}\chi\phi(iy^2 G)(z - H)^{-1}.
\end{aligned} \tag{162}$$

The third term on the r.h.s. of the last equation is bounded by (H2) and (H3). To handle the second term we commute $\mathrm{d}\Gamma(a)$ to the right. We get

$$
\begin{aligned}
-\frac{2i}{\pi} \int dxdy\, \partial_{\bar{z}}\tilde{\chi}_1 \chi(z-H)^{-1}\mathrm{d}\Gamma(a)(z-H)^{-1} = {} & -\frac{2i}{\pi}\int dxdy\, \partial_{\bar{z}}\tilde{\chi}_1\chi(z-H)^{-2}\mathrm{d}\Gamma(a) \\
& + \frac{2}{\pi}\int dxdy\, \partial_{\bar{z}}\tilde{\chi}_1\,(z-H)^{-2}\chi\mathrm{d}\Gamma(|\nabla\omega|^2)(z-H)^{-1} \\
& - \frac{2}{\pi}\int dxdy\, \partial_{\bar{z}}\tilde{\chi}_1(z-H)^{-2}\chi\phi(iaG)(z-H)^{-1}.
\end{aligned}
$$

The first term on the r.h.s. of the last equation is proportional to $\chi\chi_1'd\Gamma(a)$, where $\chi_1'$ is the first derivative of $\chi_1$, and thus vanishes, since $\chi_1$ is constant on supp $\chi$. The other two terms on the r.h.s. of the last equation are bounded. It follows, by (162), that $\chi\mathrm{d}\Gamma(y^2)\chi_1 = \chi\mathrm{d}\Gamma(y^2) +$ bounded. Similarly we find, that $\chi'\mathrm{d}\Gamma(a)\chi_1 = \chi'\mathrm{d}\Gamma(a) +$ bounded. These two results imply, by (161), that

$$
\|\mathrm{d}\Gamma(y^2)\chi\varphi\| \le C\left\{\|\mathrm{d}\Gamma(y^2)\varphi\| + \|\mathrm{d}\Gamma(a)\varphi\| + \|\varphi\|\right\}. \tag{163}
$$

The Lemma now follows using Lemma 37, part (iv), to estimate the second term on the r.h.s. of the last equation. $\qquad\square$

**Lemma 36.** *Assume hypotheses (H1), (H2) and (H3) are satisfied. Suppose moreover that $\varphi \in D(\mathrm{d}\Gamma(y^2)) \cap D(N)$ and that $\chi \in C^\infty(\mathbb{R})$ with supp $\chi \subset (-\infty, \Sigma)$. Then we have*

$$
\|\mathrm{d}\Gamma(y^2)e^{-iHt}\chi(H)\varphi\| \le C(\|\mathrm{d}\Gamma(y^2+1)\varphi\| + t^2\,\|\varphi\|),
$$

*for all $t \ge 1$.*

*Proof.* We begin by noting, that

$$
\begin{aligned}
e^{iHt}\mathrm{d}\Gamma(y^2)e^{-iHt}\chi(H) - \mathrm{d}\Gamma(y^2)\chi(H) &= \int_0^t ds\, e^{iHs}\left[iH, \mathrm{d}\Gamma(y^2/t^2)\right]\chi(H)e^{-iHs} \\
&= 2\int_0^t ds\, e^{iHs}\,\mathrm{d}\Gamma(a)\,\chi(H)e^{-iHs} - \int_0^t ds\, e^{iHs}\,\phi(iy^2G)\,\chi(H)e^{-iHs},
\end{aligned} \tag{164}
$$

where $a = [i\omega, y^2/2] = 1/2(\nabla\omega \cdot y + y \cdot \nabla\omega)$. The second integral on the r.h.s. of the last equation is bounded, with norm of order $t$, because, by (H2) and (H3), $\|\phi(iy^2G)\chi\| < \infty$. To handle the first integral on the r.h.s. of the last equation use the expansion

$$
\begin{aligned}
2\int_0^t ds\, e^{iHs}\mathrm{d}\Gamma(a)\chi(H)e^{-iHs} &= 2\mathrm{d}\Gamma(a)\chi(H) + 2\int_0^t ds\, \int_0^s dr\, e^{iHr}\left[iH, \mathrm{d}\Gamma(a)\right]\chi(H)e^{-iHr} \\
&= 2\mathrm{d}\Gamma(a)\chi(H) + 2\int_0^t ds\, \int_0^s dr\, e^{iHr}\mathrm{d}\Gamma(|\nabla\omega|^2)\chi(H)e^{-iHr} \\
&\quad - 2\int_0^t ds\, \int_0^s dr\, e^{iHr}\phi(iaG)\chi(H)e^{-iHr}.
\end{aligned} \tag{165}
$$

Since $|\nabla\omega|$ is bounded and $\|\phi(iaG)\chi(H)\| < \infty$, both the integrals on the r.h.s. of the last equation are bounded and of order $t^2$. This implies, by (164), that

$$
\begin{aligned}
\|\mathrm{d}\Gamma(y^2)e^{-iHt}\chi(H)\varphi\| &\leq \|\mathrm{d}\Gamma(y^2)\chi(H)\varphi\| + 2t\,\|\mathrm{d}\Gamma(a)\chi(H)\varphi\| + Ct^2\,\|\varphi\| \\
&= \|\mathrm{d}\Gamma(y^2)\chi(H)\varphi\| + 2t^2\,\|\mathrm{d}\Gamma(a/t)\chi(H)\varphi\| + Ct^2\,\|\varphi\|.
\end{aligned}
\tag{166}
$$

By Lemma 37, part (iv), with $y$ replaced by $y/t$, we have

$$
\|\mathrm{d}\Gamma(a/t)\chi\varphi\| \leq \|\mathrm{d}\Gamma(y^2/t^2 + 1)\chi\varphi\| \leq \|\mathrm{d}\Gamma(y^2/t^2)\chi\varphi\| + C\,\|\varphi\|.
\tag{167}
$$

The Lemma now follows inserting the last equation into (166) and applying Lemma 35. □

**Lemma 37.** *(i) For any operator $A$*

$$
A^2 + (A^*)^2 \leq AA^* + A^*A.
$$

*(ii) If $a = 1/2(y\cdot\nabla\omega + \nabla\omega\cdot y)$ and $\omega$ satisfies (H1) then*

$$
a^2 \leq \mathrm{const}(y^2 + 1).
$$

*(iii) If $a$ is a symmetric operator in $\mathfrak{h}$ then*

$$
\mathrm{d}\Gamma(a)^2 \leq N\mathrm{d}\Gamma(a^2)
$$

*(iv) With $a$ as in (ii) we have*

$$
\mathrm{d}\Gamma(a)^2 \leq \mathrm{const}\,\mathrm{d}\Gamma(y^2 + 1)^2.
$$

*Proof.* (i) This follows from $0 \leq (iA - iA^*)^2 = -A^2 - (A^*)^2 + AA^* + A^*A$.

(ii) By (i), $(\nabla\omega\cdot y + y\cdot\nabla\omega)^2 \leq 2[(\nabla\omega\cdot y)(y\cdot\nabla\omega) + (y\cdot\nabla\omega)(\nabla\omega\cdot y)]$ where

$$
\begin{aligned}
(y\cdot\nabla\omega)(\nabla\omega\cdot y) &\leq \|\nabla\omega\|_\infty^2\, y^2 \\
(\nabla\omega\cdot y)(y\cdot\nabla\omega) &\leq 2(\|\nabla\omega\|_\infty^2 + \|\Delta\omega\|_\infty^2)(y^2 + 1)
\end{aligned}
$$

(iii) It suffices to prove this for states $\varphi_n$ in $\mathrm{Ran}\,\chi(N = n)$. For such vectors $\langle\varphi_n, \mathrm{d}\Gamma(a)^2\varphi_n\rangle = \|\mathrm{d}\Gamma(a)\varphi_n\|^2 \leq (\sum_{i=1}^n \|a_i\varphi_n\|)^2 \leq n\sum_{i=1}^n \|a_i\varphi_n\|^2 = \langle\varphi_n, N\mathrm{d}\Gamma(a^2)\varphi_n\rangle$, where $a_i$ denotes the operator $a$ acting on the $i$-th boson.

(iv) This follows from (ii), (iii), and $N \leq \mathrm{d}\Gamma(y^2 + 1)$.

□

# G    Some Technical Parts of Theorem 15

**Lemma 38.** *There exists a constant $C = C(\lambda, u)$ such that*

$$\pm(J_{nk}S_n)''(y,t) \leq Cn^2 \binom{n}{k}^{1/2} S_n''(y,t)$$

*for $|y| \leq 2\lambda t$ and $ut^{1-\delta} \geq 2$. Here $y = (y_1, \ldots, y_n),\ y_i \in \mathbb{R}^d$.*

*Proof.* We drop the combinatorial factor in the definition of $J_{nk}$ for convenience. Let $S_0(y) = \sum_{i=1}^n S_0(y_i)$. From (51) and the scaling factor $1/ut$ in the arguments of $J_{nk}$ we get $tS_n''(y,t) = S_0''(y/t^\delta)$ and

$$
\begin{aligned}
t(J_{nk}S_n)''(y,t) = {} & J_{nk}(z) \cdot S_0''(w) + \alpha^{-2} J_{nk}''(z) S_0(w) \\
& + \alpha^{-1} \nabla J_{nk}(z) \otimes \nabla S_0(w) + \alpha^{-1} \nabla S_0(w) \otimes \nabla J_{nk}(z)
\end{aligned}
\tag{168}
$$

where $w = yt^{-\delta}$, $z = y/ut$, $\alpha = ut^{1-\delta}$ and hence $w = \alpha z$, $\alpha \geq 2$ and $|z| \leq 2\lambda/u$. Proving the lemma thus amounts to showing that the r.h.s. of (168) (and its negative) are bounded by const $\times n^2 S_0''(w)$. The first term in (168) enjoys the bound. To estimate the other three terms we need some "$N$-body geometry".

For each $a \in \{0, \infty\}^n$, let $1_a : \mathbb{R}^{nd} \to \mathbb{R}^{nd}$ denotes the orthogonal projection onto the subspace $\{w \in \mathbb{R}^{nd} | w_i = 0 \text{ if } a_i = 0\}$ and let $w_a = 1_a w$. Set

$$\chi_a(\sigma, w) = \prod_{r:a_r=0} \chi(w_r^2/2 \leq \sigma) \prod_{s:a_s=\infty} \chi(w_s^2/2 > \sigma).$$

Then $\sum_a \chi_a \equiv 1$. Hence, by (48) and since $\chi_a(\sigma, w)\, \chi(w_i^2/2 > \sigma) = \chi_a(\sigma, w)\delta_{a_i,\infty}$,

$$S_0(w) = \int d\sigma\, m(\sigma) S_0(\sigma, w) \tag{169}$$

$$S_0(\sigma, w) = \sum_{i=1}^n \chi(w_i^2/2 > \sigma)(w_i^2/2 - \sigma) = \sum_a \chi_a(\sigma, w)(w_a^2/2 - \sigma_a) \tag{170}$$

where $\sigma_a = \sigma \cdot \#\{i : a_i = \infty\}$. Furthermore

$$\nabla S_0(\sigma, w) = \sum_a \chi_a(\sigma, w) w_a \tag{171}$$

$$S_0''(\sigma, w) \geq \sum_a \chi(\sigma, w) 1_a \tag{172}$$

in the sense that $\nabla S_0(w) = \int d\sigma\, m(\sigma) \sum_a \chi_a(\sigma, w) w_a$ and $S_0''(w) \geq \int d\sigma\, m(\sigma) \sum_a \chi_a(\sigma, w) 1_a$. By (169) it suffices to estimate the last three terms of (168) for $S_0(w)$ replaced by $S_0(\sigma, w)$ uniformly in $\sigma$.

If $w_i^2/2 \leq 1$ and $\alpha \geq 2$ then $|z_i| \leq 1$ and hence $\partial^\beta j_\varepsilon(z_i) = 0$ for all derivatives of non-zero order $\beta$ and $\varepsilon \in \{0, \infty\}$. Hence for $w$ such that $\chi_a(\sigma, w) \neq 0$

$$J_{nk}''(z) = 1_a J_{nk}''(z) 1_a \tag{173}$$

$$\nabla J_{nk}(z) = 1_a \nabla J_{nk}(z). \tag{174}$$

From (170), (173) and (172) we get

$$\pm a^{-2} J''_{nk}(z) S_0(\sigma, w) \leq a^{-2} w^2 / 2 \, \|J''_{nk}(z)\| \sum_a \chi_a(w) 1_a$$

$$\leq C n^2 S''_0(\sigma, w).$$

The last two terms in (168) are estimated similarly using (171), (174) and (172). □

**Lemma 39.**

$$\mathcal{D}^m \partial^\alpha \left( S_{nk}(y, t) - J_{nk}(y, t) \frac{y^2}{t} \right) = n^{m+1} \binom{n}{k}^{1/2} O(t^{-(1+m) + \delta(2 - |\alpha|)})$$

*Proof.* The difference

$$S_{nk}(y, t) - J_{nk}(y, t) y^2 / 2t = J_{nk}(y, t) t^{-1 + 2\delta} \sum_{i=1}^n [a(y_i t^{-\delta}) + b] \qquad (175)$$

is of the form $t^{-c} f(y/t) h(y t^{-\delta})$ where $y = (y_1, \ldots, y_n)$ and $f, h \in C^\infty \cap L^\infty(\mathbb{R}^{nd})$. Under differentiation (175) becomes a sum of terms of the same form where in each term the exponent of $t$ is decreased by at least $\delta$ or 1 for differentiation w.r. to $y$ or $t$ (or $\mathcal{D}$) respectively. This accounts for the power in $t$. From the explicit form the right hand side of (175) and (82) we see that every derivative w.r. $y_i$ (at most) doubles the number of terms, while every derivative w.r. to $t$ multiplies it by $n + 1$. □

**Lemma 40.** *For $|y|/t \leq 2\lambda$ we have*

$$D_0^2 S_{nk} = (\nabla \Omega - V) \cdot S''_{nk} (\nabla \Omega - V)$$
$$+ (\mathcal{D} \nabla S_{nk}) \cdot (\nabla \Omega - V) + (\nabla \Omega - V) \cdot (\mathcal{D} \nabla S_{nk})$$
$$+ \mathcal{D}_V^2 S_{nk} + n^2 \binom{n}{k}^{1/2} O(t^{-1-\delta})$$

*where $V = y/t \in \mathbb{R}^{nd}$.*

*Proof.* We drop the factor $\binom{n}{k}^{1/2}$ since it appears in all terms on both sides of the equation. By definition of $D_0$,

$$D_0^2 S_{nk} = [i\Omega, [i\Omega, S_{nk}]] + 2[i\Omega, \partial S_{nk}/\partial t] + \frac{\partial^2 S_{nk}}{\partial t^2}. \qquad (176)$$

To evaluate the double commutator we use $S_{nk} = J_{nk} S_n$ and the Leibnitz rule and get for $|y/t| \leq 2\lambda$

$$[i\Omega, [i\Omega, S_{nk}]] = [i\Omega, [i\Omega, J_{nk}]] S_n + 2[i\Omega, J_{nk}][i\Omega, S_n] + J_{nk}[i\Omega, [i\Omega, S_n]]$$
$$= \nabla \Omega \cdot [J''_{nk} S_n + \nabla J_{nk} \otimes \nabla S_n + \nabla S_n \otimes \nabla J_{nk} + J_{nk} S''_n] \nabla \Omega + n^2 O(t^{-1-\delta})$$
$$= \nabla \Omega \cdot S''_{nk} \nabla \Omega + n^2 O(t^{-1-\delta}) \qquad (177)$$

where we used $[i\Omega, J_{nk}] = \nabla\Omega \cdot \nabla J_{nk} + nO(t^{-2})$, $[i\Omega, S_n] = \nabla\Omega \cdot \nabla S_n + nO(t^{-1})$ (and the same with the order in the dot products reversed) as well as $[i\Omega, [i\Omega, S_n]] = \nabla\Omega \cdot S_n'' \nabla\Omega + nO(t^{-1-\delta})$ (see Lemma9), $[i\Omega, [i\Omega, J_{nk}]] = \nabla\Omega \cdot J_{nk}'' \nabla\Omega + n^2 O(t^{-3})$ and $|\partial^\alpha S_n| = O(t^{1-|\alpha|})$ for $|y/t| \leq 2\lambda$ and $|\alpha| \leq 2$. In a similar way one shows that

$$\begin{aligned}
[i\Omega, \partial S_{nk}/\partial t] &= \nabla\Omega \cdot \nabla \frac{\partial S_{nk}}{\partial t} + n^2 O(t^{-2}) \\
&= \nabla \frac{\partial S_{nk}}{\partial t} \cdot \nabla\Omega + n^2 O(t^{-2}).
\end{aligned} \tag{178}$$

The lemma now follows from (176), (177) and (178) as Lemma 9 did from analogous equations. $\qquad\square$